\documentclass{jfm}
\usepackage{amsmath,amssymb,color,graphicx,overpic,color,rotating,dashrule,float,bm,multicol,setspace,enumerate}
\usepackage{epstopdf}
\usepackage{natbib}
\usepackage{hyperref}
\usepackage{mathrsfs}
\usepackage{mathtools}
\usepackage{graphicx}
\usepackage{textcomp}
\usepackage{setspace}
\usepackage{listings}
\usepackage{array}
\usepackage{comment}
\usepackage{tikz}
\usetikzlibrary{calc,arrows,automata,topaths}

\newcommand{\defeq}{\vcentcolon=}

\newcommand{\mat}[1]{{\mathsfbi{#1}}}

\newcommand{\mL}{\mat{L}}
\newcommand{\mM}{\mat{M}}

\newcommand{\bx}{{ \bm{x}}}

\newcommand{\bbf}{{ \bm{f}}}

\newcommand{\bg}{{ \bf g}}

\newcommand{\bu}{{ \bm{u}}}

\newcommand{\Laplace}{\Delta}

\DeclareMathOperator*{\argmin}{\arg\!\min}
\DeclareMathOperator*{\argmax}{\arg\!\max}

\newcommand{\ip}[2]{\left\langle #1, #2\right\rangle}

\shorttitle{The shape of resolvent modes }
\shortauthor{S. T. M. Dawson and B. J. McKeon}
\graphicspath{{FiguresPaper/}{../Figures/}{../Figures/Scalar/}{../Figures/ScalarDNS/}{../FiguresAiry/}}
\title{ On the shape of resolvent modes in wall-bounded turbulence}

\author{Scott T. M. Dawson\aff{1}
 \corresp{\email{sdawson5@iit.edu}},
  Beverley J. McKeon\aff{1}}
\affiliation{\aff{1}{Graduate Aerospace Laboratories, California Institute of Technology, Pasadena, CA 91125, USA}}

\begin{document}

\maketitle

\begin{abstract}

The resolvent formulation of the Navier--Stokes equations gives a means for the characterisation and prediction of features of  turbulent flows---such as statistics, structures and their nonlinear interactions---using the singular value decomposition of the resolvent operator based on the appropriate turbulent mean, following the framework developed by \cite{mckeon2010resolvent}.  
This work will describe a methodology for approximating leading resolvent (i.e., pseudospectral) modes for shear-driven turbulent flows using prescribed analytic functions.  We will demonstrate that these functions, which arise from the consideration of wavepacket pseudoeigenmodes of simplified linear operators \citep{trefethen2005wavepacket}, in particular give an accurate approximation of the class of nominally wall-detached modes that are centred about the critical layer. 
 Focussing in particular on modelling wall-normal vorticity modes, we present a series of simplifications to the governing equations that result in scalar differential operators that are amenable to such analysis. We validate our method on a model operator related to the Squire equation, and show for this simplified case how  wavepacket pseudomodes relate to truncated asymptotic expansions of Airy functions. 
 We demonstrate that the leading wall-normal vorticity response mode for the full Navier--Stokes equations may be accurately approximated by considering a second order scalar operator, equipped with a non-standard scalar inner product. 
 Using this method, optimal mode shapes may be found by finding the appropriate root of a polynomial. In addition, the variation in mode shape as a function of wavenumber and Reynolds number may be captured  by evolving a low dimensional differential equation in parameter space. 
    This characterisation provides a theoretical framework for understanding the origin of observed structures, and allows for rapid estimation of dominant resolvent mode characteristics without the need for operator discretisation or large numerical computations.   We explore regions of validity for this method, and in particular find that it remains accurate even when the modes have some amount of ``attachment" to the wall.  In particular, we demonstrate that the region of validity contains the regions in parameter space where large-scale and very-large-scale motions typically reside. 
 We relate these findings to classical lift-up and Orr amplification mechanisms in shear-driven flows. 
\end{abstract}

\section{Introduction}

The identification of pertinent structures that arise in the transition towards, and as coherent features within, turbulent wall-bounded flows has been the focus of much research over the past several decades.  
Qualitatively, such analysis includes identification and classification of empirically-observed structures, such as near-wall streaks \citep{kline1967structure}, hairpin structures  \citep{theodorsen1952mechanisms,head1981new} and their grouping in large-scale motions \citep{zhou1999mechanisms,guala2006large}, and very large-scale motions or superstructures \citep{kim1999very,guala2006large,hutchins2007evidence}. For more details concerning the (sometimes debated) properties, taxonomy and dynamics of such structures, see reviews such as \cite{robinson1991coherent}, \cite{smits2011high}, and \cite{jimenez2018coherent}, and references within.

On a quantitative level, a starting point for the prediction of coherent structure comes from consideration of properties of the governing equations, most typically in linearised form.  Features emergent in wall-bounded turbulent flows often bear little resemblance to modes identified from classical stability analysis \citep{drazin2004hydrodynamic}, where two-dimensional modes are predicted to be the least stable by Squire's theorem, and mean-linearised flows often have only stable eigenvalues \citep{reynolds1967stability,delAlamo2006linearAmp,cossu2009optimal}.  
Perhaps the most important realisation in the study of such linear operators is the fact that their non-normality can result in high amplification (in either the time or frequency domain), which cannot be predicted from their spectra alone \citep{boberg1988onset,butler1992optimal,reddy1993energy,trefethen1993science,schmid1994optimal,mckeon2010resolvent,bamieh2001energy,jovanovic2005componentwise,Schmid:2007,hwang2010linear,schmid2012book}.  
Operator nonnormality is responsible for both finite-time energy growth of the linear system from a from a given initial condition, and the amplification resulting from continual forcing (be it stochastic or harmonic).  Indeed, these two notions are mathematically related via the Kriess constant of the operator \citep{Schmid:2007}.

While most initial works considered laminar base flows (with one notable exception being \cite{Farrell93}, who considered mean-linearised equations using stochastic forcing), more recent developments have used similar methods for prediction of turbulent features by linearizing about states computed from turbulent data and/or models (which is most often a mean state).  
For example, \cite{delAlamo2006linearAmp} and \cite{cossu2009optimal} have demonstrated transient growth of near wall-streaks and large scale motions for turbulent channel and zero-pressure-gradient boundary layers respectively, while \cite{schoppa2002coherent} has showed that a turbulent mean with the addition of low-speed streaks can give rise to the growth of structures in the near-wall region of turbulent channel flow.
 
 The characterisation of the nonlinear Navier--Stokes equations in the frequency domain as the linear resolvent operator acting on the nonlinear ``forcing" terms, as developed by \cite{mckeon2010resolvent} has been particularly fruitful for elucidating operator-based predictions of structure in wall-bounded turbulence, including very large scale motions and their scaling \citep{mckeon2010resolvent}, and hairpin structures \citep{sharma2013resolvent}. 
 \cite{mckeon2017engine} summarizes further developments in this area. 
 Note that this approach has also been applied in other contexts, such as in the study of cavity \citep{gomez2016reduced,Qadri:PRF17}, airfoil \citep{yeh2018resolvent}, and jet \citep{garnaud13jet,jeun2016jet,towne2018spectral,schmidt2018spectral} flows. 
 
While such developments are recent and ongoing, many of the underlying physical mechanisms in shear flows have been understood for at least several decades. These include the Orr mechanism \citep{orr1907stability,jimenez2013linear}, which amplifies upstream-leaning disturbances while tilting them towards the downstream direction, and the lift-up mechanism \citep{landahl1980note,landahl1975wave}, in which wall-normal disturbances lead to large streamwise responses. 
Streamwise inhomogeneity provides an additional mechanism for amplification of disturbances \citep{Chomaz-05,hack2017algebraic}, which, as with the Orr and lift-up mechanisms, arises through different aspects of non normality of the underlying linear operator \citep{symon2018normal}.
 
Recent results concerning the pseudospectral properties of certain matrices \citep{trefethen2004wave} and linear differential operators by \citep{trefethen2005wavepacket} have revealed criteria by which pseudomodes exist that are  localised in both space and spatial frequency, and are highly amplified by the associated resolvent operator. This method of analysis has been used, for example, in the analysis of swept wing flow by \citet{obrist2010algebraically,obrist2011algebraically}.

 An alternative route to quantitatively define coherent structure in turbulent flows is to consider data collected from simulations or experiments.  The proper orthogonal decomposition \citep{Lumley1967}, which computes spatial modes of highest energy in a dataset, is perhaps the most ubiquitous such method.  The original formulation computes a set of energetically optimal modes for each temporal frequency. This method was recently shown by \cite{towne2018spectral} to give modes equivalent to resolvent response modes, under the assumption that the true forcing to the system, arising from the nonlinear term in the Navier--Stokes equations, results in uncorrelated resolvent response mode expansion coefficients.  
 Note that quantitatively correct prediction of characteristics of wall-bounded turbulence using resolvent-based approaches can be improved by considering the properties of the nonlinear forcing terms, which are dependent on second order turbulence statistics \citep{zare2017colour}, or through the addition of an eddy viscosity term to the linearised dynamics \citep{illingworth2018estimating}.  
 
 Connections may also be made between resolvent analysis and the dynamic mode decomposition \citep{schmid2008,Schmid2010DMD,rowley2009spectral}, which computes spatial modes from the eigendecomposition of an linear operator that best matches the evolution of the data between adjacent snapshots, as described in \cite{sharma2016koopman} and \cite{towne2018spectral}. 
 A widely used POD variant computes modes without filtering into temporal frequencies, which gives a basis that can be used to project the Navier--Stokes equations onto to obtain a reduced-order model \citep{berkooz1993pod,holmes2012pod}. Note in particular that this has been used to describe the dynamics of coherent structures in wall-bounded turbulent flows, such as those associated with the near-wall cycle \citep{aubry1988turbbl}. 
 Reviews of modal decomposition techniques, and their use in reduced-order modelling, may be found respectively in \cite{Taira:2017aa} and \cite{rowley2017arfm}. 

With the exception of very simple systems, the identification of mode shapes typically requires resorting to numerical methods, applied either to the discretised (and most-often linearised) governing equations, or to data collected from experiments or numerical simulations of their evolution.  
In this work, we describe methods to approximate mode shapes and amplification mechanisms without requiring the formation of discretised operators, for shear flows with arbitrary mean velocity profiles.
 In section \ref{sec:math}, we provide a review of the mathematical concepts that underpin our analysis. 
 Section \ref{sec:BL} presents the formulation and sample results for resolvent analysis of a turbulent boundary layer, and presents a sequence of simplifications to the governing equations that retain the correct features of the leading resolvent response mode.  
 In Section \ref{sec:shape}, we detail a procedure for estimating mode shapes by solving an optimisation problem using a prescribed template function, which may be derived either from approximations to exact solutions of the given operator, or from the wavepacket pseudomode theory introduced in section \ref{sec:math}. 

\section{Mathematical preliminaries: The resolvent and pseudospectra of a linearised operator}
\label{sec:math}

This section presents  material on the pseudospectral analysis of a mean-linearised system, which will provide background for the analysis in later sections. Section \ref{sec:nlres} introduces the resolvent formulation of a nonlinear system. The singular value decomposition of the resolvent and its connection with pseudospectra is discussed in Section \ref{sec:svd}. Section \ref{sec:packet} discusses the underlying theory behind the existence of wavepacket pseudospectral modes, which will be related to our subsequent analysis.

\subsection{The resolvent form of a nonlinear dynamical system}
\label{sec:nlres}
We  begin by considering a nonlinear dynamical system 
\begin{equation}
\label{eq:NLsys}
\dot\bu = \bg(\bu).
\end{equation}
Let $\bu_0$ denote the temporal mean of the state of the system, where we are assuming that the dynamics are statistically stationary. Expressing the system state as $\bu(t) = \bu_0 + \bu'(t)$, we may rewrite equation \ref{eq:NLsys} as
\begin{equation}
\label{eq:NLsys2}
\dot\bu' = \bg(\bu_0+ \bu') = \left.\frac{\partial \bg}{\partial \bu}\right|_{\bu_0}\bu'+ \bbf(\bu),
\end{equation}
where we have linearised about the mean state, but retained the full dynamics of the system with the remaining nonlinear dynamics $\bbf(\bu)$. 
Taking a Fourier transform in time, equation \ref{eq:NLsys2} may be expressed as  
$$\left(-i\omega  -  \left.{\frac{\partial \bg}{\partial \bu}}\right|_{\bu_0}\right) \hat\bu' = \widehat{\bbf(\bu)},$$
where $\hat\cdot$ denotes a Fourier-transformed function.
 The mean-subtracted state of the system may then be expressed by
\begin{equation}
\label{eq:NLsysRes}
\hat\bu' =\left(-i\omega  -  \left.\frac{\partial \bg}{\partial \bu}\right|_{\bu_0}\right)^{-1} \widehat{\bbf(\bu')} = \mathcal{H}_\omega\widehat{\bbf(\bu')},
\end{equation}
where we refer to $\mathcal{H}_\omega$ as the associated resolvent operator for this system for a given frequency $\omega$, where we are assuming here that this inverse exists (i.e., that $i\omega$ is not an eigenvalue of $\left.\frac{\partial \bg}{\partial \bu}\right|_{\bu_0}$). It is important to note that we have not made any approximations to the nonlinear system at this point, and have only made the assumption that the system is statistically stationary in time with a well-defined mean. 

\subsection{The singular value decomposition of the resolvent operator}
\label{sec:svd}
From equation \ref{eq:NLsysRes}, the properties of the mean-subtracted state $\bu'$ will depend both on the nature of the nonlinear term $\hat\bbf$, and the properties of the linear operator $\mathcal{H}_\omega = (-i\omega + \mathcal{L})^{-1}$, where following on from section \ref{sec:nlres}, we let $\mathcal{L} = -  \left.\frac{\partial \bg}{\partial \bu}\right|_{\bu_0}$. 
In particular, if $\mathcal{H}_\omega$ amplifies a small number of directions, or ``modes" to a much larger degree than all others, then so long as these directions are excited to some extent by $\hat\bbf$, this can allow prediction of the dominant features of $\bu'$ by studying only $\mathcal{H}_\omega$. 

More precisely, we consider the singular value decomposition  of the resolvent operator
\begin{equation}
\label{eq:svdinf}
\mathcal{H}_\omega = \sum_{j = 1}^\infty \sigma_j \psi_j\phi_j^*, 
\end{equation}
with $\sigma_k > \sigma_{k+1}$ for all $k$. 
For the remainder of this work, we will consider $\mathcal{H}_\omega$ to be a discretised operator, with a corresponding singular value decomposition (SVD) defined in the same manner as \ref{eq:svdinf}, but with a finite sum of modes. 
The SVD of a linear operator requires the definition of an inner product (or more precisely, inner products on both the input and output spaces), which prescribes how the adjoint is prescribed and computed (and also
induces the norms used in defining various properties of the SVD). For discrete operators, we may characterise an inner product by a positive definite weighting matrix $\mM$ for which
\begin{equation}
\ip{ \bx_1}{ \bx_2} = \bx_1^{\bar T} \mM \bx_2, 
\end{equation}
where $\cdot^{\bar T}$ denotes the conjugate transpose.  
If $\mM$ characterises the inner product on the spaces containing both the forcing and response functions of a finite-dimensional linear operator $\mL$ (which we consider the discretisation of a linear operator $\mathcal{L}$), the adjoint $\mL^*$ satisfies
 $ \ip{ \mL \bx_1}{ \bx_2} = \ip{ \bx_1}{ \mL^*\bx_2}$, from which one may show that  $\mL^* = \mM^{-1} \mL^{\bar T} \mM$.
  Throughout this paper we will largely work with (infinite-dimensional) operators acting on a continuous domain, with the understanding that numerical computation of the SVD is performed on a finite-dimensional discrete approximation. 

For some aspects of this work, it will be convenient to think of the leading singular values and vectors of $\mathcal{H}_\omega $ as defined in the following manner:
\begin{align} 
\sigma_1 & = \max_{\|\phi\| = 1}\| \mathcal{H}_\omega  \phi \| \defeq \| \mathcal{H}_\omega  \|,\\
\psi_1 &= \argmax_{\psi :\  \|\psi\| = 1}\|  \mathcal{H}^*_\omega  \psi \|, \\
\phi_1 &= \argmax_{\phi :\ \|\phi\| = 1}\|  \mathcal{H}_\omega\phi\|,
\end{align} 
where here and throughout we use the spectral norm when taking the norm of an operator.
Note that we also have the relationship $ \psi_1 = \sigma_1^{-1}  \mathcal{H}_\omega  \phi_1,$ which may also be rearranged as  $ \phi_1 = \sigma_1  \mathcal{H}^{-1}_\omega  \psi_1$. In other words, $\phi_1$ gives the shape of the forcing which gives rise to the largest amplification (a factor of $\sigma_1$) upon the application of the operator $\mathcal{H}_\omega$, the result of which is the (scaled by $\sigma_1$) response mode $\psi_1$. 
It will also be useful to recognise that these singular values and vectors are related to the smallest singular values and vectors of 
$\mathcal{H}_\omega^{-1}= \left(-i\omega  + \mathcal{L}\right)$,
 by
\begin{align} 
\label{eq:Ressig}
\sigma_1 &= \left(\min_{\|\psi\| = 1}\| \mathcal{H}^{-1}_\omega  \psi \|\right)^{-1},  \\
\psi_1 &= \argmin_{\psi :\  \|\psi\| = 1}\|  \mathcal{H}^{-1}_\omega  \psi \|, \label{eq:minpsi}\\
\phi_1 &= \argmin_{\phi :\ \|\phi\| = 1}\|  \left(\mathcal{H}_\omega^{-1}\right)^*\phi\|.\label{eq:minphi}
\end{align} 
In this work, we will make use of these definitions, that allow us to  define leading singular values and vectors as solutions to an optimisation problem.
Roughly speaking, we will search for analytic functions that become as small as possible upon the action of $(-i\omega+\mathcal{L}) = \mathcal{H}^{-1}_\omega$, 
and reason from equation \ref{eq:minpsi} they will be close approximations to leading resolvent response modes. 

 These ideas may be  formalised as follows.
For a linear operator $\mathcal{L}$ and some $\epsilon > 0$, define the $\epsilon$-pseudospectrum as a set $\Lambda_\epsilon(\mathcal{L}) \subset \mathbb{C}$ satisfying 
\begin{equation}
\Lambda_\epsilon(\mathcal{L})  = \left\{ z : (\mathcal{L} + \mathcal{E})\bu = z\bu, \text{ for some } \bu \text{ and  } \mathcal{E}, \text{ with } \|\mathcal{E}\| \leq \epsilon \right\}.
\end{equation}
Here $\mathcal{E}$ is an operator mapping between the same spaces as $\mathcal{L}$.  
Note in particular that we have the equivalent definition of pseudospectra based on the norm of the resolvent:
$$z \in \Lambda_\epsilon(\mathcal{L})\backslash\Lambda(\mathcal{L}) \iff \|(-z + \mathcal{L})^{-1}\| \geq \epsilon^{-1},$$
where here the $\backslash$ operator refers to set exclusion.  
In particular, for any $z \in \mathbb{C}$, {$\epsilon_z = \min \{ \epsilon : z \in \Lambda_\epsilon(\mathcal{L}) \} = \sigma_1^{-1}(z),$} where $\sigma_1(z)$ is the largest singular value of  $(-z +\mathcal{L})^{-1}$.
Furthermore, the corresponding pseudoeigenvector $ \bu$ satisfying $ (\mathcal{L} + \mathcal{E})\bu = z\bu$, for some $\mathcal{E}$ with $\|\mathcal{E}\| = \epsilon_z$ is the left singular vector of $(-z + \mathcal{L})^{-1}$ corresponding to $\sigma_1(z)$. 
Note that the presence of the operator norm in the definition means that, unlike when looking at the spectrum, the definition of pseudospectra requires that we work in a vector space equipped with a norm, which the $\epsilon$-pseudospectrum associated with an operator is dependent upon.

\subsection{Conditions for the existence of wavepacket resolvent modes}
\label{sec:packet}
 Here we briefly describe the conditions under which a linear differential operator permits wavepacket pseudomodes corresponding to small $\epsilon$ (or equivalently, large resolvent norm).  More complete description of this phenomenon, as well as the related proofs, may be found in \cite{trefethen2005spectra} and \cite{trefethen2005wavepacket}. 
 For some (small) parameter $h > 0$, we may define a family of scaled differential operators
 \begin{equation}
 \label{eq:genDiffOp}
  \mathcal{D}^j_h = (i h)^j \frac{d^j}{(dy)^j},
  \end{equation}
 which we assume to act on a closed finite interval. We may then assemble any arbitrary $n$-th order differential operator $\mathcal{L}_h$ acting on a scalar variable $ u(y)$ by
 \begin{equation}
 \label{eq:packetD}
( \mathcal{L}_h u)(y) = \sum_{j = 0}^n c_j(y)  \mathcal{D}^j_h u (y).
 \end{equation}
 Defining a test function $v_h(y) = \exp(-i k y/h)$ for a complex scalar $k$, we have
 \begin{equation}
 (  \mathcal{L}_h v_h)(y) = \sum_{j = 0}^n c_j(y) k^j v_h (y) = f(y,k) v_h(y),
 \end{equation}
 where we refer to $f(y,k)$ as the symbol corresponding to the family of differential operators $\mathcal{L}_h$.
 We will consider potential pseudoeigenvalues $\lambda  = f(y_*,k_*)$. 
  The symbol $f(y,k)$ is said to satisfy the twist condition for real $k$ if we have
  \begin{equation}
  \label{eq:twist}
  \text{Im} \left( \frac{\partial_y f }{\partial_k f}\right) > 0.
  \end{equation}
  When the twist condition is satisfied for some $(y_*,k_*)$, one it can be a pseudomode $\psi(y; y_*,k_*)$ (of unit norm) that has phase variation matching $v_h$ close to $y_*$, satisfying both
  \begin{equation}
  \| (-\lambda + \mathcal{L}_h)  \psi(y; y_*,k_*)\| \leq M^{-1/h}
  \end{equation}
 for some $M > 1$, 
  and also that 
 \begin{equation}
 \label{eq:Loc}
 \frac{|\psi(y; y_*,k_*)|}{\max|\psi(y; y_*,k_*)|} \leq C \exp\left(-b(y-y_*)^2/h\right),
 \end{equation}
 for some $b,C  > 0$. 
 In other words, we have modes that are spatially localised near $y_*$, have localised spatial frequency $k_*$, and are ``asymptotically good" pseudoeigenfuctions. Note that the region of $\mathbb{C}$ where the twist condition is satisfied is independent of the choice of norm. 
 Note also that this theory developed in  \cite{trefethen2005wavepacket} builds upon earlier observations of certain classes of equations by \cite{davies1999pseudo,davies1999semi}.
 More generally, these ideas are closely related to classical WKBJ expansions, 
  which have also been utilised recently by \cite{leonard2016afmc} in the context of studying approximate inviscid solutions for channel flow.
  
  In this work, using these ideas as justification and inspiration, for a given temporal frequency (where $\lambda = i \omega$), we will assume that the leading resolvent response mode may be closely approximated by a function that is localised in the wall-normal direction, both spatially (as for functions satisfying equation \ref{eq:Loc}) and in the frequency domain (as is the case for the test functions $v_h$). When this assumption is justified, it allows for a reformulation of resolvent analysis in terms of finding the spatial width and frequency of a template function $\psi$ that minimises $\| (-\lambda + \mathcal{L})\psi\|$, and thus are close to the true leading resolvent response mode (from equation \ref{eq:minpsi}).

\section{The behaviour of leading resolvent modes in wall-bounded turbulence }
\label{sec:BL}
This section will apply the concepts introduced in section \ref{sec:math} to explore methods by which the shape of resolvent modes may be approximated. In this section and elsewhere, we will largely focus on a boundary layer configuration, which is assumed to be approximately homogeneous in the streamwise direction (as well as being homogeneous in the spanwise direction), though the analysis holds for any (approximately) parallel shear-driven turbulent flow.

In section \ref{sec:form} we introduce a resolvent formulation of the Navier--Stokes equations. The setup is similar to that developed in \cite{mckeon2010resolvent}, though following more closely the formulation used in \cite{rosenberg2018efficient}. 
Following this, in section \ref{sec:simp} we will present a sequence of simplifications to the governing equations, that will render them amenable to application of wavepacket pseudomode theory. 
We will additionally present some sample results to motivate these developments, as well as those later presented in section \ref{sec:shape}.

\subsection{A resolvent formulation of the Navier--Stokes equations}
\label{sec:form}
We will restrict attention to flows which are (approximately) homogenous in the streamwise $(x)$ and spanwise $(z)$ directions, with a mean streamwise velocity $(u,v,w) = (U,0,0)$ that varies in the wall normal $(y)$ direction. 
We will consider the incompressible Navier--Stokes equations in (wall-normal) velocity-vorticity form, and will take Fourier transforms in the streamwise and spanwise directions, with wavenumbers given by $k_x$ and $k_z$. Applying the procedure detailed in section \ref{sec:nlres} gives 
\begin{equation}
\label{eq:ResolventNSE}
\begin{pmatrix}
\hat v \\
\hat \eta \end{pmatrix} = 
\underbrace{\begin{pmatrix}
-i\omega  + \Laplace^{-1} \mathcal{L}_{os} & 0 \\
i k_z U_y &-i\omega  + \mathcal{L}_{sq}
\end{pmatrix}^{-1}}_{\mathcal{H}_{\bm{k}}}
\begin{pmatrix}
\hat f_v \\
\hat f_\eta \end{pmatrix}.
\end{equation}
 where $\hat v$ and $\hat \eta = i k_z \hat u - i k_x \hat w$ are the Fourier-transformed wall-normal velocity and vorticity fields, $U_y$ is the wall-normal gradient of the streamwise velocity, and 
 the Laplacian $\Laplace = \partial_{yy} - k_\perp^2$, with $k_\perp^2 = k_x^2 +k_z^2$. The Orr-Sommerfeld (OS) and Squire (SQ) operators are given respectively by
\begin{align}
\mathcal{L}_{os} &= i k_x U \Laplace - i k_x U_{yy} - \frac{1}{Re} \Laplace^2, \label{eq:os} \\
\mathcal{L}_{sq} &= ik_x U - \frac{1}{\Rey}\Laplace. \label{eq:sq}
\end{align}
Further details concerning the equivalence of this formulation to one using primitive variables (in particular the implicit restriction of the forcing to the solenoidal component) is given in \cite{rosenberg2018efficient}. The resolvent operator $\mathcal{H}_{\bm{k}}$ is now parametrised by the spatiotemporal wavevector $\bm{k} = (\omega,k_x,k_z)$.

The SVD of the resolvent operator is computed using an inner product that is proportional to kinetic energy for a given set of spatial wavenumbers \citep{gustavsson1986excitation,butler1992optimal}, defined by
\begin{equation}
\label{eq:IP}
\ip{(\hat v_1,\hat\eta_1)}{(\hat v_2,\hat\eta_2)} = \frac{1}{k^2} \int_y \left(- \bar {\hat v}_1\Laplace \hat v_2 + \bar{\hat\eta}_1\hat\eta_2\right)dy,
\end{equation}
where the overbar denotes complex conjugation.

In later sections, we will consider the following two scalar inner products arising from the components of \ref{eq:IP}:
\begin{align}
\label{eq:IPscalar}
\ip{\hat z_1}{\hat z_2} &= \frac{1}{k_\perp^2} \int_y \bar{\hat z}_1\hat z_2dy, \\
\label{eq:IPscalarLap}
\ip{\hat z_1}{\hat z_2}_\Laplace &= -\frac{1}{k_\perp^2} \int_y  \bar {\hat z}_1\Laplace \hat z_2 dy.
\end{align}
That is to say, when considering scalar operators, we will assume that the inner product is the ``standard" one unless using a $\Laplace$-subscript. We will also refer to $\ip{\cdot}{\cdot}_\Laplace $ as the Laplacian inner product.

\begin{figure}
 \centering {
\includegraphics[width= 0.32\textwidth]{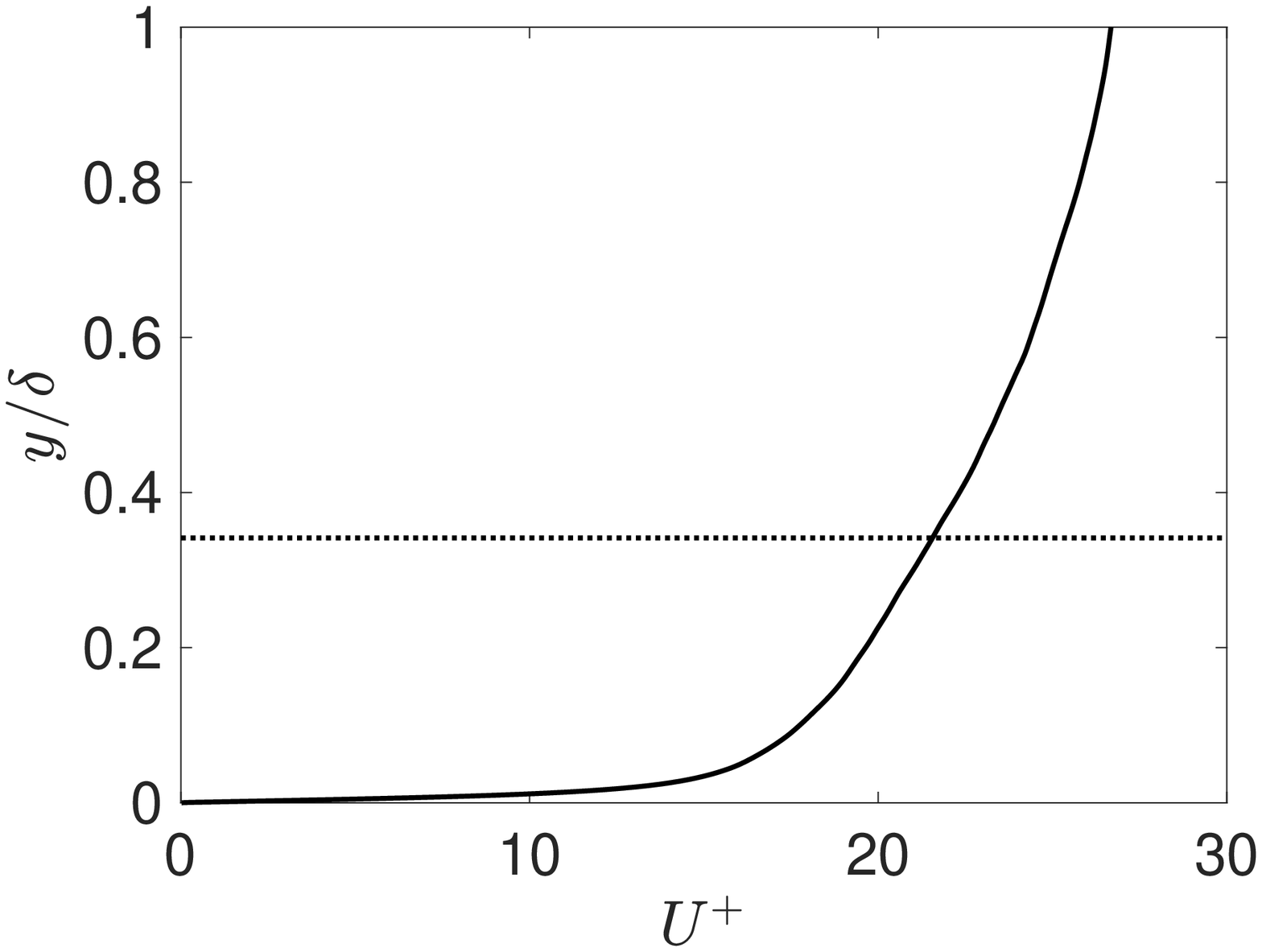}
\includegraphics[width= 0.32\textwidth]{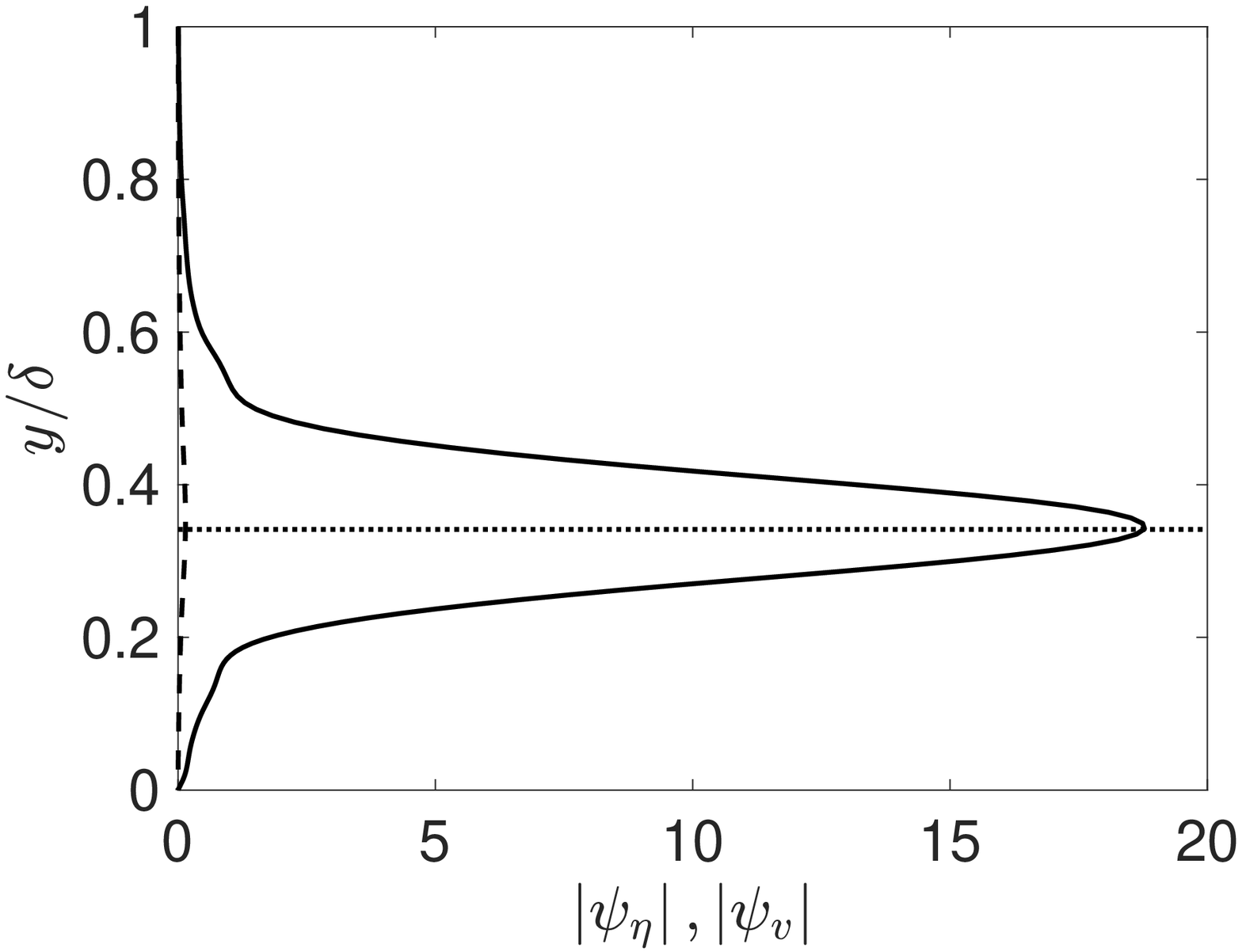} 
\includegraphics[width= 0.32\textwidth]{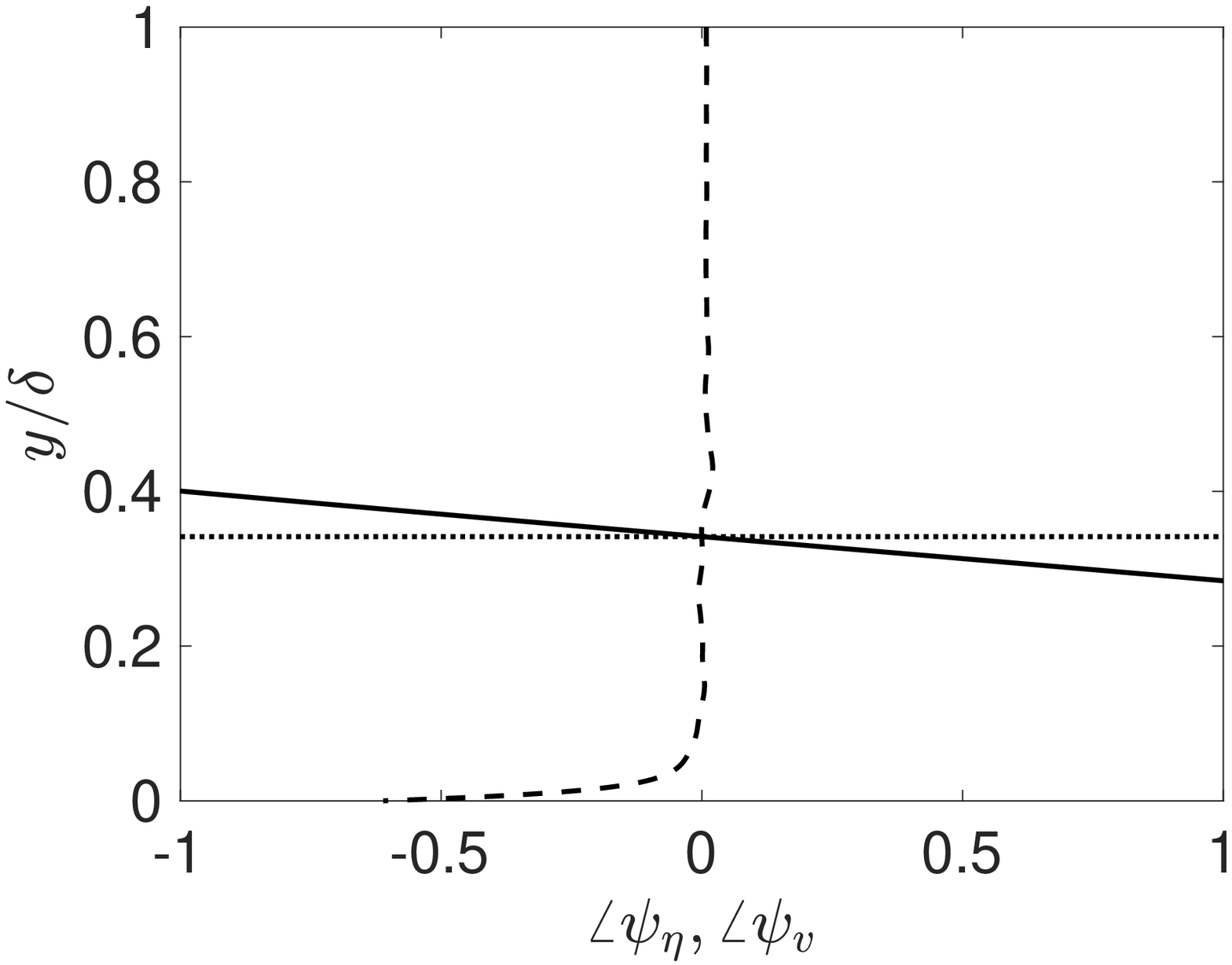} 
}
\caption{ (a) Mean velocity profile for a turbulent boundary layer at $Re_\tau = 900$ ($-$) and critical layer location corresponding to a wavespeed $c = 0.8U_\infty$ ($\cdot\cdot$). (b) Amplitude and (c) phase of wall normal velocity ($-\cdot$) and vorticity (-) for leading resolvent response modes with $k_x  = \pi/2 $, $k_z = 2\pi$  and $c = 0.8U_\infty$.}
\label{fig:Modes1} 
\end{figure}

We will now present some sample results that will motivate many of the developments in the remainder of this paper. 
We consider a turbulent boundary layer with friction Reynolds number $\Rey_\tau = 900$. Mean velocity profiles are obtained from the DNS data of \citet{wu2017transitional}, by averaging data at a single streamwise location. 
Figure \ref{fig:Modes1} shows leading resolvent response mode shapes for this system. The wavenumbers $k_x  = \pi/2 $, and $k_z = 2\pi$  and wavespeed ($c = 0.8U_\infty$)  have been chosen to be consistent with the typical size of large-scale motions in zero-pressure-gradient boundary layers (e.g.,~\cite{monty2009comparison,kovasznay1970large,cantwell1981organized,saxton2017coherent}).  
Here and throughout, the resolvent operator is discretised using a Chebyshev collocation method, utilising the toolbox of \citet{weideman2000matlab}.

We observe that the mode is dominated by the wall-normal vorticity component, which is centred on and localised around the critical layer, and has an approximately linear variation in phase within this region.  These observations, which are typical of modes that are ``detached" from the wall, suggests that this numerically computed mode resembles a wavepacket mode as described in section \ref{sec:packet}. 
Here and throughout, we consider a mode to be detached from the wall if its shape is not substantially effected by the wall's presence. 
In section \ref{sec:shape} we will seek to predict the shape of this mode without explicitly computing an SVD (or indeed, without explicitly forming a discretised resolvent operator). 
Making this procedure tractable, however, will require simplification of the governing equations, which are described in section \ref{sec:simp}. 

\subsection{Simplifications to the Navier--Stokes operator for resolvent mode approximation}
\label{sec:simp}
\begin{figure}
 \centering {
\includegraphics[width= 0.45\textwidth]{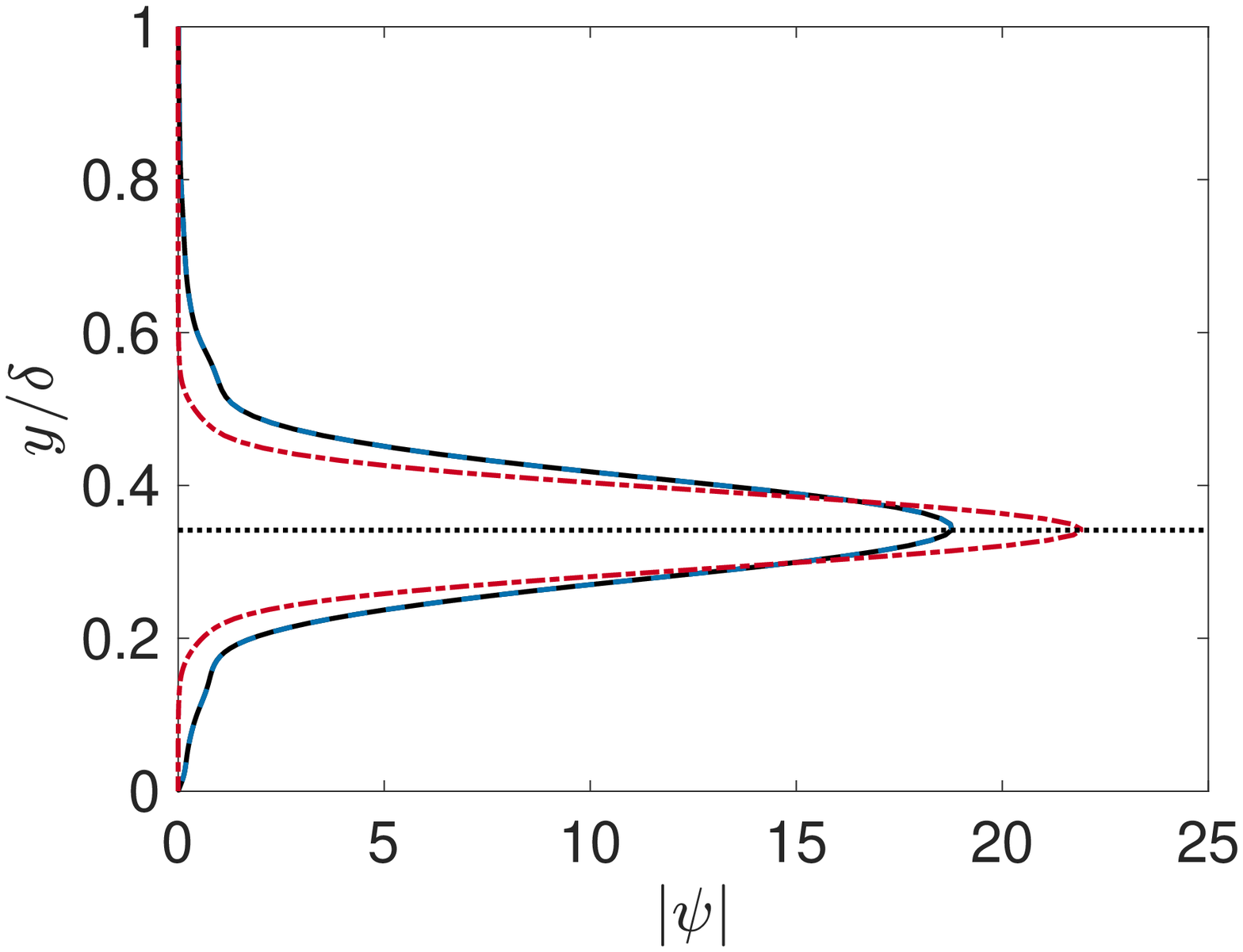} 
\includegraphics[width= 0.45\textwidth]{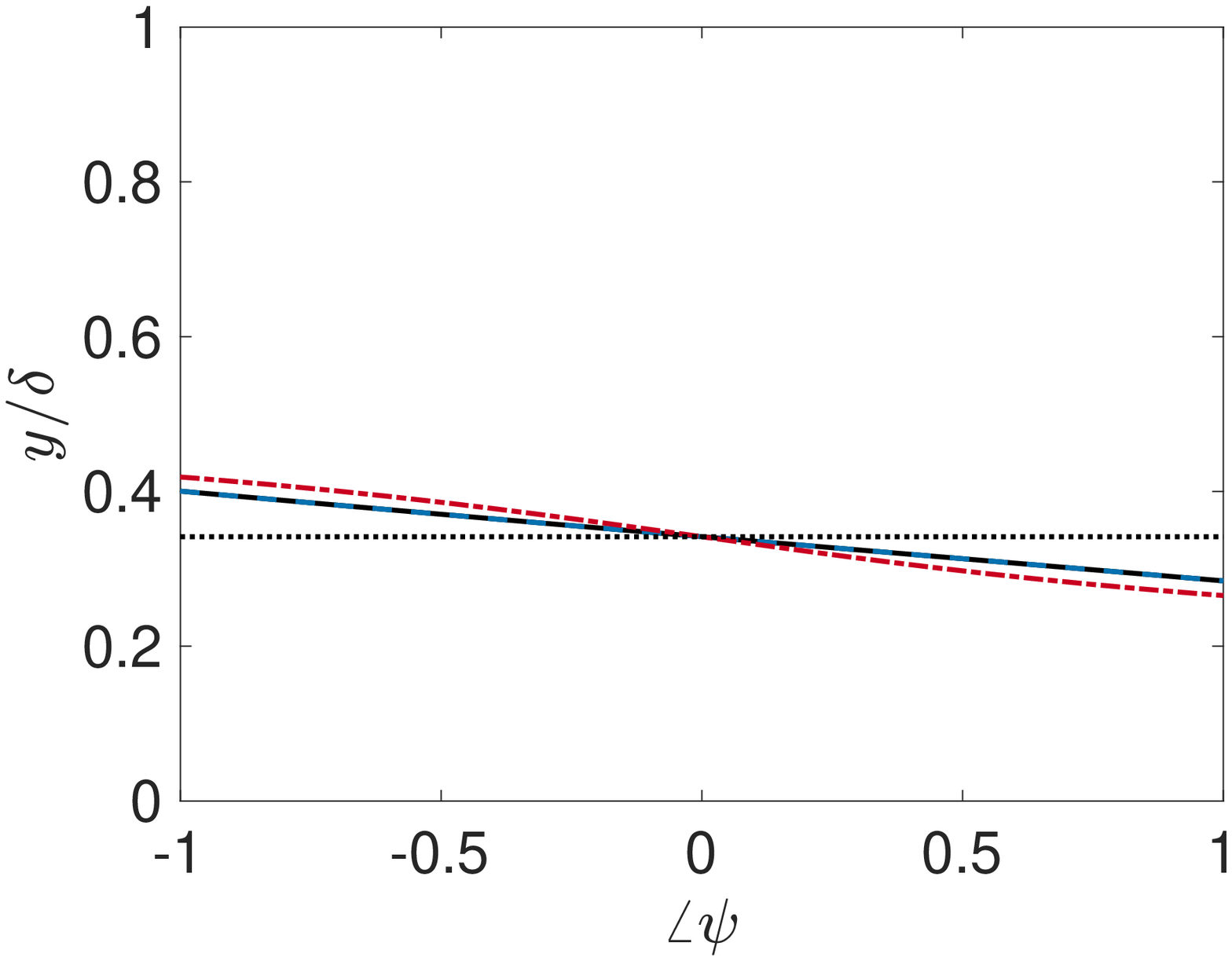} 
}
\caption{ (a) Amplitude and (b) phase of wall normal vorticity response mode computed from full system ($-$), SQ subsystem ($-\cdot$), and OS subsystem ($--$), with the same parameters as in figure \ref{fig:Modes1}.}
\label{fig:Modes2}
\end{figure}

Following \cite{rosenberg2018efficient}, rather than studying the full system we may start by separating the resolvent modes into those being forced by $f_v$ and $f_\eta$ separately. 
In particular, defining scalar resolvent operators
\begin{align}
\mathcal{H}_{vv} &= (-i\omega  + \Laplace^{-1} \mathcal{L}_{os})^{-1}, \\
\mathcal{H}_{\eta\eta} &= (-i\omega  +  \mathcal{L}_{sq})^{-1},
\end{align}
the Orr-Sommerfeld (OS) and Squire (SQ) modes may be computed by taking SVD's of the reduced resolvent operators given as
\begin{align}
\label{eq:ResolventOSSQ}
\begin{pmatrix}
\hat v \\
\hat \eta_{os} \end{pmatrix} &= 
\underbrace{\begin{pmatrix}
\mathcal{H}_{vv}  & 0 \\
i k_z \mathcal{H}_{\eta\eta} U_y \mathcal{H}_{vv}  & 0 
\end{pmatrix}}_{\mathcal{H}_{os,\bm{k}}}
\begin{pmatrix}
\hat g_v \\
 0 \end{pmatrix}, \\
\begin{pmatrix}
 0 \\
\hat \eta_{sq} \end{pmatrix} &= 
\underbrace{\begin{pmatrix}
0 & 0 \\
0 & \mathcal{H}_{\eta\eta}
\end{pmatrix}}_{\mathcal{H}_{sq,\bm{k}}}
\begin{pmatrix}
 0 \\
\hat g_\eta \end{pmatrix}.
\end{align}
While the modes computed from different subsystems are no longer orthogonal, \cite{rosenberg2018efficient} demonstrated they can provide a basis that better captures dynamical features of the system.  Our interest in this decomposition is primarily concerned with using this decomposition to simplify the required analysis.
Figure \ref{fig:Modes2} shows the $\eta$-component of the response modes of the OS and SQ subsystems for the same parameters considered in figure \ref{fig:Modes1}. We observe in particular that the OS operator gives the same mode shape as the full system, with the SQ mode also sharing the same qualitative characteristics. This suggests that the ability to predict and understand the shape of the vorticity response of the full system can be reduced to studying only the OS subsystem.

In order to make the ensuing analysis more tractable, we will require additional simplifications.
Note first that the components of the OS and SQ sub-operators may be written in scalar form as
\begin{align}
\hat v &= {H}_{vv} \hat g_v \\
\label{eq:scalarOS}
\hat\eta_{os} &= \mathcal{H}_{\eta v} \hat g_v, \\ 
\hat\eta_{sq} &= \mathcal{H}_{\eta\eta}\hat g_\eta,
\end{align}
where $\mathcal{H}_{\eta v} = (i k_z \mathcal{H}_{\eta\eta} U_y \mathcal{H}_{vv})$ is the off-diagonal term in $\mathcal{H}_{os,\bm{k}}$ in equation \ref{eq:ResolventOSSQ}.

\begin{figure}
 \centering {
\includegraphics[width= 0.45\textwidth]{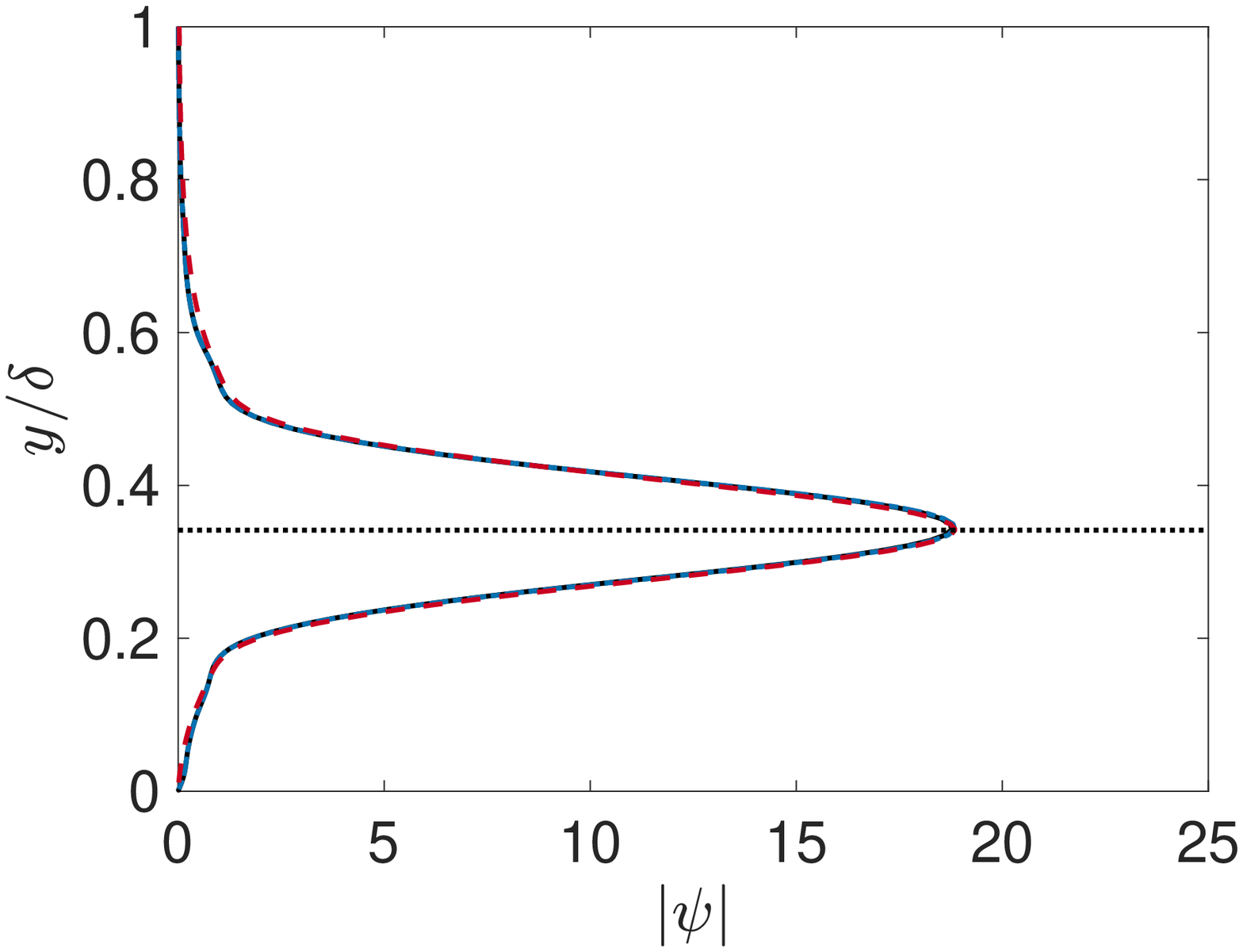} 
\includegraphics[width= 0.45\textwidth]{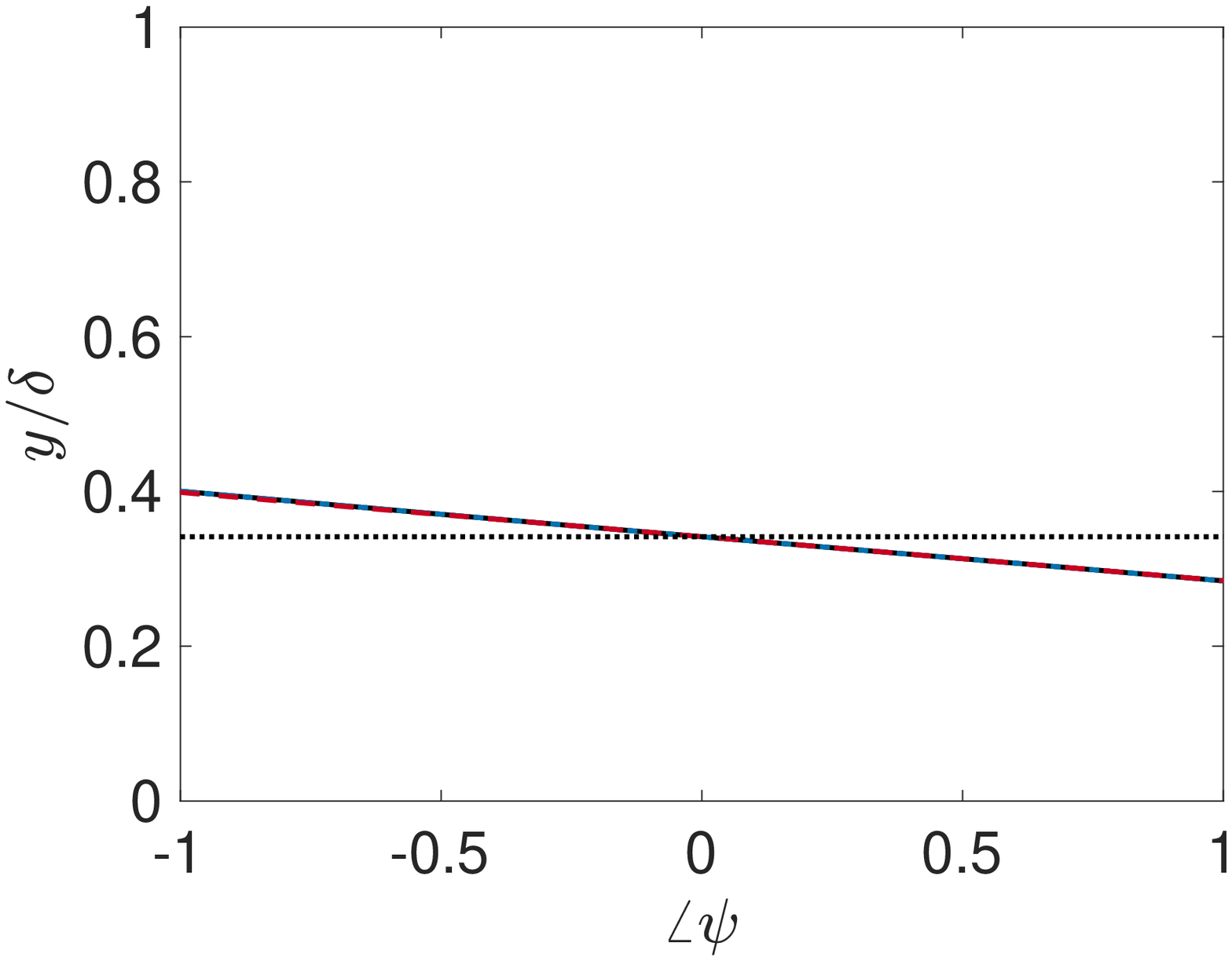} 
}
\caption{ (a) Amplitude and (b) phase of wall normal vorticity response mode computed from the OS subsystem ($-$), the scalar off-diagonal term $\mathcal{H}_{\eta v}$ of OS subsystem ($-\cdot$), and its simplified scalar approximation computed using $\mathcal{H}_{\eta\eta}$ with a Laplacian inner product ($--$), all with the same parameters as in figure \ref{fig:Modes1}.}
\label{fig:Modes3}
\end{figure}

When considering the SVD of these subsystems, it is important that we use the appropriate inner product on both the input and output spaces. 
Since the response is energetically dominated by wall-normal vorticity, it is reasonable that a close approximation to the $\eta$ response of the full system (which itself is approximated by the OS response) can be obtained just by considering the scalar operator in equation \ref{eq:scalarOS}. 
Figure \ref{fig:Modes3} shows that this assumption is indeed valid for the sample parameters considered in this section.
As well as simplifying the analysis, reducing the full resolvent operator to a scalar operator brings us closer to being able to apply the wavepacket pseudomode theory discussed in section \ref{sec:packet}.
An additional requirement for the direct application of this theory is that the operator under consideration be representable as a differential operator in the form of equation \ref{eq:packetD}. The operator in equation \ref{eq:scalarOS} does not satisfy this property, as the OS component contains the inverse Laplacian. 

Motivated by this, we now present an approximating simplification to \ref{eq:packetD}, which we believe to be novel. 
This development will take advantage of the similarity between the OS and SQ operators, and will consider variations of the ``standard" inner products discussed in section \ref{sec:form}. 
We start by taking advantage of the localised nature of the resolvent response modes, and assume that the mean velocity profile may be linearised about the critical layer location, $y_c$. Under this assumption,  we have 
\begin{equation}
\label{eq:Hvvlin}
\mathcal{H}_{vv} = \Laplace^{-1} \mathcal{H}_{\eta\eta} \Laplace.
\end{equation}
This means that the operator that we are attempting to simplify is 
\begin{align}
 \mathcal{H}^{{Lin}}_{\eta v } &= (ik_z U_{y_c}) \mathcal{H}_{\eta\eta}\mathcal{H}_{vv} \\
& = (ik_z U_{y_c}) \mathcal{H}_{\eta\eta} \Laplace^{-1} \mathcal{H}_{\eta\eta} \Laplace,
\end{align}
where $U_{y_c}$ is the wall-normal gradient in the mean velocity profile at the critical layer location, $y_c$, and the $Lin$ superscript denotes that the mean velocity profile has been linearised about $y_c$. Figures \ref{fig:Modes2} and \ref{fig:Modes3} show that the resolvent response modes of $\mathcal{H}_{\eta\eta} $ do not quantitatively match those of $ \mathcal{H}_{\eta v } $. This is because the leading forcing mode of $\mathcal{H}_{\eta\eta} $ does not coincide with the mode most amplified by $\mathcal{H}_{vv}$.  
Given that the leading resolvent modes are dependent on the choice of inner product, it is reasonable to ask if a different choice of inner product could be used in the SVD of $\mathcal{H}_{\eta\eta} $, such that the leading forcing mode approximately coincides with the leading response of  $\mathcal{H}_{vv}$. Physically, what we are seeking to do is to find an inner product weighting that will most amplify the lift-up mechanism (at the expense of optimising the Orr mechanism in isolation). 
 
 Let $\mathcal{H}^*_{\eta\eta}$ denote the adjoint of $\mathcal{H}_{\eta\eta}$ with respect to the standard scalar inner product, equation \ref{eq:IPscalar}.  If we instead consider the ``Laplacian" inner product, we obtain
$$ \mathcal{H}^{*,\Laplace}_{\eta\eta} = \Laplace^{-1} \mathcal{H}^*_{\eta\eta} \Laplace.$$
This has the same form as equation \ref{eq:Hvvlin}, except with the (standard) adjoint of $\mathcal{H}_{\eta\eta}$, which amounts to taking the complex conjugate of the critical layer term. 
 Note also that $\mathcal{H}_{\eta\eta}$ with the regular inner product has leading forcing and response modes (i.e., right and left singular vectors) that differ only in the direction of the phase change. We also then have 
 $$ \left(\mathcal{H}^{*}_{\eta\eta} \right)^{*,\Laplace} = \Laplace^{-1} \mathcal{H}_{\eta\eta} \Laplace,$$
 which is identical to equation \ref{eq:Hvvlin}, meaning in particular that both share the same resolvent forcing and response modes. 
 If we are to weight this operator explicitly to compute an SVD in the Laplacian scalar norm, we obtain
 \begin{equation}
\mathcal{H}^{W,\Laplace}_{vv} =\left[ \left(\mathcal{H}^{*}_{\eta\eta} \right)^{*,\Laplace}\right]^{W,\Laplace}  = \Laplace^{-1/2} \mathcal{H}_{\eta\eta} \Laplace^{1/2}.
\end{equation}
 
 Therefore, computing this SVD should give a leading forcing mode for  $\mathcal{H}_{\eta\eta}$ that coincides with the leading response mode of $\mathcal{H}_{vv}$, thus minimising the ``projection loss" of the total amplification.  
 In effect, we are modifying the inner product used for  $\mathcal{H}_{\eta\eta}$ such that it's leading forcing mode aligns with the leading response mode of $\mathcal{H}_{vv}$. 
This analysis therefore relies on the assumption that the total amplification is dominated by the lift-up, rather than the Orr, mechanism.
  With this assumption, this trick allows us to compute optimal response mode shapes by only considering the Squire operator, which is a differential operator of the general form given in equation \ref{eq:genDiffOp}, rather than the full OS operator, which is not.  It will be shown later that this assumption holds except for large $k_\perp^2$, in which case the Laplacian approaches a constant, and so $\mathcal{H}_{\eta\eta}$ and $\mathcal{H}_{vv}$  converge to the same operator.  
 Figure \ref{fig:Modes3} shows that this modification to the inner product allows us to closely match the wall-normal vorticity component of the response mode of the full Navier--Stokes system, for the sample parameters chosen.  We will discuss reasons for the success of this approximation, and in particular the relative balance between optimising the Orr and lift-up mechanisms, in section \ref{sec:shape}. 
 
\section{Predicting the shape of  resolvent modes}
\label{sec:shape}
This section will present a method that allows for the prediction of resolvent mode shapes, focusing in particular on the wall-normal vorticity component. 
The main idea will be to assume the existence of a mode that is localised in both wall-normal location, and wall-normal spatial frequency, and then find the parameters which result from maximum amplification of the resolvent operator, or equivalently, minimisation of the action of its inverse, on a given function. We will start by predicting the shape of a model operator (which is closely related to the Squire operator) in sections \ref{sec:Airy}, which relates mode shapes to asymptotic expansions of Airy functions, and \ref{sec:SquirePred}, which approximates mode shapes under the assumption of the existence of wavepacket pseudomodes. This latter approach is then extended to consider the shape of modes for the full incompressible Navier--Stokes system in section \ref{sec:OSpred}, using the approximations presented in section \ref{sec:simp}. We will validate our method first on laminar Couette flow with a linear velocity profile in section \ref{sec:OSpred}, before returning to the turbulent boundary layer configuration in section \ref{sec:BLpred}.

\subsection{ Relationship between wavepacket resolvent modes and Airy functions}
\label{sec:Airy}
Before applying wavepacket pseudomode theory more generally, in this section we focus on a model (Airy) operator that is equivalent to the Squire operator with a linearised mean velocity profile. 
This analysis will show how numerically computed resolvent response modes relate to exact analytical solutions to simplified governing equations. 
 Linearising the mean velocity profile about the critical layer, the Squire equation reduces to
\begin{align}
(-i\omega + \mathcal{L}^{Lin}_{sq}) &=  ik_x U_{y_c} (y-y_c)  - {\Rey}^{-1}\Laplace.
\end{align} 
For clarity, we let $ R =k_x U_{y_c} Re$, and consider the closely-related operator
\begin{align}
\mathcal{T}  = (i k_xU_{ y_c})^{-1}(-i\omega + \mathcal{L}^{Lin}_{sq}) &= - (iR)^{-1}\frac{d^2}{d y^2} + \left(y - y_c +  (iR)^{-1} k_\perp^2\right)\nonumber\\
& =  - (iR)^{-1}\frac{d^2}{d y^2} + (y -y_c -\omega_c i )\nonumber\\
& =  - (iR)^{-1}\frac{d^2}{d y^2} + (y -\lambda),
\end{align}
where $\lambda = y_c +\omega_c i$, and $\omega_c = R^{-1}{k_\perp^2}$, and we are assuming that $k_x \neq 0$. 
Note that we are generally interested in performing resolvent analysis along the axis of neutral stability (i.e., for real-valued frequencies with no growth or decay), but here we are incorporating the viscous dissipation term $\omega_c$ into $\lambda$, so in general $\lambda$ is complex for this analysis. 
$\mathcal{T}$ is a complex Airy-type operator, and is identical to the operator considered, for example, in \cite{reddy1993pseudospectra} as a model for studying the pseudospectra of the Orr-Sommerfeld operator. 
The equation $\mathcal{T}u = 0$ has a general solution that can be expressed by two independent Airy functions, such as 
$$u(y) = c_1 Ai[z] +  c_2 Ai[e^{2\pi i/3} z],$$
with
\begin{equation}
\label{eq:zy}
z = (iR)^{1/3}\left(y-\lambda\right).
\end{equation}
Note that distinct solutions are obtained through the choice of contour in the complex plane when applying standard Laplace transform methods. 

\begin{figure}
 \centering {
\includegraphics[width= 0.45\textwidth]{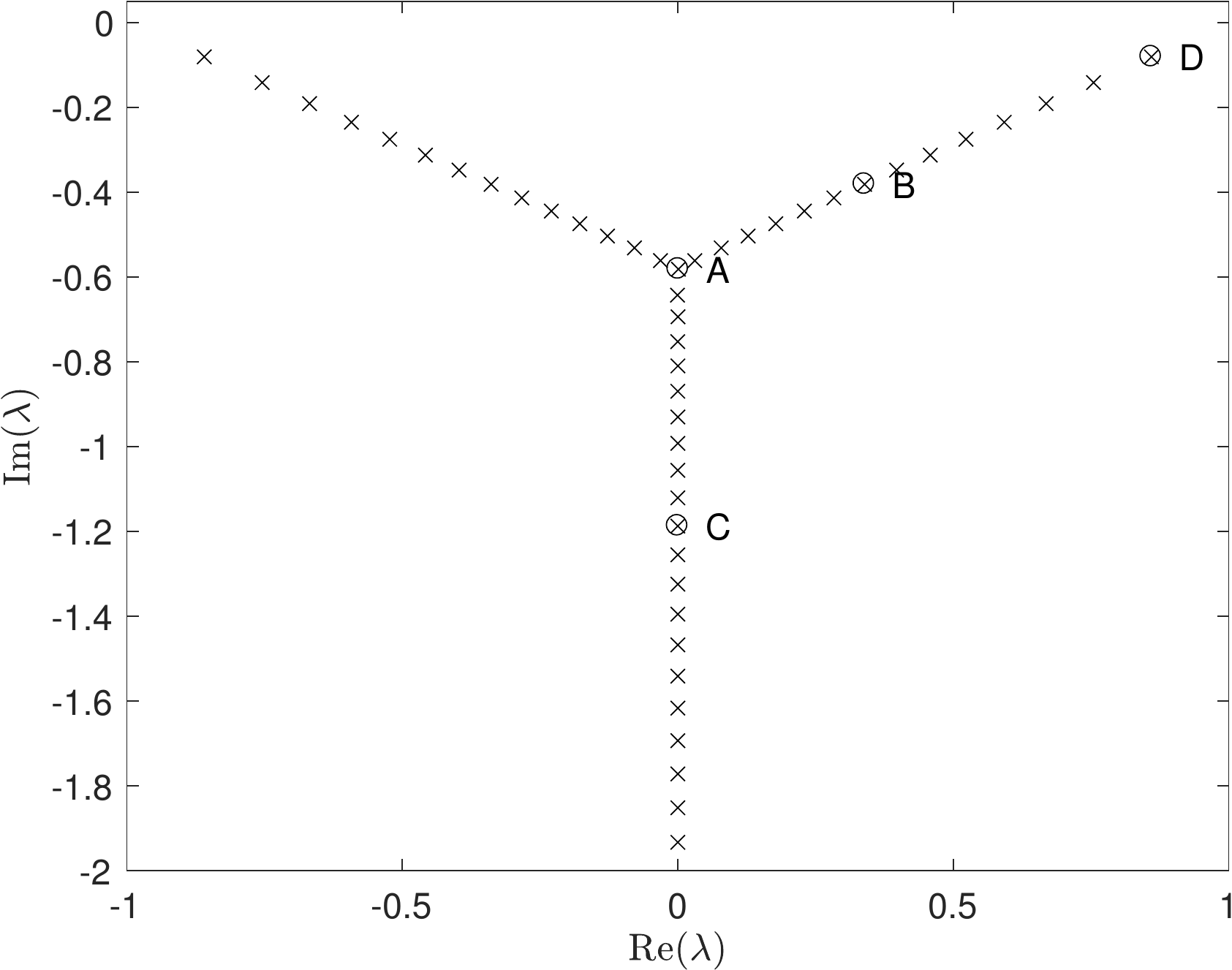}
\includegraphics[width= 0.45\textwidth]{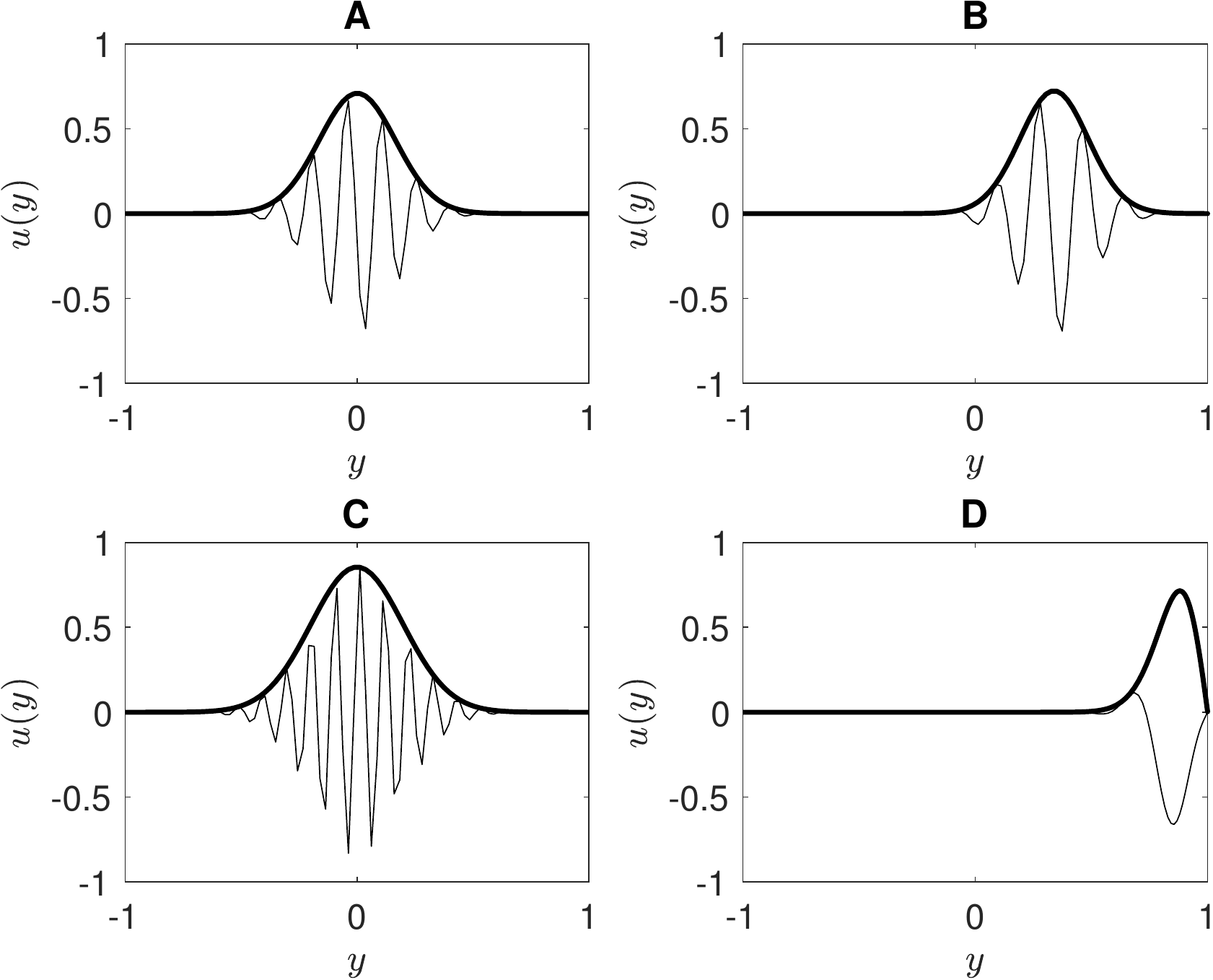}
}
\caption{Spectrum (left) and selected eigenfunctions (right) of $\mathcal{T}$ with Dirichlet boundary conditions on the domain $y\in[-1,1]$, with $R = 3000$. Absolute value and real component of eigenfunctions are shown with thick and thin lines, respectively.}
\label{fig:EigsT}
\end{figure}

\begin{figure}
 \centering {
\includegraphics[width= 0.45\textwidth]{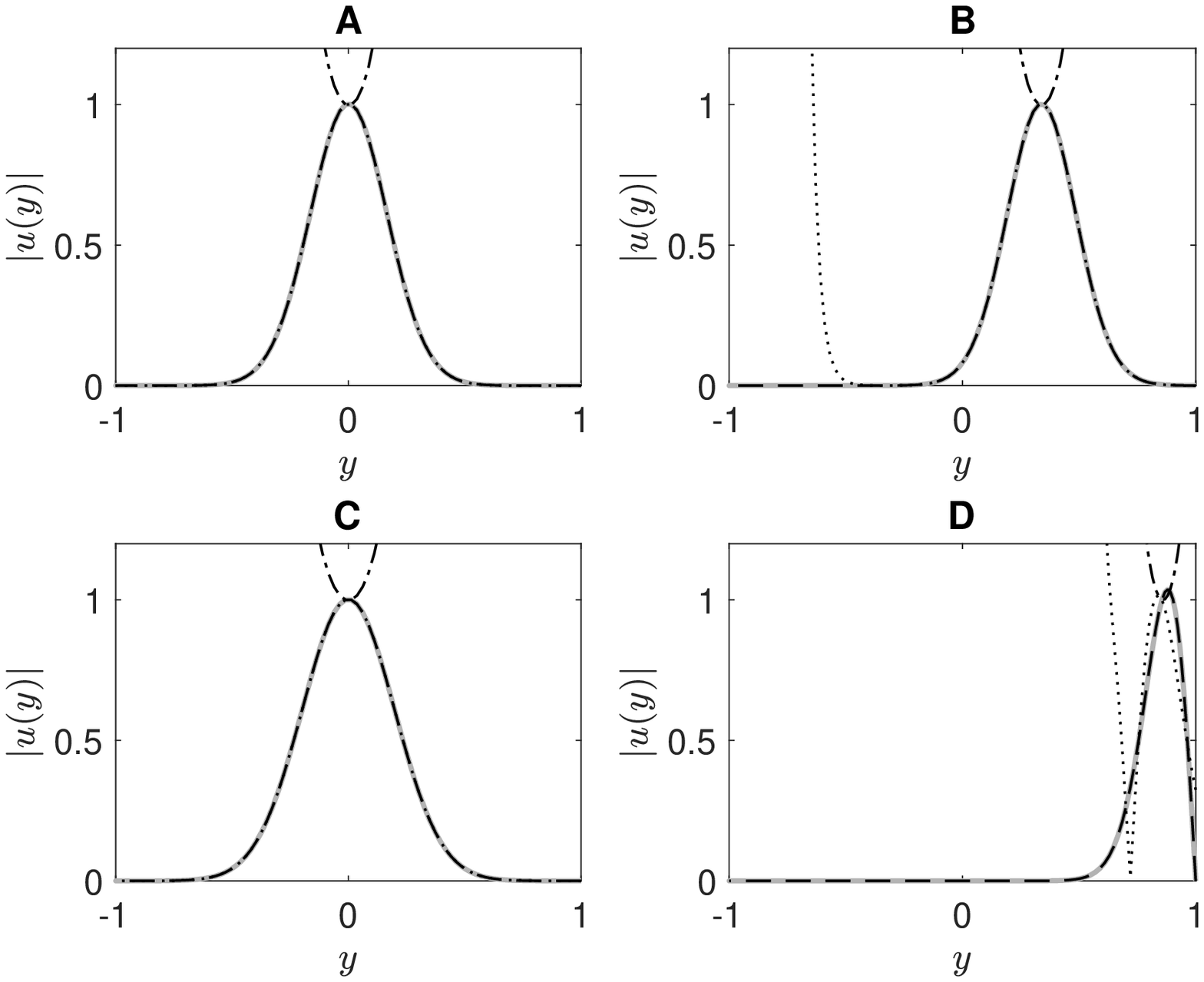}
\includegraphics[width= 0.45\textwidth]{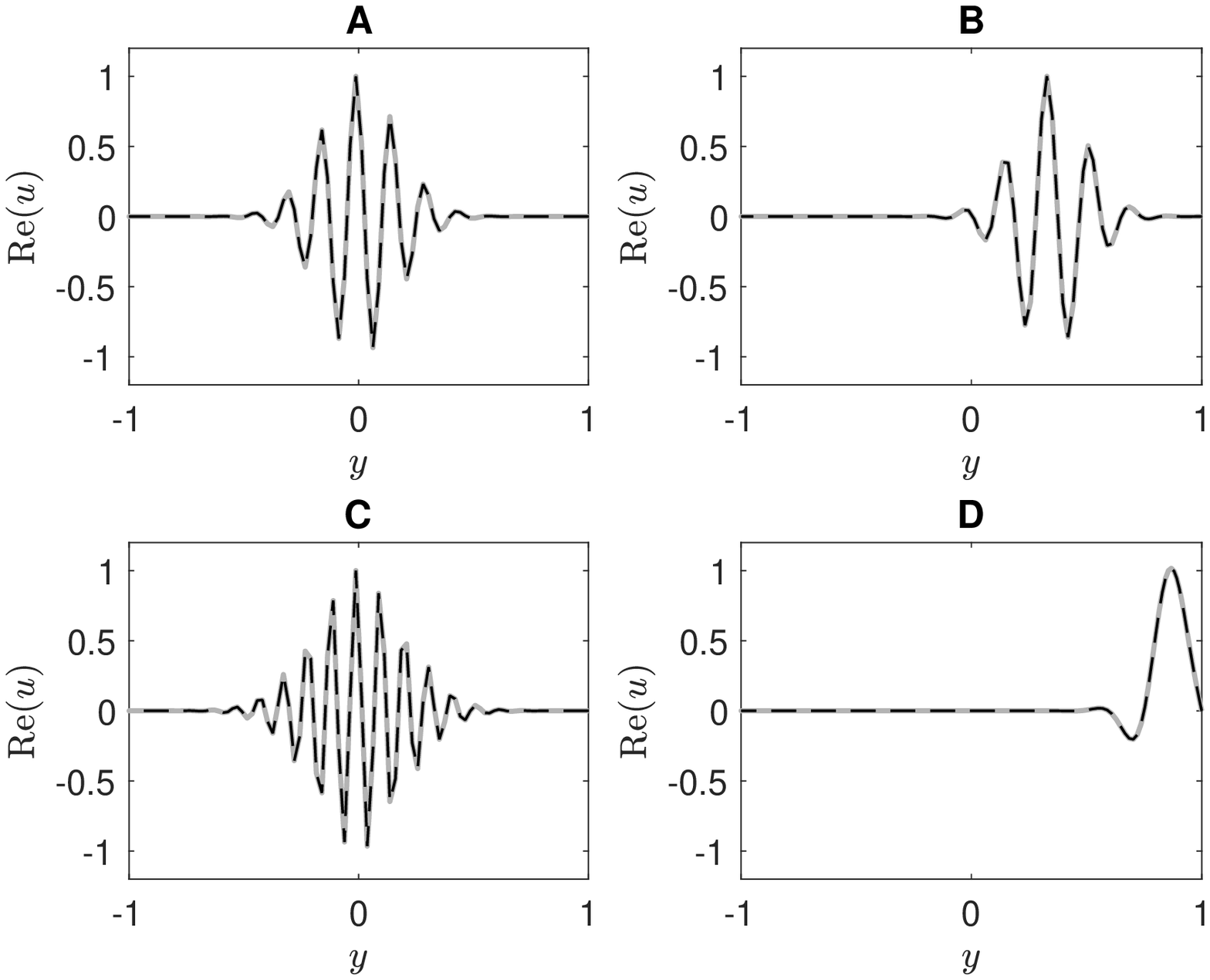}
}
\caption{Comparison between absolute value (left) and real component (right) of numerically computed eigenfunctions (grey solid lines) of $\mathcal{T}$ with $R  = 3000$, and those obtained from  analytical solutions of the Airy equation $\Ai[z]$, $\Ai[e^{2\pi i/3} z]$, and $\Ai[e^{-2\pi i/3} z]$. The analytic solution $\Ai[e^{2\pi i/3} z]$ closely matches the numerical solution for all cases, and is the only analytic solution shown in the right subplot. }
\label{fig:AiryEigs}
\end{figure}

\begin{figure}
 \centering {
(a)\includegraphics[width= 0.45\textwidth]{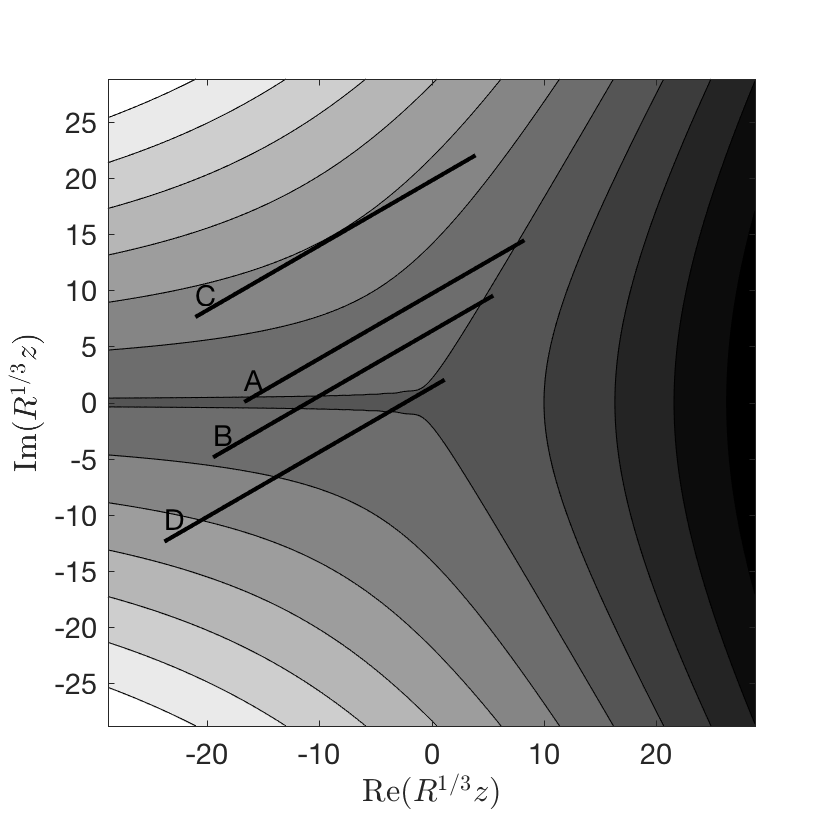}
(b)\includegraphics[width= 0.45\textwidth]{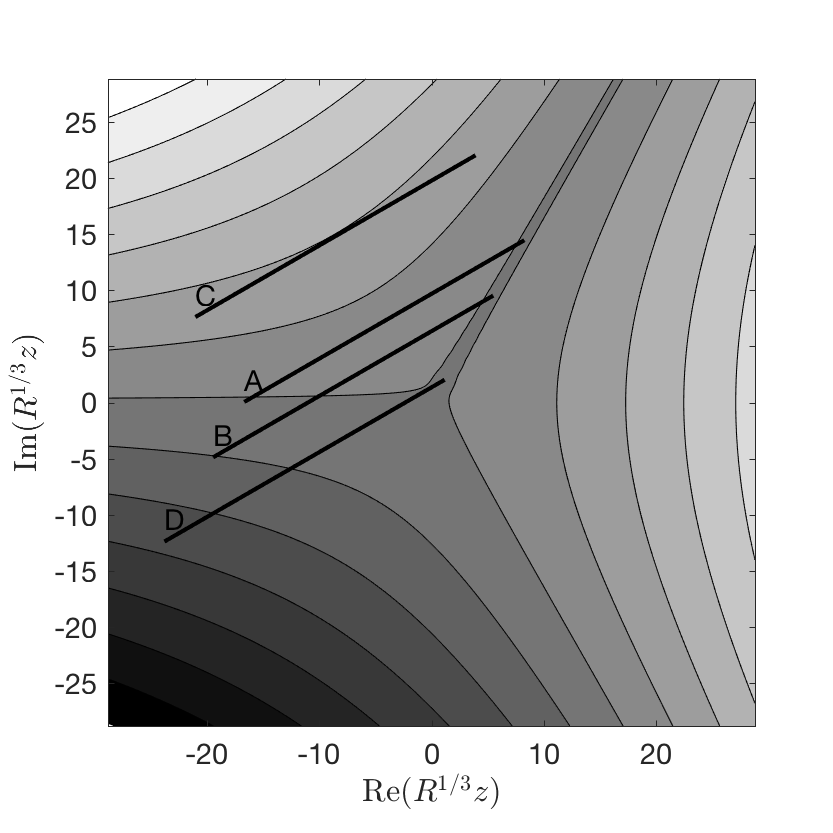}
}
\caption{Contour plots of the magnitude of solutions $\Ai[z]$ (left) and $\Ai[e^{2\pi i/3} z]$ (right) to the Airy equation, with intervals overlaid representing the domains of the eigenfunctions considered in figures \ref{fig:EigsT} and \ref{fig:AiryEigs}. 14 logarithmically-spaced contours between contours are $10^{-70}$ (black) and $10^{60}$ (white) are shown.}
\label{fig:AiryContour}
\end{figure}

Shown in figure \ref{fig:EigsT} are eigenvalues and selected eigenfuctions of $\mathcal{T}$ with Dirichlet boundary conditions on a finite domain $y \in [-1,1]$, with $R = 3000$. 
In figure \ref{fig:AiryEigs}, these numerical eigenfunctions are compared to $\Ai[z]$, $\Ai[e^{2\pi i/3} z]$, and $\Ai[e^{-2\pi i/3} z]$, and it is seen that  $\Ai[e^{2\pi i/3} z]$ (when appropriately-scaled) closely matches the numerically computed eigenfunctions on the finite domain. 
Note that if we had chosen eigenvalues on the left branch (i.e., with negative real component), then the $\Ai[z]$ solution would have been accurate for all cases. 
This can be seen more clearly by overlaying the $y$-domain on the amplitude of the Airy functions in the complex plane, as shown in figure \ref{fig:AiryContour}. 

  On an infinite domain (and when acting on square-integrable functions) $\mathcal{T}$ has an empty spectrum, and $\epsilon$-pseudospectra that are independent of location on the real axis (that is, the boundaries of the pseudospectrum for various $\epsilon$ are horizontal lines). 

As discussed in \cite{reddy1993pseudospectra}, a solution to $\mathcal{T}u= 0$ which is within $\epsilon$ of a function that satisfies the appropriate boundary conditions will be an $\epsilon$-pseudoeigenfunction.
We  consider the pseudospectra of $\mathcal{T}$, again for the finite domain $y \in [-1,1]$. 
As discussed in section \ref{sec:svd}, this amounts to computing the leading singular values and vectors of the resolvent of $\mathcal{T}$ for each $\lambda \in \mathbb{C}$ of interest. In particular, we may seek analytic approximations of these resolvent (optimal pseudospectral) modes using the same method as for eigenfunctions.  
In other words, for a given $\lambda \in \mathbb{C}$, we are interested in finding a function which is as close as possible to an eigenmode of $\mathcal{T}$.
Noting that the Airy functions are continuous, the fact that there are regions of $\mathbb{C}$ for which they are very close to satisfying the boundary conditions of the finite domain suggest that these regions correspond to very large values of resolvent norm $\sigma_1$. 

 \begin{figure}
 \centering{
\includegraphics[width= 0.45\textwidth]{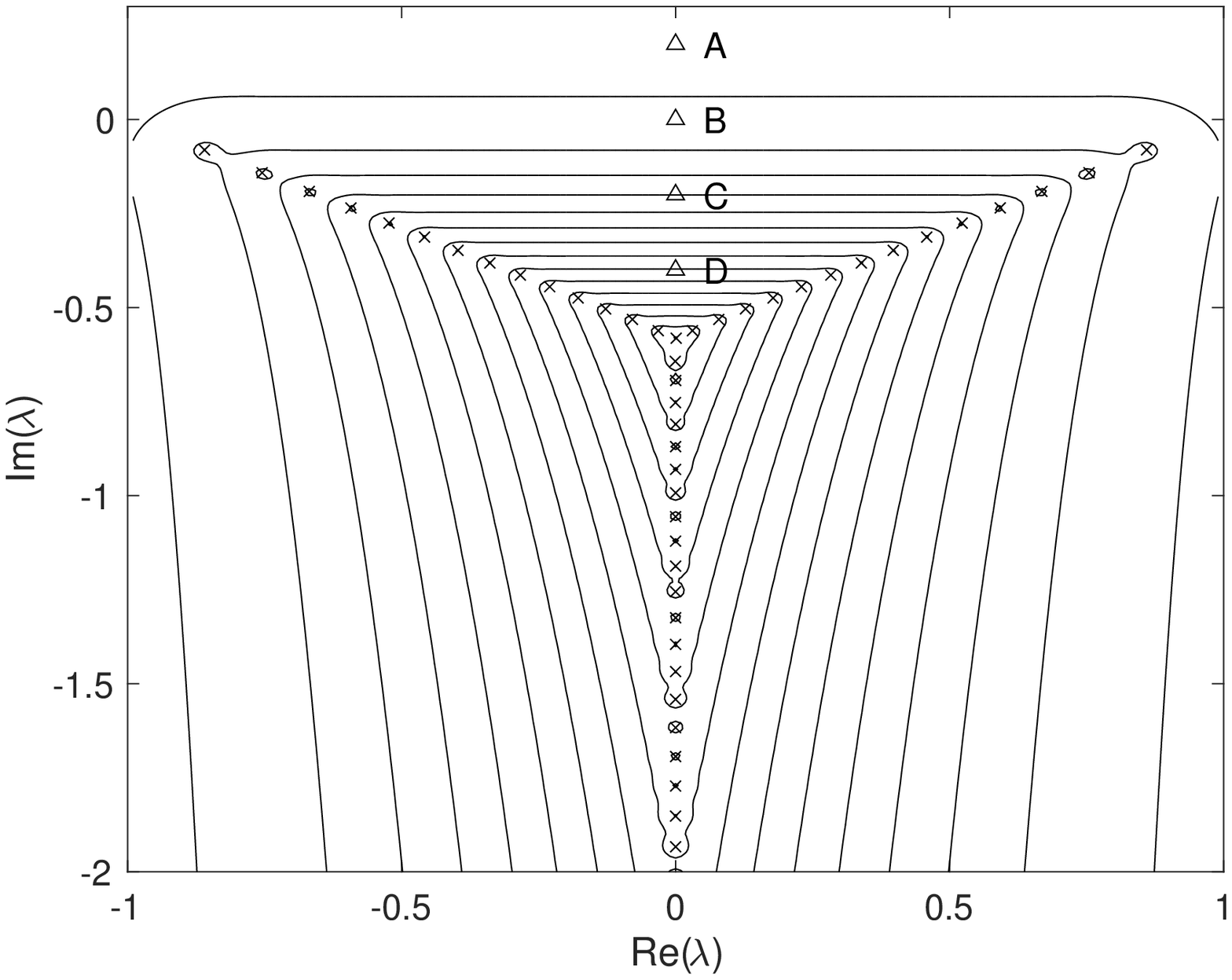}
\includegraphics[width= 0.45\textwidth]{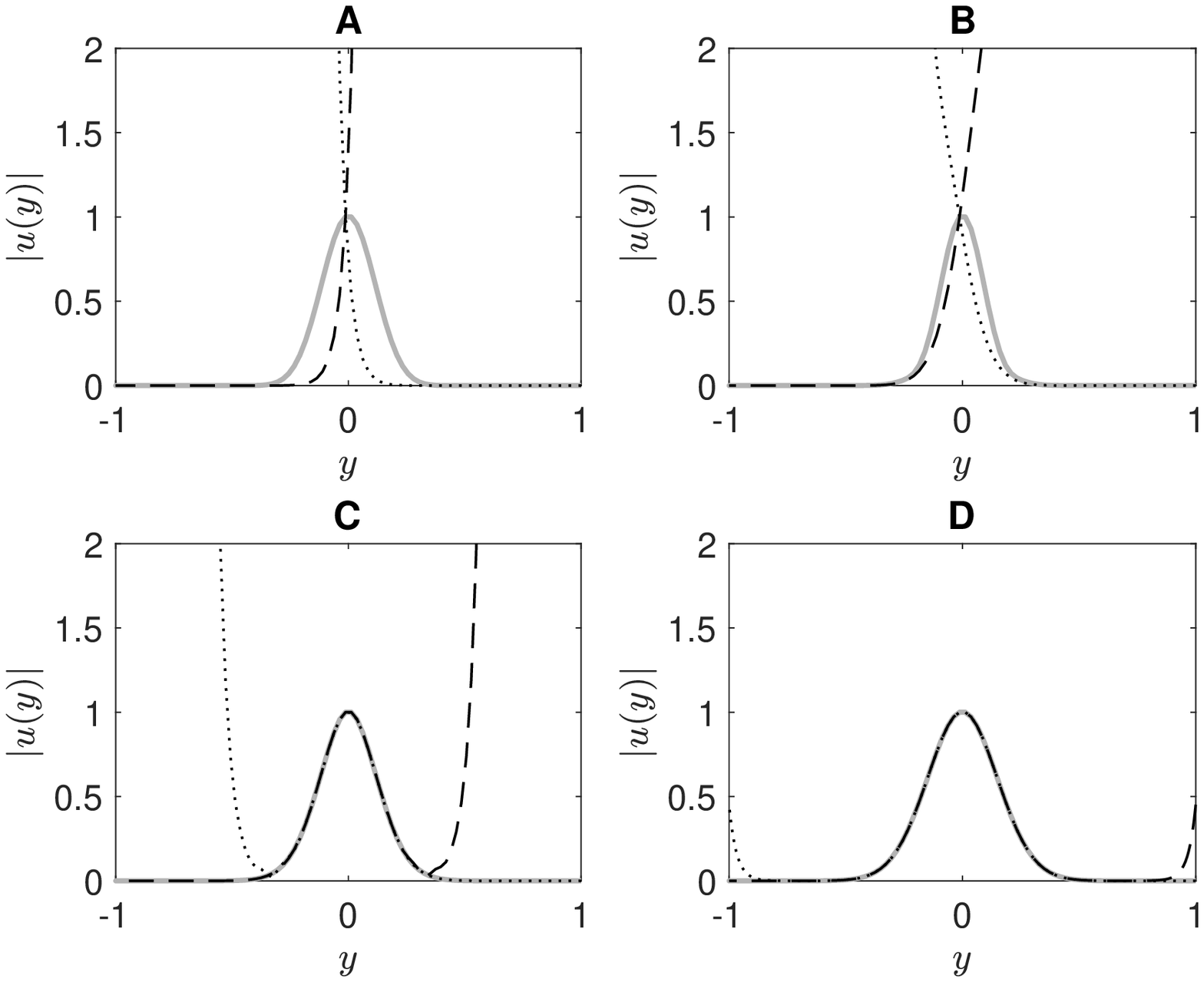}
}
\caption{(left) $\epsilon$-pseudospectra of $\mathcal{T}$ (with parameters equal to those for figures \ref{fig:EigsT} and \ref{fig:AiryEigs}),  with contour levels corresponding to $\epsilon \in \{10^{-15},10^{-14},\cdots,10^{-1}\}$, and (right) amplitudes of resolvent modes for locations in $\mathbb{C}$ as indicated on the left subplot. The right subplot compares numerically computed modes (grey) with Airy functions $\Ai[z]$ ($\cdot\cdot$), and $\Ai[e^{2\pi i/3} z]$ ($--$). }
\label{fig:pseudo}
\end{figure}

Figure \ref{fig:pseudo} shows $\epsilon$-pseudospectra  for $\mathcal{T} $, along with optimal pseudoeigenmodes (i.e., resolvent modes) and Airy function approximations thereof, for specified locations  in the centre of the domain. 
In the central region, the pseudospectral properties of $\mathcal{T}$ resemble that of the operator on an infinite domain, where resolvent modes, which are ``detached" from the boundary, are invariant to horizontal translation within this region. 
Moreover, resolvent modes within this region will resemble translated versions of eigenmodes along the inclined branches at the same vertical position. 
We observe that, as $\epsilon$ increases and we approach the real axis, the analytical Airy functions become less accurate, owing to their growth in magnitude occurring closer and closer to the critical layer, eventually saturating the component resembling the numerically computed response mode. 
As discussed in  \cite{reddy1993pseudospectra}, these Airy functions can be used to compute constructive lower bounds to the pseudospectrum of $\mathcal{T} $, though these bounds are  inaccurate whenever the Airy functions diverge far away from $ y = \Real(\lambda)$. 
The fact that $\omega_c  = R^{-1}k_\perp^2$ means that the most physically-relevant region of the complex plane is in the upper half-plane, which means that the Airy functions themselves are not close approximations of physically-relevant resolvent modes. 

Away from the origin, one may approximate Airy functions by simpler expressions that do not involve integrals. The derivation of such asymptotic approximations involves finding a location in the complex that has a dominant contribution to the contour integral. 
We may derive the approximation  (e.g., \citet{olver2014asymptotics,vallee2010airy})
\begin{equation}
\text{Ai}(z) \approx \frac{1}{2}\sqrt{\pi}z^{-1/4} \exp\left(-\frac{2}{3} z^{3/2}\right) \sim \exp\left(-\frac{2}{3} z^{3/2}\right), 
\end{equation}
which is accurate for large $|z|$, where here $z = \gamma (y-\lambda)$, where (with reference to equation \ref{eq:zy}) we have $\gamma = (iR)^{1/3}$, $\lambda = ik_\perp^2/R  = \omega_c i $, and we assume for now that $y_c = 0$. 
Expanding the exponential   term gives
$$ \exp\left(-\frac{2}{3} z^{3/2}\right) = \exp\left[-\frac{2}{3} (-\lambda \gamma)^{3/2} \left(1 - \frac{3}{2}\frac{y}{\lambda} + \frac{3}{8}\left(\frac{y}{\lambda}\right)^2 - \dots\right)\right],$$
and thus 
\begin{align}
\text{Ai}(z) &\sim C \exp\left[ (-\lambda\gamma)^{3/2}\frac{y}{\lambda}  - \frac{1}{4} (-\lambda\gamma)^{3/2}\left(\frac{y}{\lambda}\right)^2 - \cdots \right],
\end{align}
where we must take care when dealing with multi-valued roots.  In particular, for a bounded approximation, we require that 
$$ \Real\left[(-\lambda\gamma)^{3/2}\lambda^{-1}  \right] = 0.$$
If $\lambda = |\lambda| e^{i\theta_\lambda}$, then we find that this condition is only satisfied when $\theta_\lambda = \frac{3\pi}{2}.$ Note that this corresponds to the lower half plane region where Airy functions can give close approximations to the numerically-computed solutions. 
Assuming $y_c = 0$, so that  $\lambda = \omega_c i$, for $\omega_c < 0$  we obtain an approximation of the form
\begin{equation}
\label{eq:abAsymp}
\psi(y)  \sim \exp\left[ -i\sqrt{|\omega_c| R}y - \frac{1}{4} \sqrt{\frac{R}{|\omega_c|}} y^2\right].
\end{equation}

\subsection{Predicting wavepacket modes for a model operator}
\label{sec:SquirePred}
Section \ref{sec:Airy} demonstrated that, in certain regimes, resolvent modes may be approximated by a function of the general form
 \begin{equation}
\label{eq:guessExp}
\psi(y) =c \exp\left( ai y-b y ^2\right), 
\end{equation}
where $a \in \mathbb{R}$ and $b> 0 $ are functions of $\omega_c$ and $R$, and 
$$ c = \left(\frac{2b}{\pi}\right)^{1/4}$$ 
is a constant that gives $\psi $ unit norm with respect to the regular scalar inner product. 

This section will present an alternative analysis with this template function as a starting point. 
Applying wavepacket pseudomode theory as described in section \ref{sec:packet}, one may show that $\mathcal{T}$ satisfies the twist condition (equation \ref{eq:twist}) is satisfied within the half strip 
\begin{equation}
 \left\{ \lambda : -1<\Real(\lambda) < 1, \Imag(\lambda) < 0 \right\}.
 \label{eq:strip}
 \end{equation}
The $ \text{Imag}(\lambda) < 0$ condition is consistent with the predicted condition for a spatially localised mode from the truncated asymptotic expansion of an Airy function in equation \ref{eq:abAsymp}. 
Note that the region described in equation \ref{eq:strip} is outside the region that is most physically relevant for resolvent analysis of the Squire operator, where a positive $k_\perp^2$ and $R$ give a positive imaginary component of $\omega_c$. 
Despite this, figure \ref{fig:pseudo} shows that even outside this region, the numerically computed pseudomode still maintains the same spatially localised structure. 
To predict the shape of resolvent modes outside this region, we will seek to directly optimize the shape parameters in the template function, equation \ref{eq:guessExp}.
In particular, we wish to find the values of $a$ and $b$ for which  $\|\mathcal{T} \psi \|$ is minimized. 
That is, we seek the minimum of the cost function
\begin{equation}
J_\mathcal{T}(a,b;R,\omega_c) = \|\mathcal{T}\psi\|^2 =   \int_{y = -\infty}^{\infty} \left(\mathcal{T} \psi \right)^* (\mathcal{T} \psi) dy,
\end{equation}
where we are assuming that our mode is ``detached" and localised, so can integrate over an infinite domain, and are assuming a standard inner product for this model operator (note that including a constant multiplicative factor of $k_\perp^{-2}$ as in equation \ref{eq:IPscalar} does not affect mode shapes or singular values). 
For simplicity, we again take $y_c = 0$.
Note that the leading resolvent singular value is related to this cost function by
$$ \sigma_1(R,\omega_c) = \left[ \min_{a,b} J_\mathcal{T}(a,b;R,\omega_c) \right]^{-1/2}.  $$
Noting that we have
\begin{align*}
 \mathcal{T}^* &= (iR)^{-1}\frac{d^2}{dy^2} + \left(y + i\omega_c\right), \\
 \psi^*(y) &= c \exp[ -ayi-b y ^2],
\end{align*}
it can be shown that 
\begin{align*}
\mathcal{T} \psi(y) &= -\frac{1}{iR} \left[ 4b^2y^2 -i(4ab+R)y - a^2 - 2b-R\omega_c\right] \psi(y), \\
\left(\mathcal{T} \psi(y)\right)^* &=  \frac{1}{iR} \left[ 4b^2y^2 +i(4ab+R)y - a^2 - 2b-R\omega_c\right] \psi^*(y).
\end{align*}
From this, we may compute
\begin{equation}
\label{eq:J}
J_\mathcal{T}(a,b;R,\omega_c) = \frac{1}{4b} + \frac{3}{R^2}b^2 
+ \frac{2(3a^2+R\omega_c)}{R^2}b + \frac{a^4+2aR+2\omega_c a^2R+\omega_c^2R^2}{R^2}.
\end{equation}

\begin{figure}
\hspace{-1.4cm}\includegraphics[width= 1.2\textwidth]{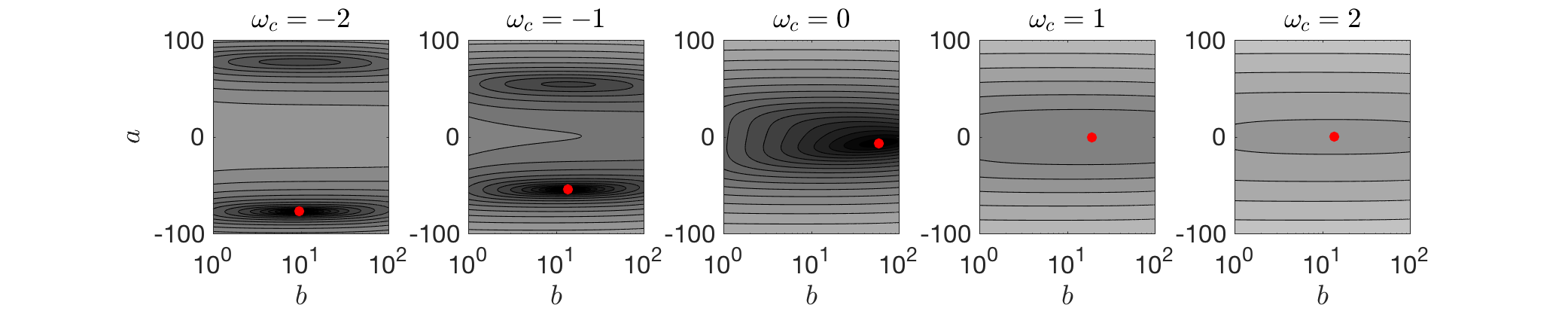}
\caption{Contours of the cost function $J_\mathcal{T}$ as a function of shape parameters $(a,b)$ for $R = 3000$ and several values of $\omega_c$, with global minima identified with filled circles. 25 contour levels spaced logarithmically between $10^{-8}$ (black) and $10^4$ (white) are used.}
\label{fig:Jcontours}
\end{figure}

Contours of this cost function for $R = 3000$ and several values of $\omega_c$ are shown in figure \ref{fig:Jcontours}.
We may analytically determine the locations of these minima as follows. This cost function is minimised when 
\begin{align}
\label{eq:Ja}
\frac{\partial J_\mathcal{T}}{\partial a} &= \frac{2}{R^2} \left( 6ab + 2a^3+ 2R\omega_c a + R \right) = 0 \\
\label{eq:Jb}
\frac{\partial J_\mathcal{T}}{\partial b} &=  -\frac{1}{4b^2}+ \frac{2}{R^2} \left( 3b+3a^2+R\omega_c \right) = 0 
\end{align}

\begin{figure}
 \centering {
(a)\includegraphics[width= 0.5\textwidth]{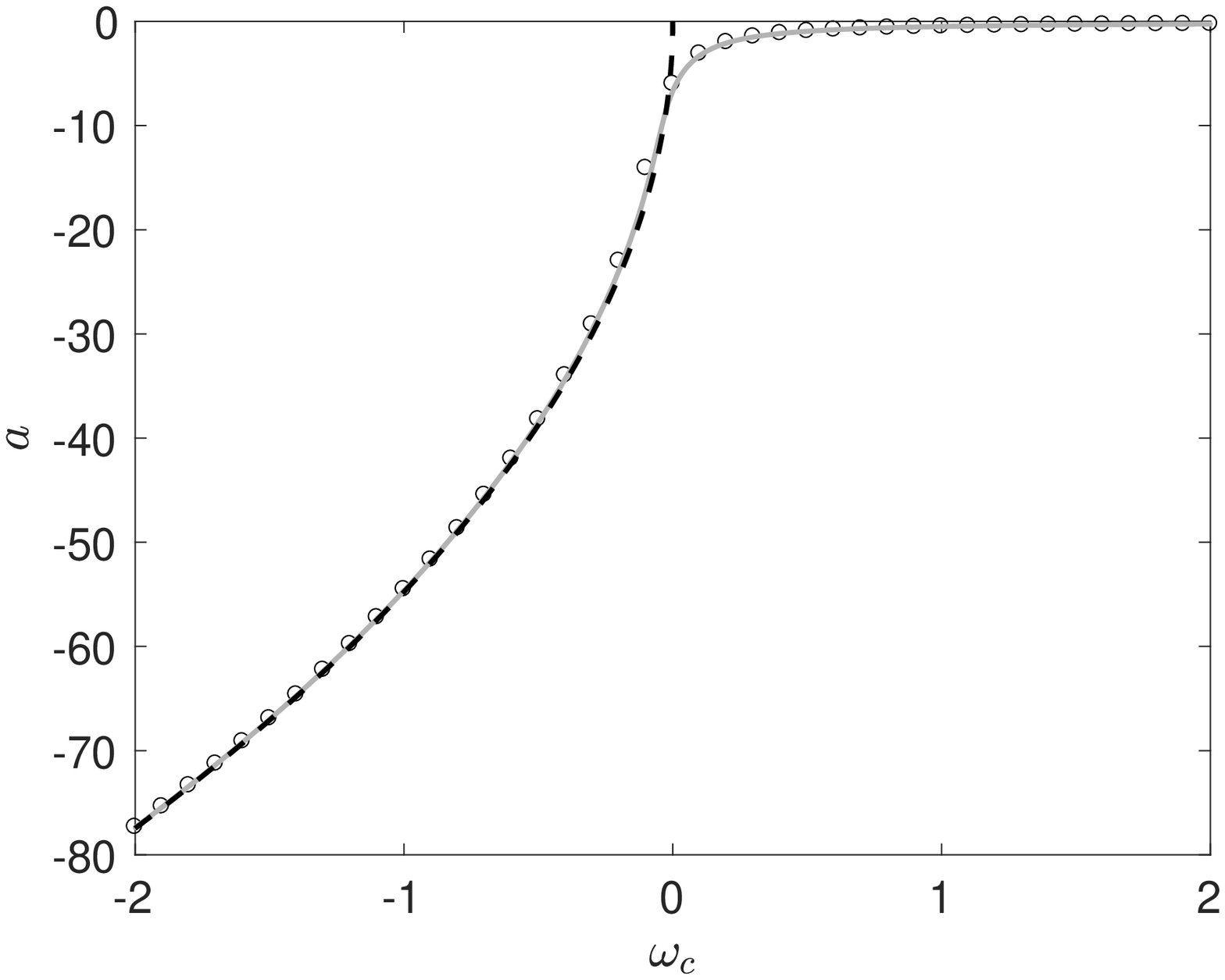}
(b)\includegraphics[width= 0.5\textwidth]{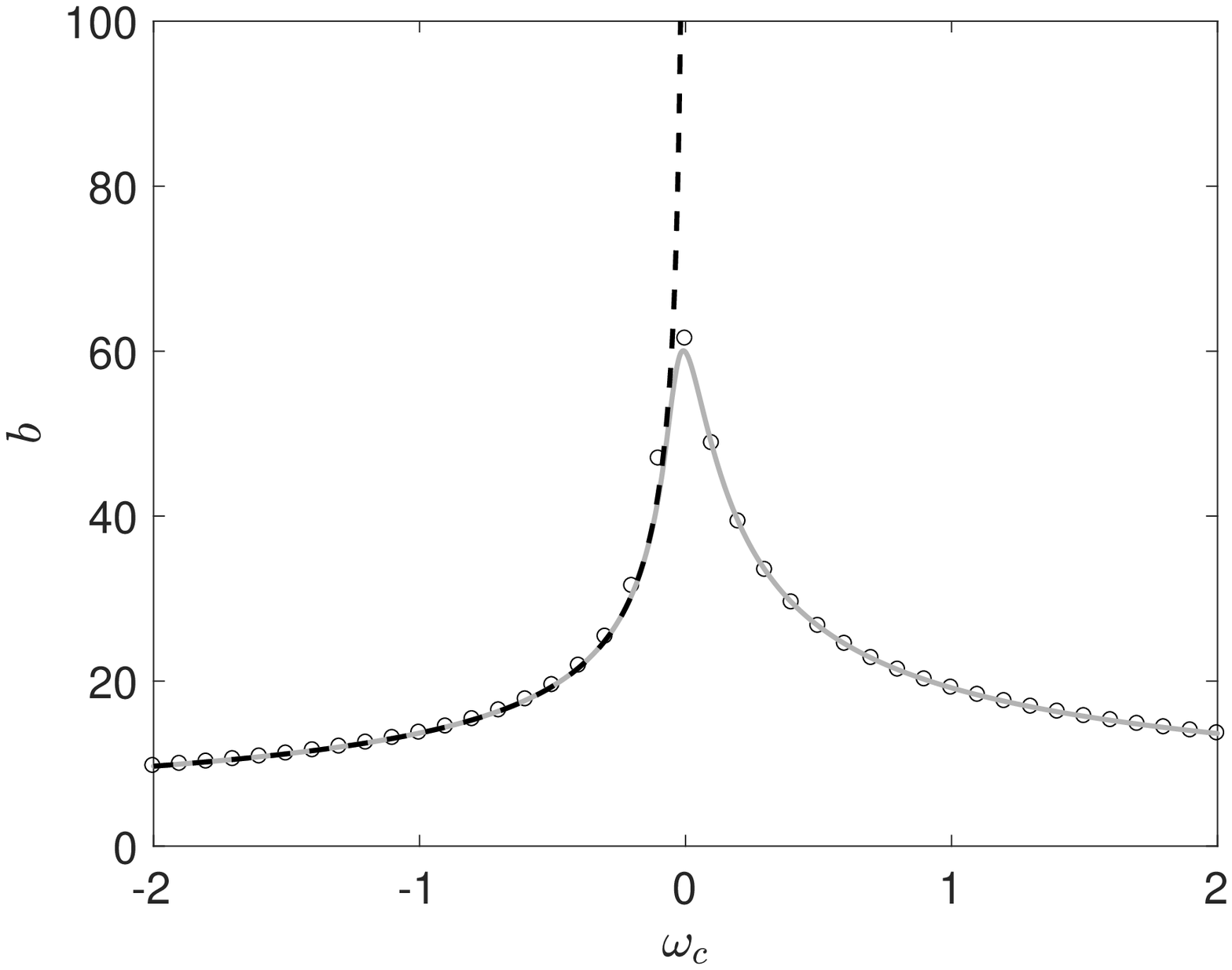}
(c)\includegraphics[width= 0.48\textwidth]{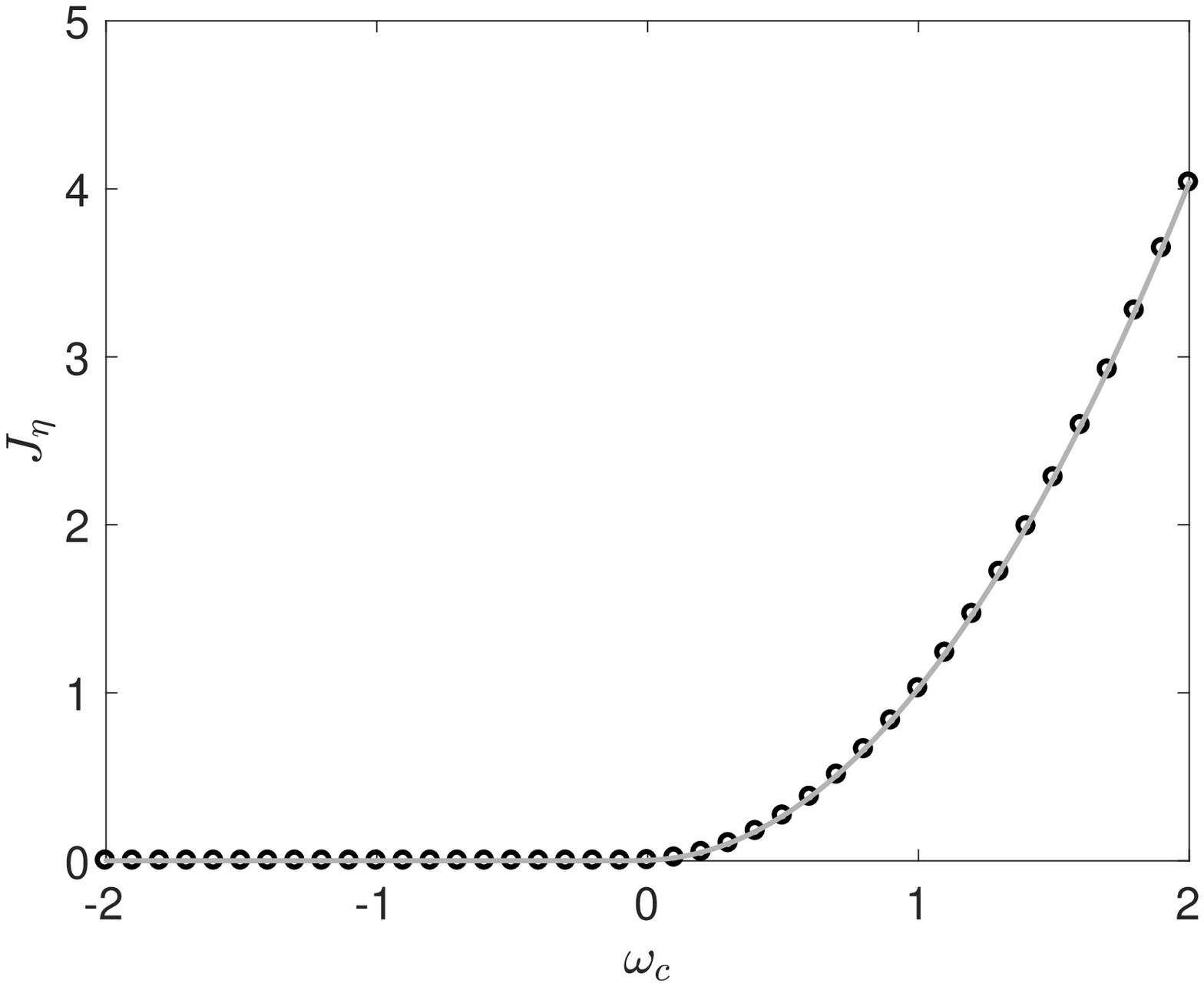}
}
\caption{Prediction of optimal shape parameters $a$ (a) and $b$ (b) as a function of  $ \omega_c = \text{Imag}(\lambda)$ with $R = 3000$ using both the truncated asymptotic Airy function approximation (equation \ref{eq:abAsymp}, black dashed lines) and minimising the cost function given in equation \ref{eq:J} (grey lines), in comparison to values fitted to numerically computed eigenfunctions on a finite domain (circles). (c) compares the cost function value to the true value ($\sigma_1^{-2}$) obtained from the resolvent norm. }
\label{fig:abevolution}
\end{figure}

The equations \ref{eq:Ja} and \ref{eq:Jb} may be solved to give the parameters $\{a,b\}$ minimising $J_\mathcal{T}$, with care taken to select the correct minimising solution.  It is also possible to obtain a parametrised family of solutions as follows. Assuming that $R$ is fixed, suppose that we seek the optimal values $\left(a(\omega_c),b(\omega_c)\right)$ for a range of values of $\omega_c$. By implicitly differentiating equations \ref{eq:Ja} and \ref{eq:Jb} with respect to $\omega_c$, the following ordinary differential equations governing the evolution of the optimal mode shape parameters may be found:
\begin{align}
\label{eq:odea}
\frac{\partial a}{\partial \omega_c} &=\frac{-R^3 a}{ 
  \omega_c R^3+ 3(a^2+b)R^2 + 12\omega_c b^3 R + 36b^3(b-a^2)},\\
\label{eq:odeb}
\frac{\partial b}{\partial \omega_c} &=  \frac{-4  b^3R (\omega_c R - 3a^2 + 3 b) }{
    \omega_c R^3+ 3(a^2+b)R^2 + 12\omega_c b^3 R + 36b^3(b-a^2)}.
   \end{align}
    Note that this approach assumes that the global minimiser of $J_\mathcal{T}$ stays on the same branch, and is smoothly continuous with $\omega_c$. 
Figure \ref{fig:abevolution} shows that the predicted values of the optimal values of parameters $a$ and $b$ closely match both the asymptotic approximations (for $\omega_c < 0$), and the values obtained from fitting the numerically computed modes.
The width parameter for this fit to the numerically-computed mode  data is found by fitting a Gaussian function to the amplitude of the computed mode using MATLAB's {\verb fit } command. The phase parameter is found by considering the gradient of the phase in a small region near the critical layer location. 
 The predicted shape parameters are obtained from evolving the equations \ref{eq:odea} and \ref{eq:odeb} from an initial condition obtained by solving equations \ref{eq:Ja} and \ref{eq:Jb} directly (at $\omega_c = 0$).  Here, and in subsequent sections, such solutions (along with the symbolic computation of integrals) are obtained using Mathematica.  Note that one could also use the asymptotic approximation \ref{eq:abAsymp} for initial conditions, where we must start from a  sufficiently large negative value of $\omega_c$ to ensure that the initial conditions are accurate. 
It is additionally shown that the value of the cost function $J_\mathcal{T}$ for these optimal shape parameters closely matches the value of the cost function that may be computed directly from the numerically computed singular value. 
In essence, this shows that the optimal mode and amplification across all functions is closely approximated by the optimal over the class of functions of the form given in equation \ref{eq:guessExp}. 

We may also consider the behaviour of the optimal shape parameters as $R$ varies.  Keeping $\omega_c$ constant, from equations \ref{eq:Ja}-\ref{eq:Jb} we obtain 
   \begin{align}
\label{eq:odeRa}
\frac{\partial a}{\partial R} &=\frac{(1+2\omega_c a)R^3 + 4a(a^2+3b)R^2  +12 b^3R-96a^3b^3}{ 
  2R\left( \omega_c R^3 +3(a^2+b)R^2 +12\omega_c b^3 R +36b^3(b-a^2)\right)},\\
\label{eq:odeRb}
\frac{\partial b}{\partial R} &=  \frac{4  b^3 \left(\omega_c^2 R^2 + 3R(-a +  \omega_c a^2 +9\omega_c b)+6a^4 + 18 b^2\right) }
{R\left( \omega_c R^3 +3(a^2+b)R^2 +12\omega_c b^3 R +36b^3(b-a^2)\right)}.
   \end{align}
   In the limit of large $R$ and for $\omega_c \neq 0$, (and assuming restrictions on the growth of $a$ and $b$ with $R$), we have the approximations
   \begin{align}
\label{eq:odeRaApprox}
\frac{\partial a}{\partial R} &\approx \frac{1}{2\omega_c R},\\
\label{eq:odeRbApprox}
\frac{\partial b}{\partial R} & \approx \frac{4  b^3\omega_c}{R^2},
   \end{align}
   from which we may infer the scalings $a \to C$, $b\propto R^{1/2}$. 
   For large $R$ but with $ R \omega_c \ll 1 $, we instead obtain
      \begin{align}
\label{eq:odeRaApprox}
\frac{\partial a}{\partial R} &\approx \frac{1}{3(a^2+b)},\\
\label{eq:odeRbApprox}
\frac{\partial b}{\partial R} &\approx -\frac{4ab^3}{R^2(a^2+b)},
   \end{align}
   which can be shown to permit a consistent asymptotic solution with $a\propto R^{1/3}$ and $b\propto R^{2/3}$. 
   This scaling of $b$ corresponds to a mode with that scales with $R^{1/3}$, which agrees with the critical layer scaling \citep{drazin2004hydrodynamic}, and can also be inferred from the transformation described in equation \ref{eq:zy}. 
    \begin{figure}
 \centering {
(a)\includegraphics[width= 0.5\textwidth]{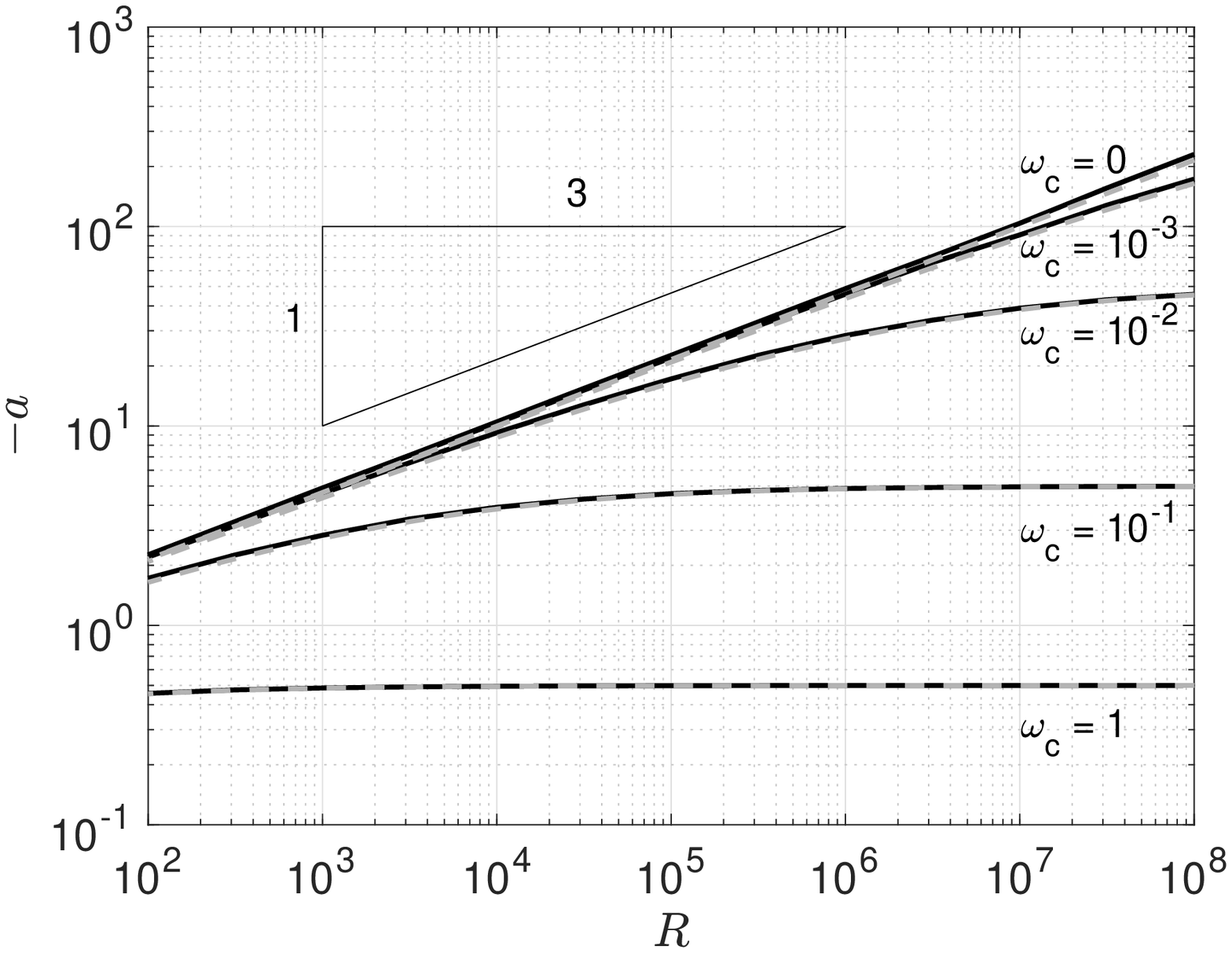}
(b)\includegraphics[width= 0.5\textwidth]{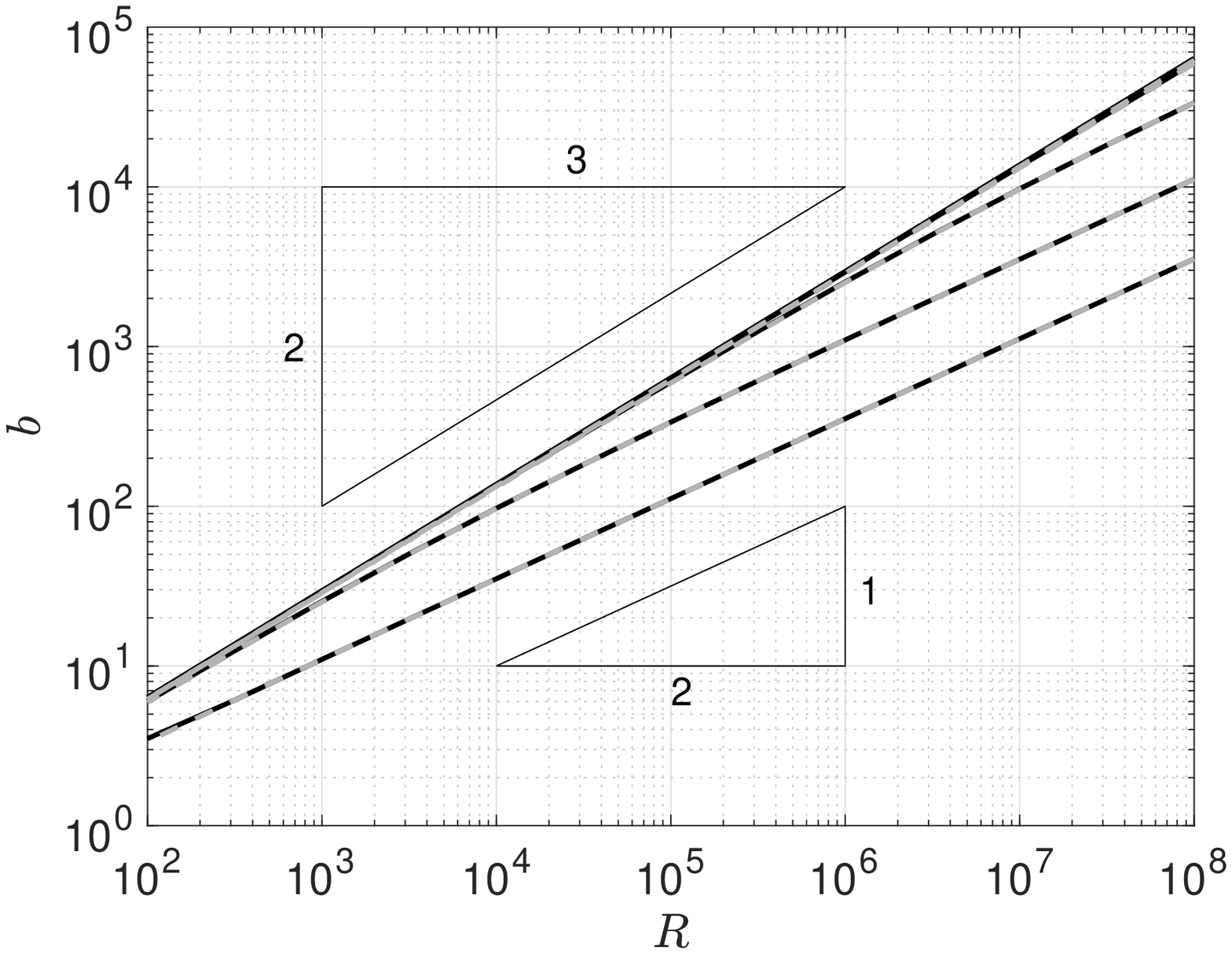}
(c)\includegraphics[width= 0.52\textwidth]{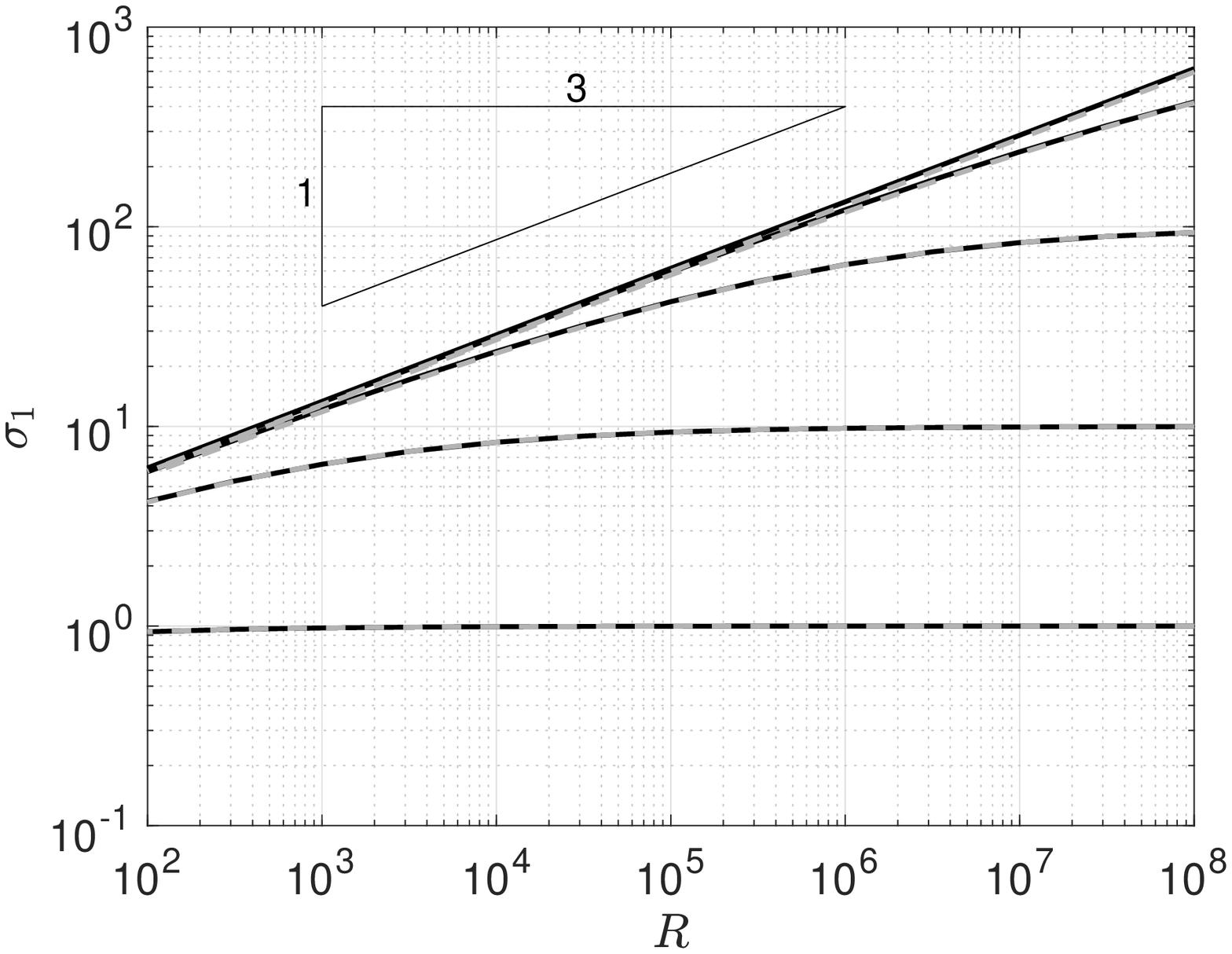}
}
\caption{Comparison between predicted ($-$) and fitted ($--$) parameters $a$ (subplot a) and $b$ (b), and associated singular value (c) as a function $R$. }
\label{fig:abevolutionR}
\end{figure}
   Figure \ref{fig:abevolutionR} shows that these scalings closely match the shape parameter trends obtained from the numerically computed modes. Moreover, these shape parameters are accurately predicted from finding optimal parameters of cost function $J_\mathcal{T}$, which can be obtained either from solving equations \ref{eq:Ja} and \ref{eq:Jb} for each $\omega_c$ and $R$, or by evolving the differential equations \ref{eq:odeRa} and \ref{eq:odeRb} over $R$. 
      It should be noted that, even for this model system, computing the leading resolvent modes directly can require substantial computational resources for large $R$, highlighting an advantage of approximations using the methods described in this section. 
   We lastly note that one could seek alternative parameterisations of the two-dimensional space of optimal mode shapes, such as by varying  $R$ while keeping $\omega_c R$  ($= k_\perp^2$) constant. 

\subsection{Predicting the shape of resolvent modes for laminar Couette flow}
\label{sec:OSpred}
In this section, we will extend the analysis in section \ref{sec:SquirePred} to consider the Navier--Stokes operator and suboperators considered in sections \ref{sec:form} and \ref{sec:simp}. We again will restrict our attention to predicting the vorticity component of the leading response mode of the resolvent operators $ \mathcal{H}_{\bm{k}}$, $ \mathcal{H}_{os,\bm{k}}$ and $ \mathcal{H}_{sq,\bm{k}}$. For the Squire suboperator, the analysis is almost identical to that of the model operator considered in section \ref{sec:SquirePred}. To predict the mode shapes of the full system, we rely on the fact that the response is dominated by the effect of the OS suboperator, which we analyse using the simplifying approximations introduced in section  \ref{sec:simp}. In particular, this  also results in a methodology similar to that used in section  \ref{sec:SquirePred}, but with a modification to use the Laplacian scalar inner product (equation \ref{eq:IPscalarLap}.

We begin by again assuming mode shapes of the form
\begin{align*}
\psi_{sq}(y) &=c_{sq} \exp\left( a_{sq}i y-b_{sq} y ^2\right), \\
\psi_{os}(y) &=c_{os} \exp\left( a_{os}i y-b_{os} y ^2\right), 
\end{align*}
but now to satisfy the unit-norm requirements of the relevant  inner products (equation \ref{eq:IPscalar}) for the SQ operator, equation \ref{eq:IPscalarLap} for the OS) we have 
$$ c_{sq} = \left(\frac{2b_{sq}}{\pi}\right)^{1/4}k_\perp, \ \ \ \  c_{os} = \left(\frac{2b_{os}}{\pi}\right)^{1/4}k_\perp(b_{os}+a_{os}^2+k_\perp^2),$$
where, with reference to the previous section, we have $k_\perp^2 = \omega_c R$. 
Using the approximation to the Orr-Sommerfeld operator introduced in section \ref{sec:simp}, and noting that $U_{y_c} = 1$, the relevant cost functions for optimizing the shape of the SQ and OS template functions are

\begin{align}
J_{sq}(a,b;R,\bm{k}) = \| U_{y_c}k_x \mathcal{T}\psi_{os}\|^2 &= \frac{U^2_{y_c}k_x^2}{k_\perp^2}  \int_{y = -\infty}^{\infty} \left(\mathcal{T} \psi_{os} \right)^{*} (\mathcal{T} \psi_{os}) dy \\
J_{os}(a,b;R,\bm{k}) = \|U_{y_c}k_x \mathcal{T}\psi_{os}\|_\Laplace^2 &= \frac{U^2_{y_c}k_x^2}{k_\perp^2}  \int_{y = -\infty}^{\infty} \left(\mathcal{T} \psi_{os} \right)^{*,\Laplace} (\mathcal{T} \psi_{os}) dy \nonumber \\
& = -\frac{U^2_{y_c}k_x^2}{k_\perp^2}\int_{y = -\infty}^{\infty} \left(\mathcal{T} \psi_{os} \right)^{*}\Laplace (\mathcal{T} \psi_{os}) dy
\end{align}

For laminar Couette flow we have $U_{y_c} = 1$ for all $y_c$ in the domain, but we keep this term so that the equations are directly applicable for arbitrary mean velocity profiles. Also note that while the cost function for the Squire modes is ``exact" given the assumed template function and a linear mean velocity profile, the OS modes are relying on the accuracy of a simplified operator (being the scalar Squire operator with a Laplacian inner product) capturing the correct behavior of the response mode. 

Substituting in the mode template functions (and dropping the subscripts on $a$ and $b$ parameters for brevity), explicit expressions for the cost functions are 
\begin{equation}
J_{sq}(a,b;R,\bm{k}) =\frac{U_{y_c}^2k_x^2}{k_\perp^2} \left[\frac{1}{4b} + \frac{3}{R^2}b^2 
+ \frac{2(3a^2+k_\perp^2)}{R^2}b + \frac{a^4+2aR+2k_\perp^2 a^2+k_\perp^4}{R^2}\right],
\label{eq:Jsq}
\end{equation}
\begin{equation}
\begin{aligned}
 J_{os}(a,b; R,\bm{k}) =& \frac{U_{y_c}^2k_x^2}{4 k_\perp^2 b\left(b+a^2 + k_\perp^2\right) R^2} \left[
 (3b+a^2+k_\perp^2)R^2 +16ab(3b+a^2+k_\perp^2)R\right. \\
 &  \left.+  4 b\left(15b^3 + (a^2+k_\perp^2)^3+9b^2(5a^2+k_\perp^2) +3b(a^2+k_\perp^2)(5a^2+k_\perp^2)\right)\right].
\end{aligned}
\end{equation}
As before, the minima of these cost functions may be found by selecting the appropriate solution to the equations
\begin{align}
\label{eq:Jsqa}
\frac{\partial J_{sq}}{\partial a} &= \frac{2U_{y_c}^2k_x^2}{k_\perp^2 R^2} \left( 6ab + 2a^3+ 2k_\perp^2 a + R \right) = 0, \\
\label{eq:Jsqb}
\frac{\partial J_{sq}}{\partial b} &=  \frac{U_{y_c}^2k_x^2}{k_\perp^2}\left[-\frac{1}{4b^2}+ \frac{2}{R^2} \left( 3b+3a^2+k_\perp^2 \right) \right]= 0, 
\end{align}
\begin{equation}
\begin{aligned}
\label{eq:Josa}
\frac{\partial J_{os}}{\partial a} &= \frac{U_{y_c}^2k_x^2}{k_\perp^2 \left(b + a^2 + 
   k_\perp^2\right)^2}\left[  -a +  \frac{4}{R} \left(3b^2+4k_\perp^2 b + (a^2+k_\perp^2)^2\right) \right.\\
   &\left. \ \ \ +  \frac{4a}{R^{2}} \left(15b^3+9b(a^2+k_\perp^2)^2 + (a^2+k_\perp^2)^3  
   + 3b^2(5a^2+9k_\perp^2\right)    \right] = 0, \\ 
   \end{aligned}
  \end{equation}
  \begin{equation}
\label{eq:Josb}
\begin{aligned}
\frac{\partial J_{os}}{\partial b} &=\frac{U_{y_c}^2k_x^2}{k_\perp^2 \left(b + a^2 + 
   k_\perp^2\right)^2} \left[ -\frac{3}{4} - \frac{1}{2b}(a^2+k_\perp^2) - \frac{1}{b^{2}}(a^2+k_\perp^2)^2 \right. + \frac{8}{R}a(a^2+k_\perp^2) \\
  & \ \ \  \left. +\frac{2}{R^2}\left(15b^3+(a^2+k_\perp^2)^2(7a^2+k_\perp^2)+9b^2 (5a^2+3k_\perp^2) + 9b(5a^2 +k_\perp^2)(a^2+k_\perp^2) \right)   \right]\\
  & = 0.
  \end{aligned}
\end{equation} 
    \begin{figure}
 \centering {
(a)\includegraphics[width= 0.45\textwidth]{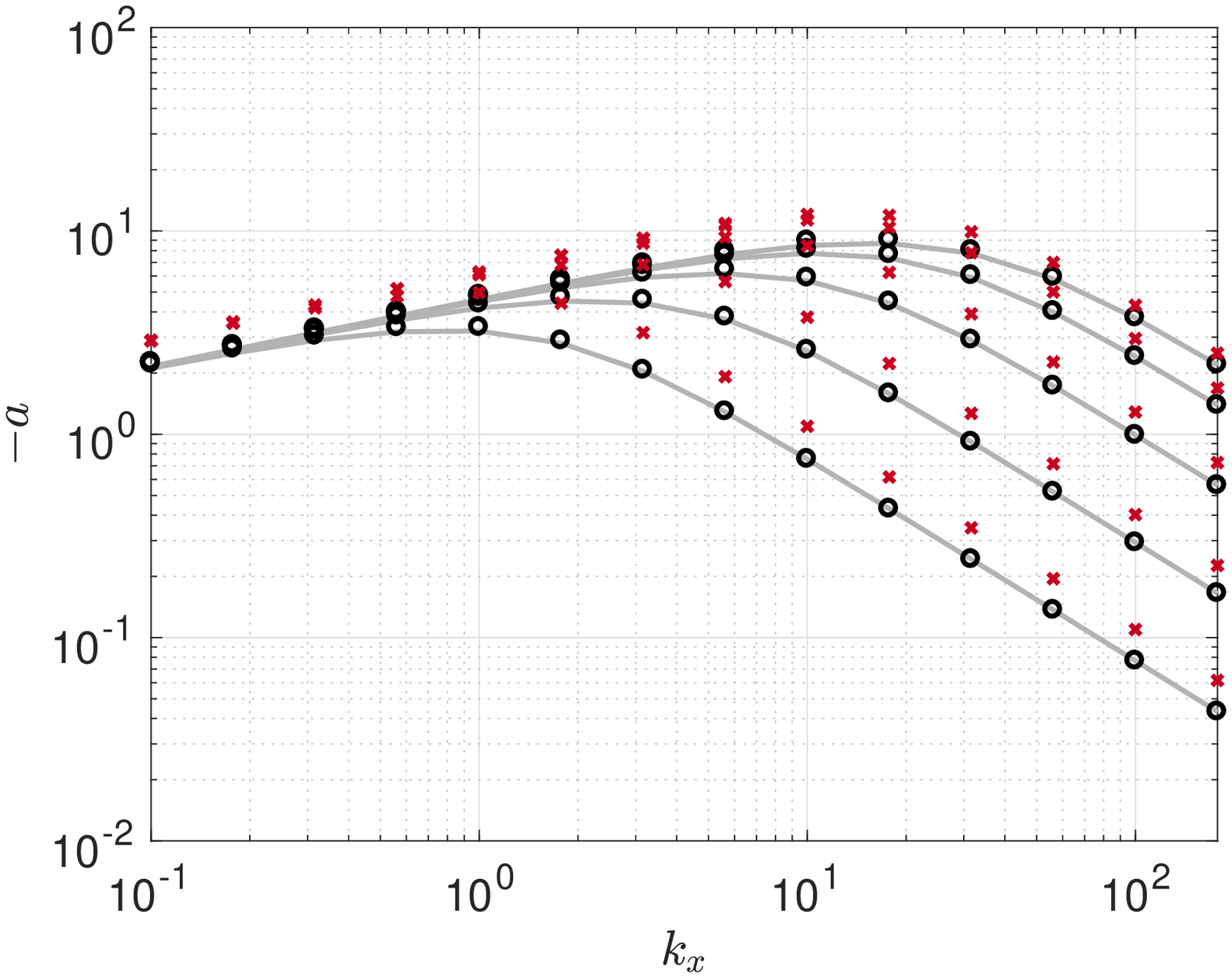}
(b)\includegraphics[width= 0.45\textwidth]{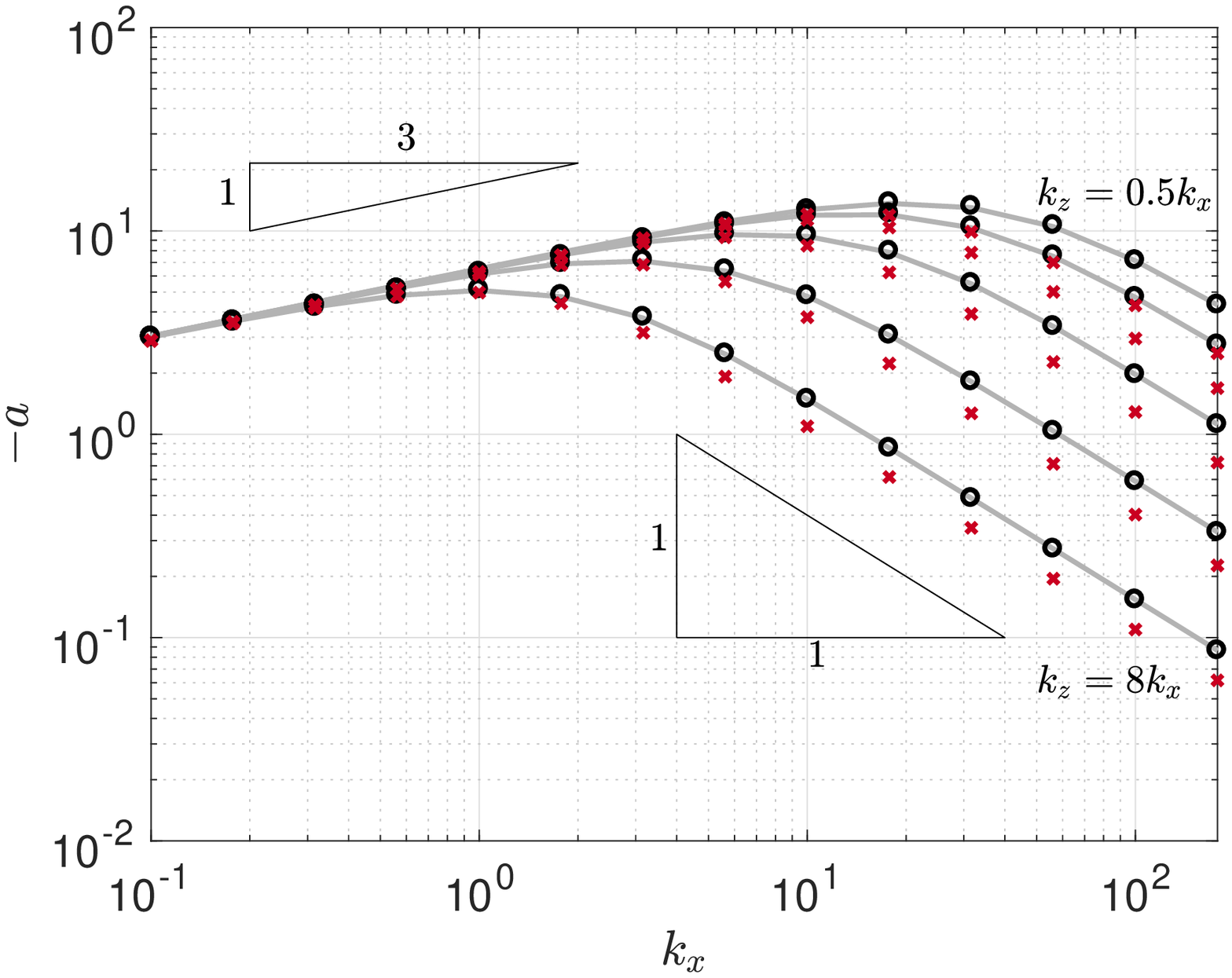}
(c)\includegraphics[width= 0.45\textwidth]{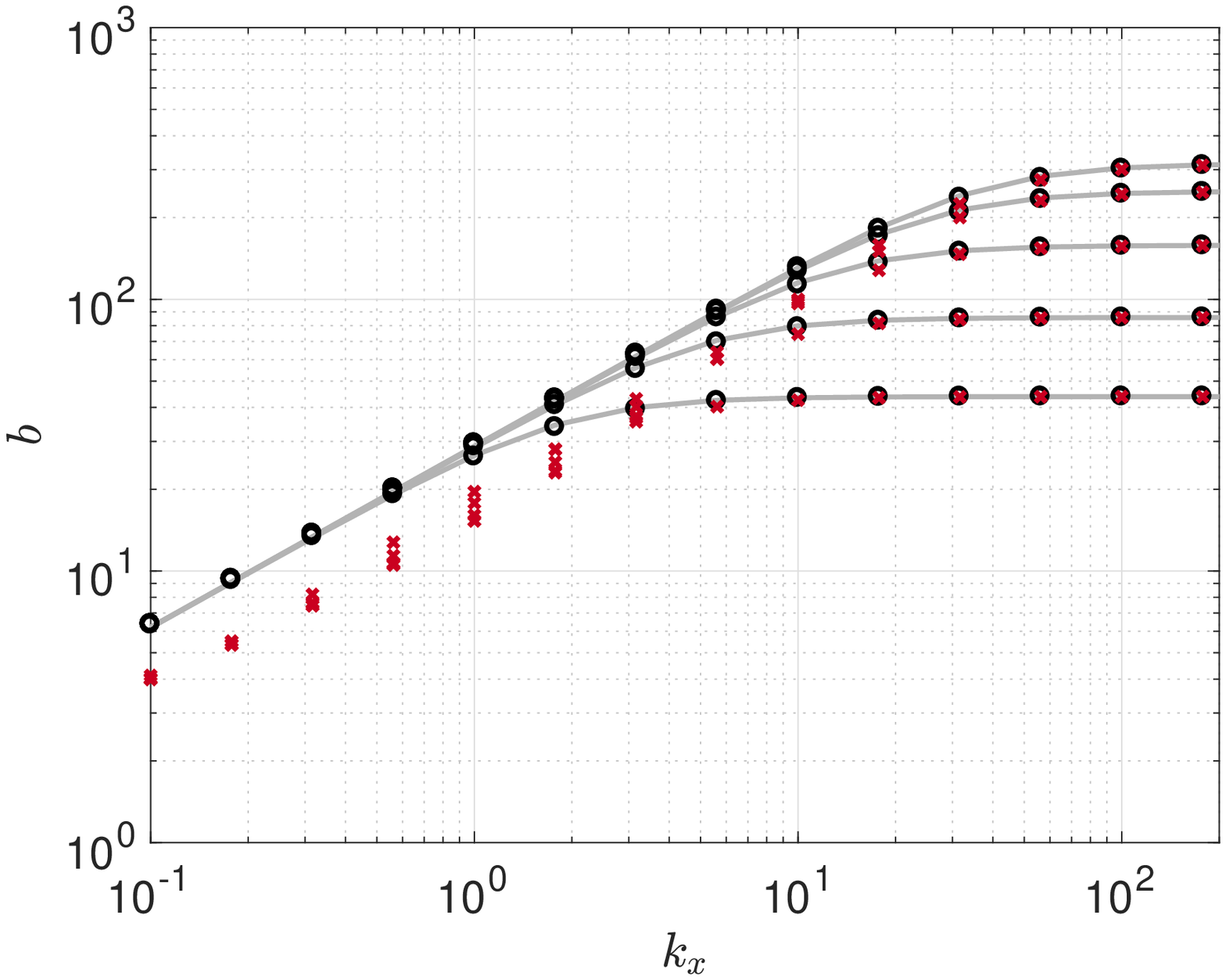}
(d)\includegraphics[width= 0.45\textwidth]{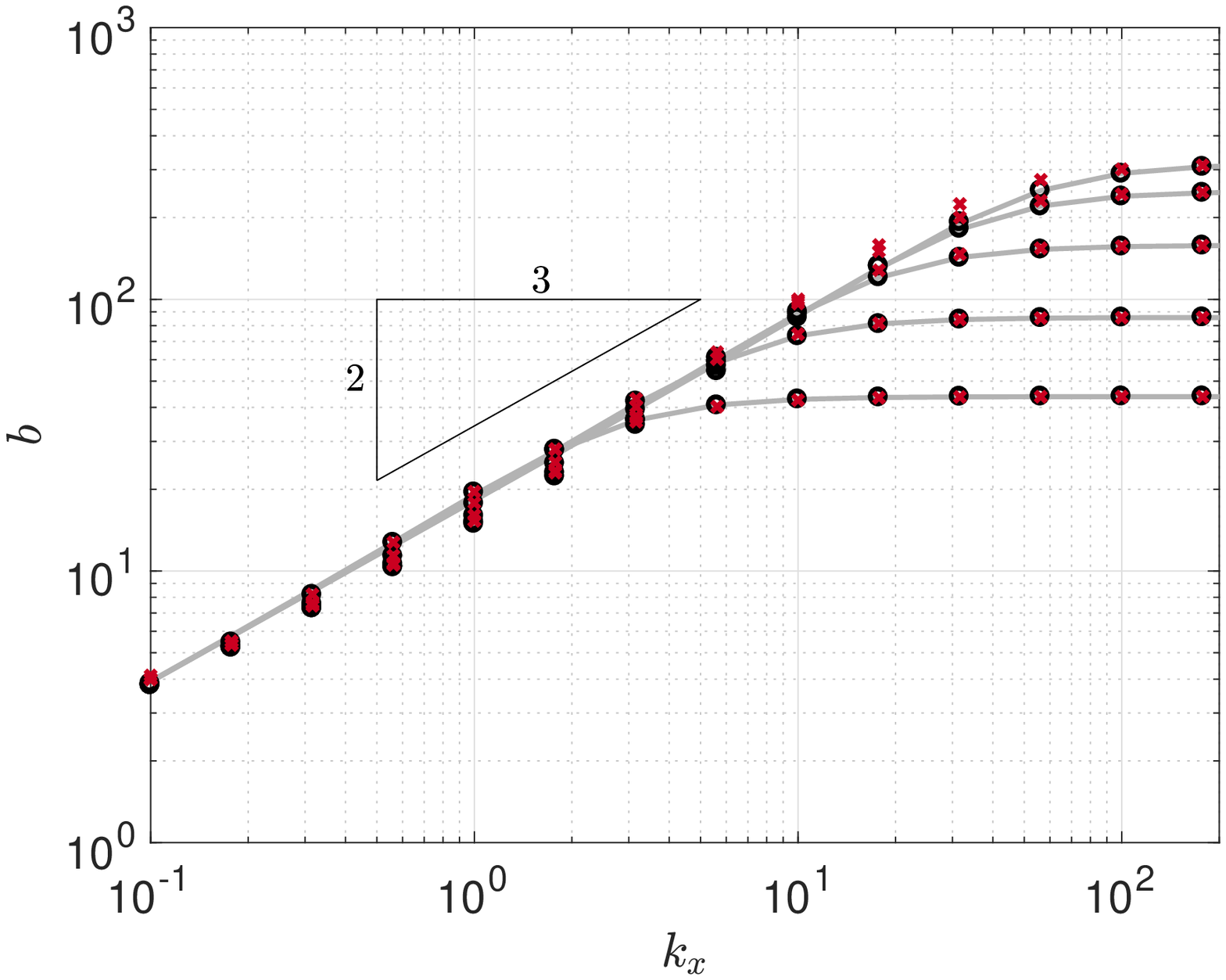}
}
\caption{Comparison between predicted ($-$) and fitted ($\circ$) mode shape parameters $a$ (subplots a,b) and $b$ (c,d)
for the Squire operator with a standard (a,c) and Laplacian (b,d) inner product, as a function of streamwise wavenumber $k_x$, for  $Re = 1000$ and spatial aspect ratios $k_z/k_x \in  \{0.5,1,2,4,8\}$. Also shown are fitted shape parameters for the $\eta$ component of the full Navier--Stokes system ($\times$'s).}
\label{fig:abkxCouette}
\end{figure}
Figure \ref{fig:abkxCouette} compares the predicted shape parameters to those obtained from fitting these parameters to numerically-computed modes, which are computed for $Re= 1000$ and various values of $k_x$ and $k_z/k_x$.  We observe in subplots (a) and (c) that the Squire mode parameters are accurately predicted from solving equations \ref{eq:Jsqa} and \ref{eq:Jsqb}. Subplots (b) and (d) show that solving equations \ref{eq:Josa} and \ref{eq:Josb} predicts the fitted parameters to the leading resolvent response mode of the Squire operator with Laplacian inner product. Furthermore, this prediction also accurately predicts the shape of the wall-normal vorticity component of leading response modes for the full Navier--Stokes system. 
Note that the scaling laws for small $k_x$ are identical to those for small $R$ observed in figure \ref{fig:abevolutionR}. The trends at high $k_x$ differ from those in figure \ref{fig:abevolutionR}, due to the fact that here we keep $k_z$ proportional to $k_x$ (and thus  $k_\perp \propto R$), whereas in figure  \ref{fig:abevolutionR} constant $\omega_c$ resulted in $k_\perp^2 \propto R$. These trends can again be inferred from studying the dominant terms in the evolution equations for the governing parameters, which can be computed from considering the partial derivatives of $J_{sq}$ and $J_{os}$ with respect to $a$ and $b$, which we omit for brevity. 

For large $k_x$, the mode shapes for both operators converge, which may be explained by the fact that the Laplacian operator  is dominated by the constant $k_\perp^2$ term in this regime. As a consequence, here analysis of the Squire operator gives accurate predictions of the behaviour of the full Navier--Stokes system, particularly for the mode width, which approaches a constant value with increasing $k_x$. There is some difference between the phase gradient parameter $a$ for the full system and for the Squire system with both the standard and Laplacian inner product (with the value for the full system lying between those for the two simplified systems),  though this difference decreases as $k_x$ increases and the phase variation decays towards zero.

\begin{figure}
 \centering {
(a)\includegraphics[width= 0.45\textwidth]{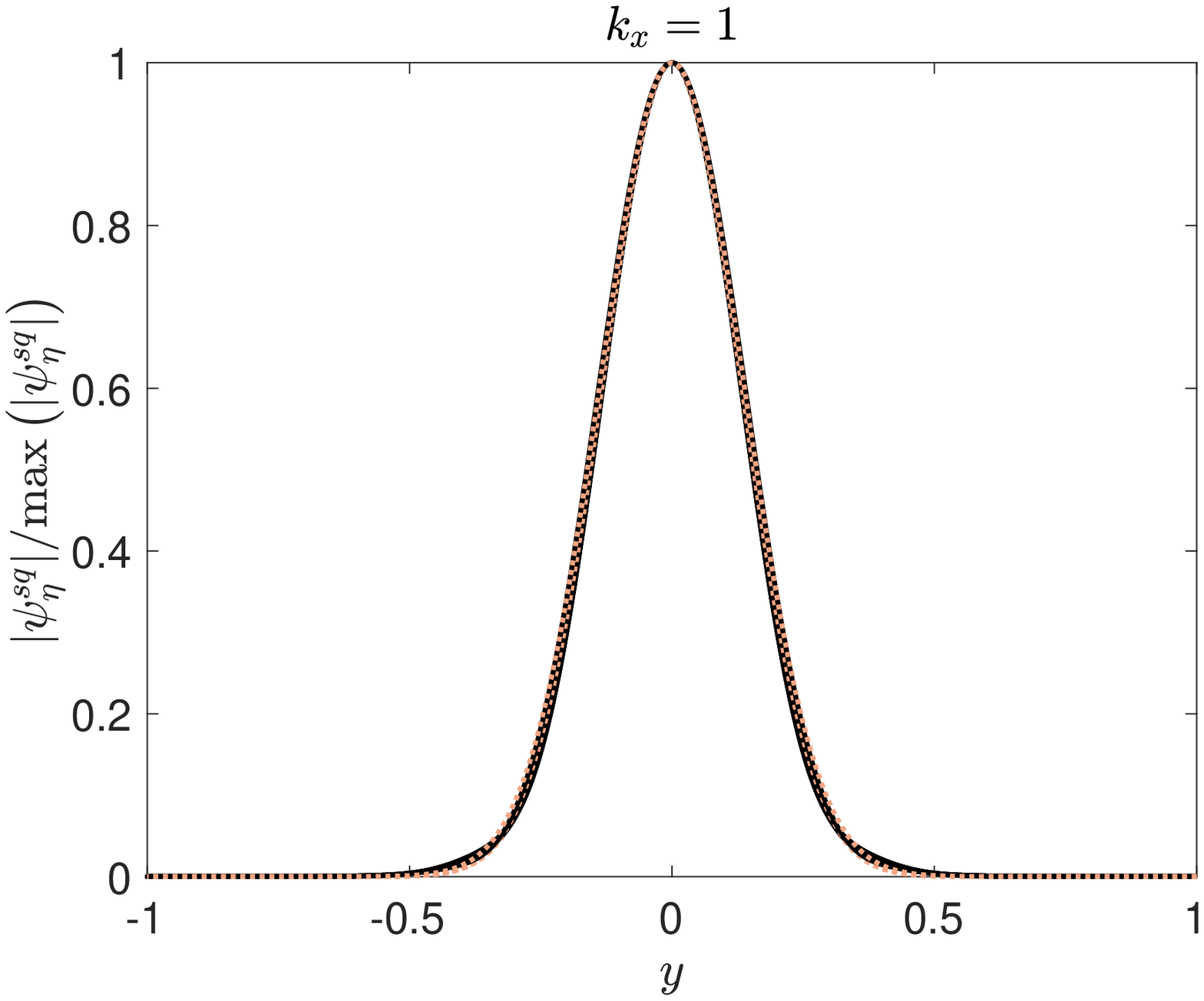}
(b)\includegraphics[width= 0.45\textwidth]{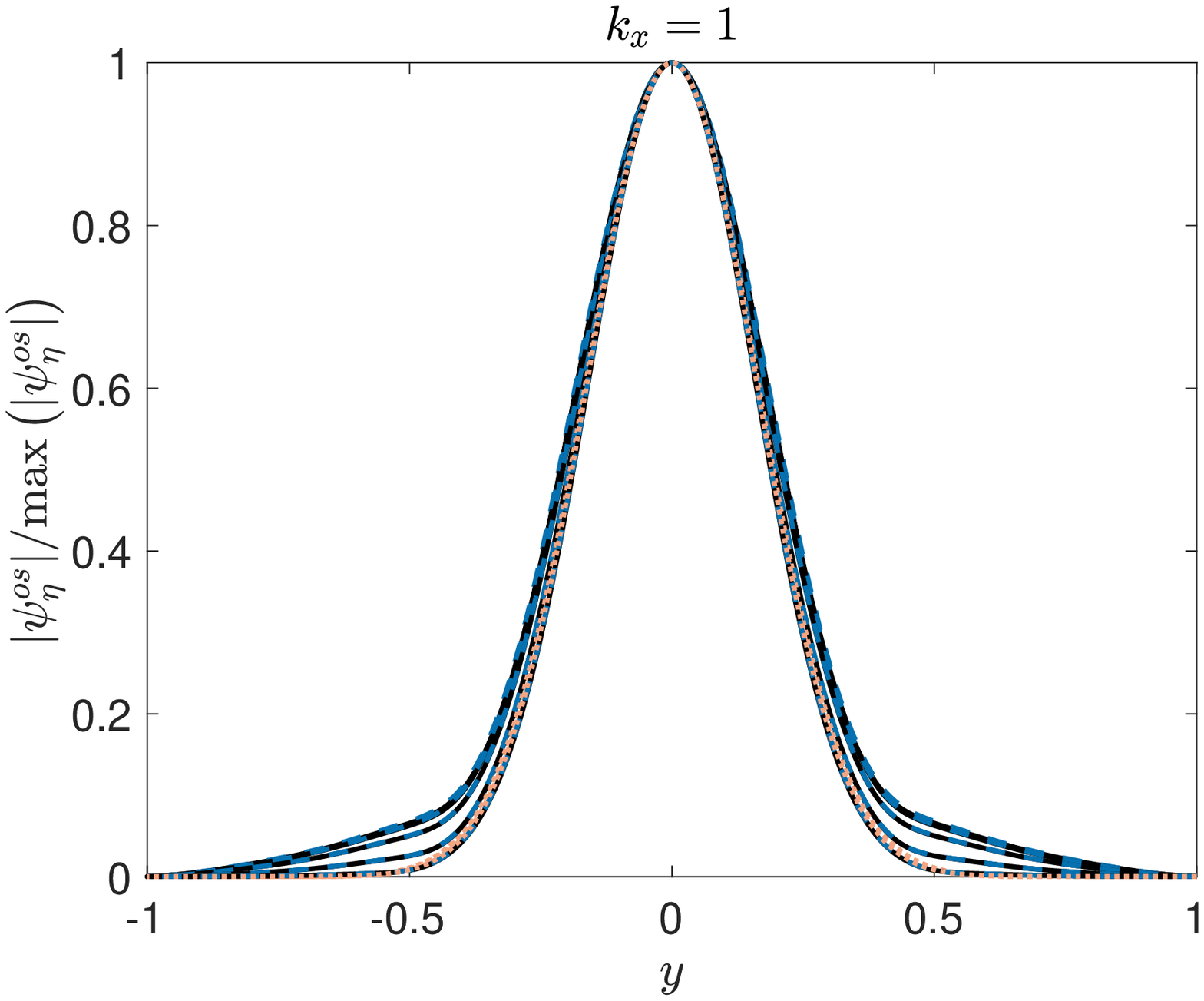}
(c)\includegraphics[width= 0.45\textwidth]{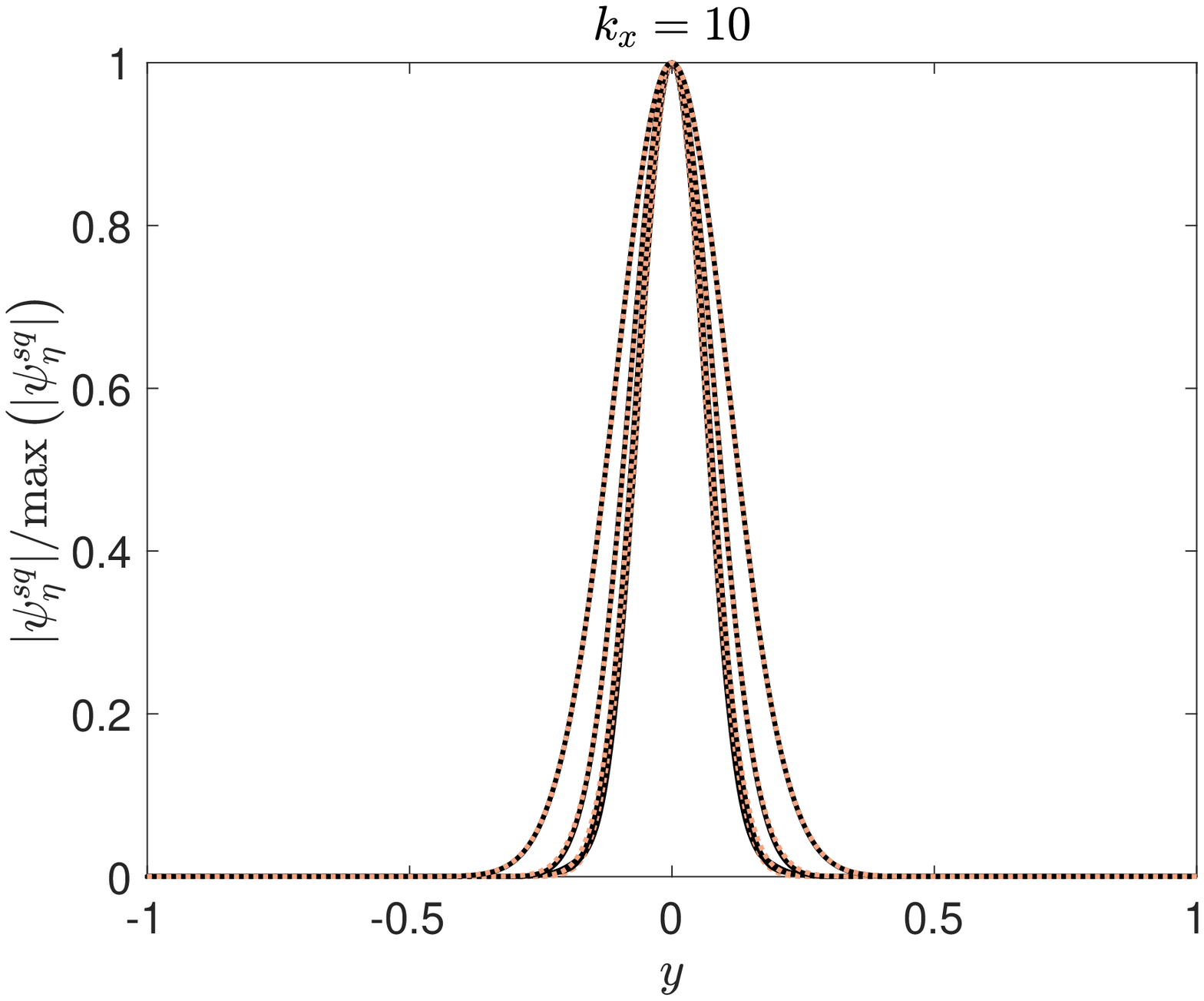}
(d)\includegraphics[width= 0.45\textwidth]{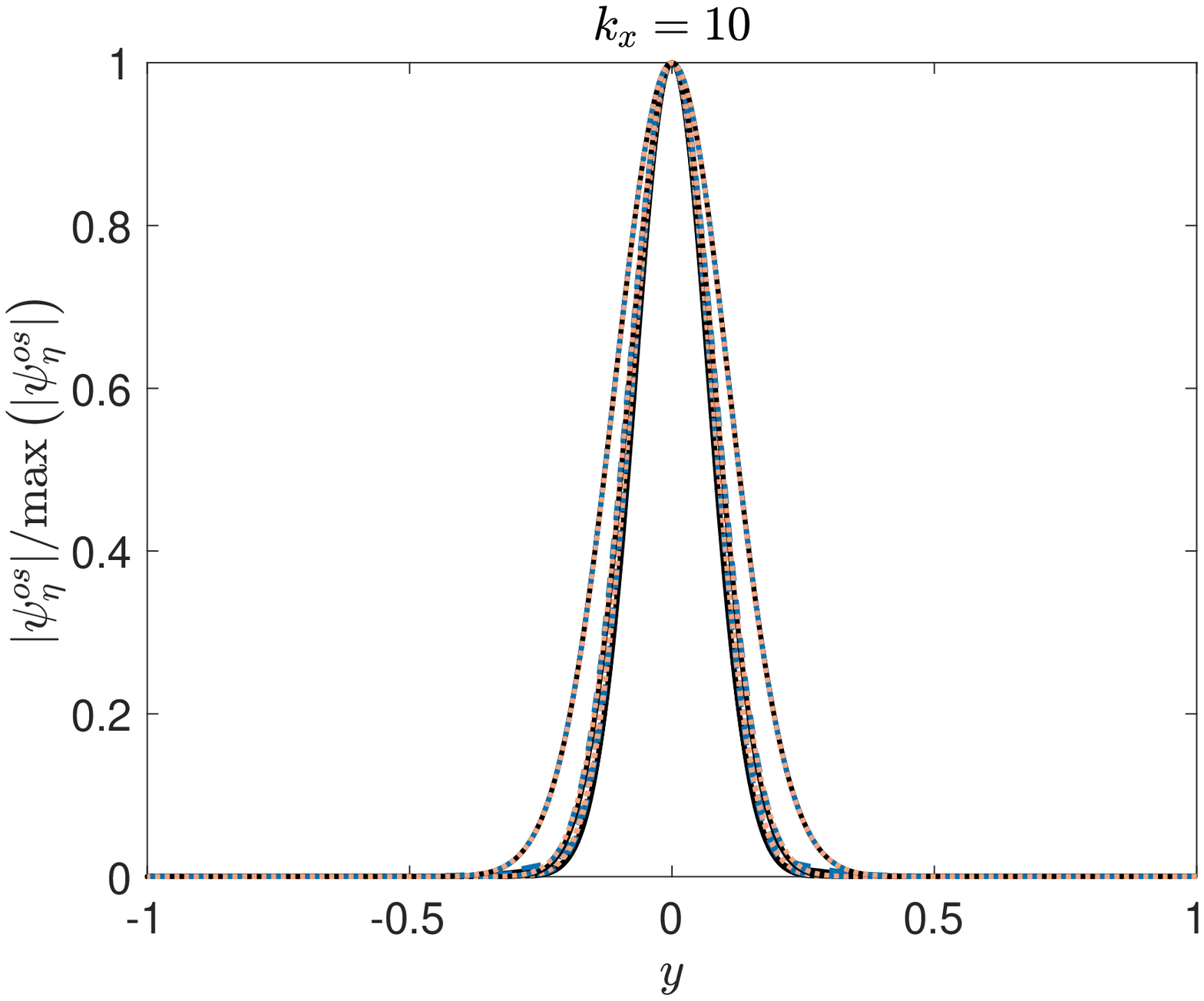}
(e)\includegraphics[width= 0.45\textwidth]{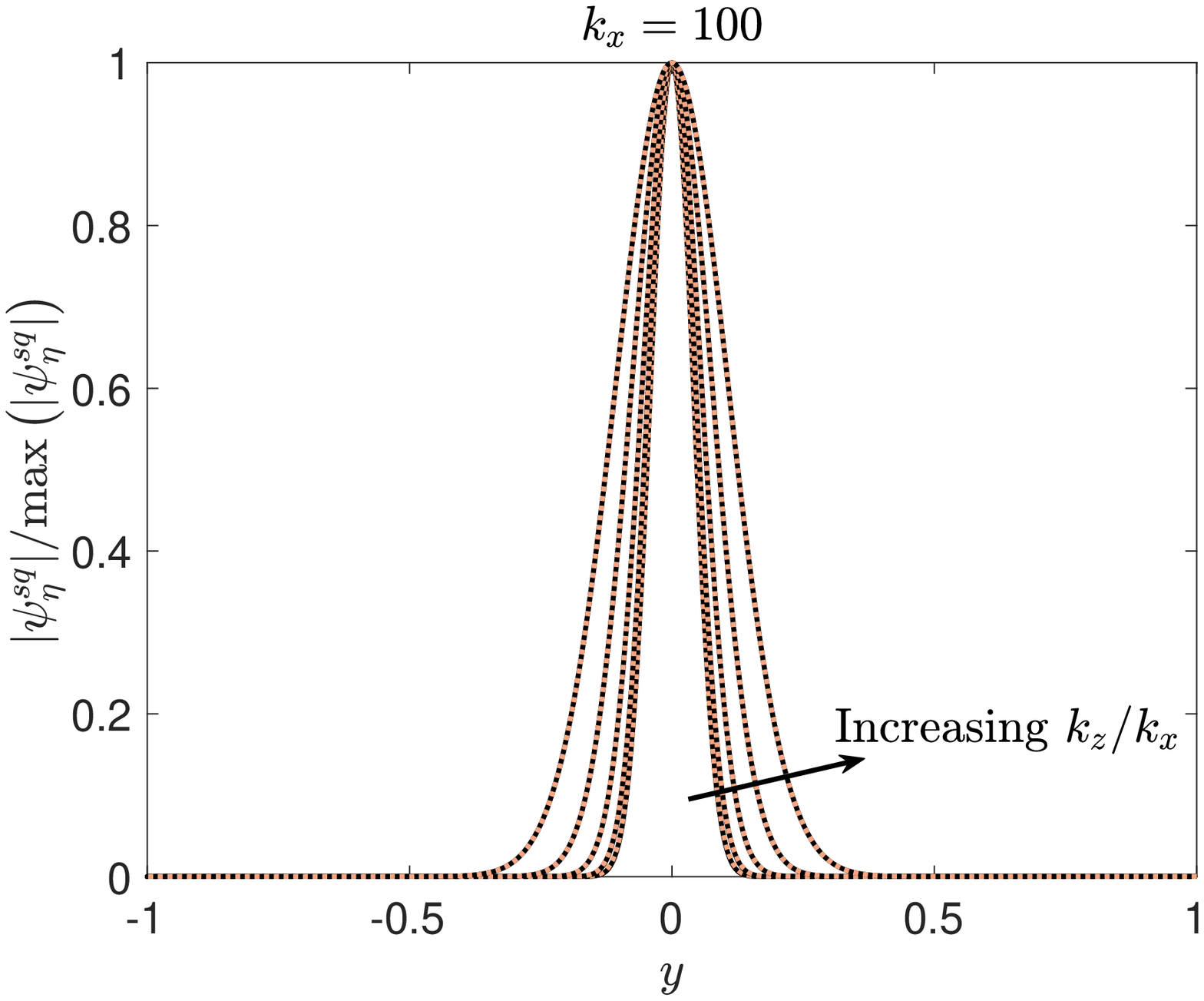}
(f)\includegraphics[width= 0.45\textwidth]{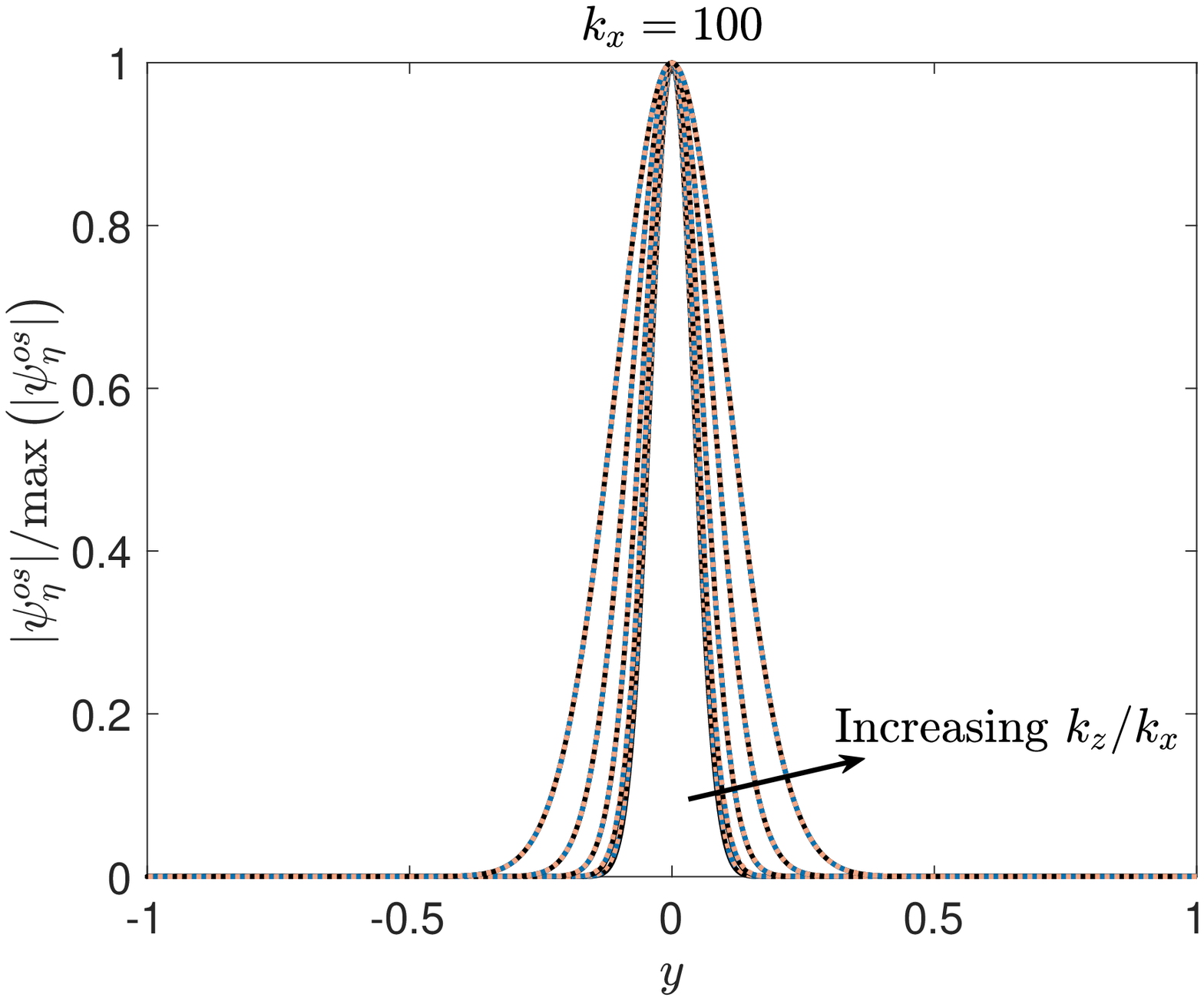}
}
\caption{Comparison between true (numerically computed, solid lines) and predicted (dotted lines) mode amplitude for laminar Couette flow with $Re = 1000$ and $k_x = 1$ (a,b) 10 (c,d) and 100 (e,f), and aspect ratios $k_z/k_x \in \{0.5,1,2,4,8\}$. Subplots (a,c,e) show results for the Squire subsystem, while subplots (b,d,f) are for the Navier--Stokes operator. The dashed lines in subplots (b,d,f) also show (numerically computed) mode amplitudes for the Orr-Sommerfeld subsystem.}
\label{fig:ModeShapesCouette}
\end{figure}

Figure \ref{fig:ModeShapesCouette} plots predicted and true mode amplitudes for several of the cases considered in figure \ref{fig:abkxCouette}. The predicted mode shapes for $\eta$ closely match the computed modes for both the Squire and full Navier--Stokes system, with the modes for the latter being very close to those of the OS subsystem. 
The largest discrepancy arises in the $k_x = 1$ case, where the tails of the mode amplitudes are significantly heavier than those of a Gaussian distribution, and extend far from the critical layer towards the wall. This phenomenon is related to a wider distribution in the $v$-component of the mode (not plotted), and gives a larger variation in fitted $b$ values for this $k_x$ for the full Navier--Stokes system, as observed in figure \ref{fig:abkxCouette}.

\begin{figure}
 \centering {
\includegraphics[width= 0.6\textwidth]{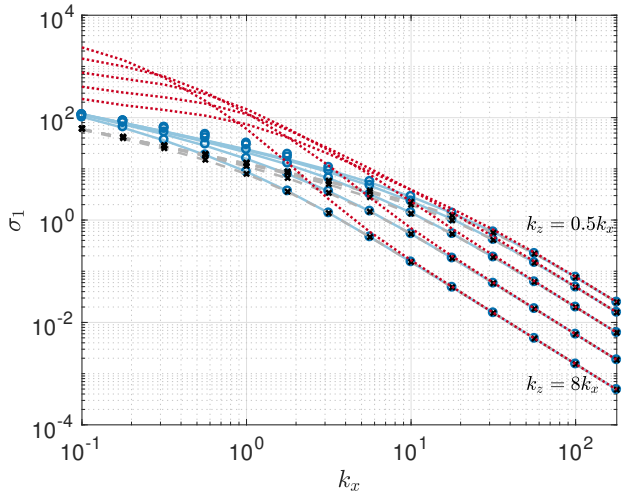}
}
\caption{Comparison between predicted and computed leading singular values for the Squire subsystem with regular (predicted $--$, computed $\times$), and Laplacian (predicted $-$, computed $\circ$) inner product, for  $Re = 1000$ and spatial aspect ratios $k_z/k_x \in  \{0.5,1,2,4,8\}$. Also shown are computed singular values for the full Navier--Stokes system ($\cdot\cdot$).}
\label{fig:sigkxCouette}
\end{figure}

Figure \ref{fig:sigkxCouette} compares the leading resolvent singular values for the Squire and Navier--Stokes systems to those estimated from the minima of the cost functions $J_{sq}$ and $J_{os}$, for the same wavenumbers considered in figure \ref{fig:abkxCouette}. The cost functions are able to accurately predict the singular values of the scalar Squire operator with both the regular and Laplacian inner product, though this is only accurate for the full Navier--Stokes system for large $k_x$ (and thus large $k_\perp^2$). In particular, these model operators are incapable of predicting the increase in singular value for high-aspect-ratio (i.e., large $k_z/k_x$) modes that is observed, though we have shown that they still accurately predict mode shapes in this regime. 

\subsection{Predicting mode shapes for a turbulent boundary layer}
\label{sec:BLpred}
\begin{figure}
 \centering {
\includegraphics[width= 0.24\textwidth]{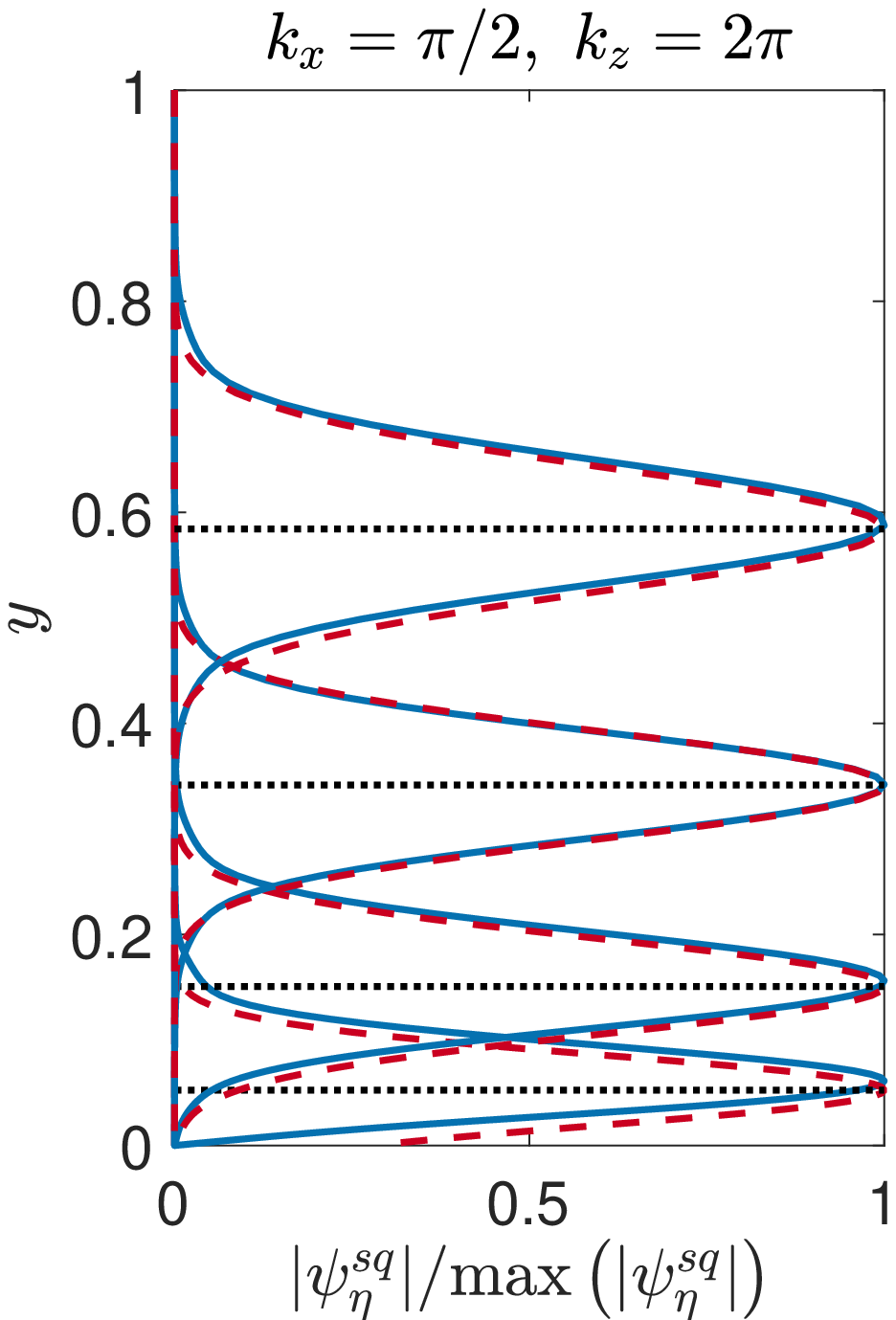}
\includegraphics[width= 0.24\textwidth]{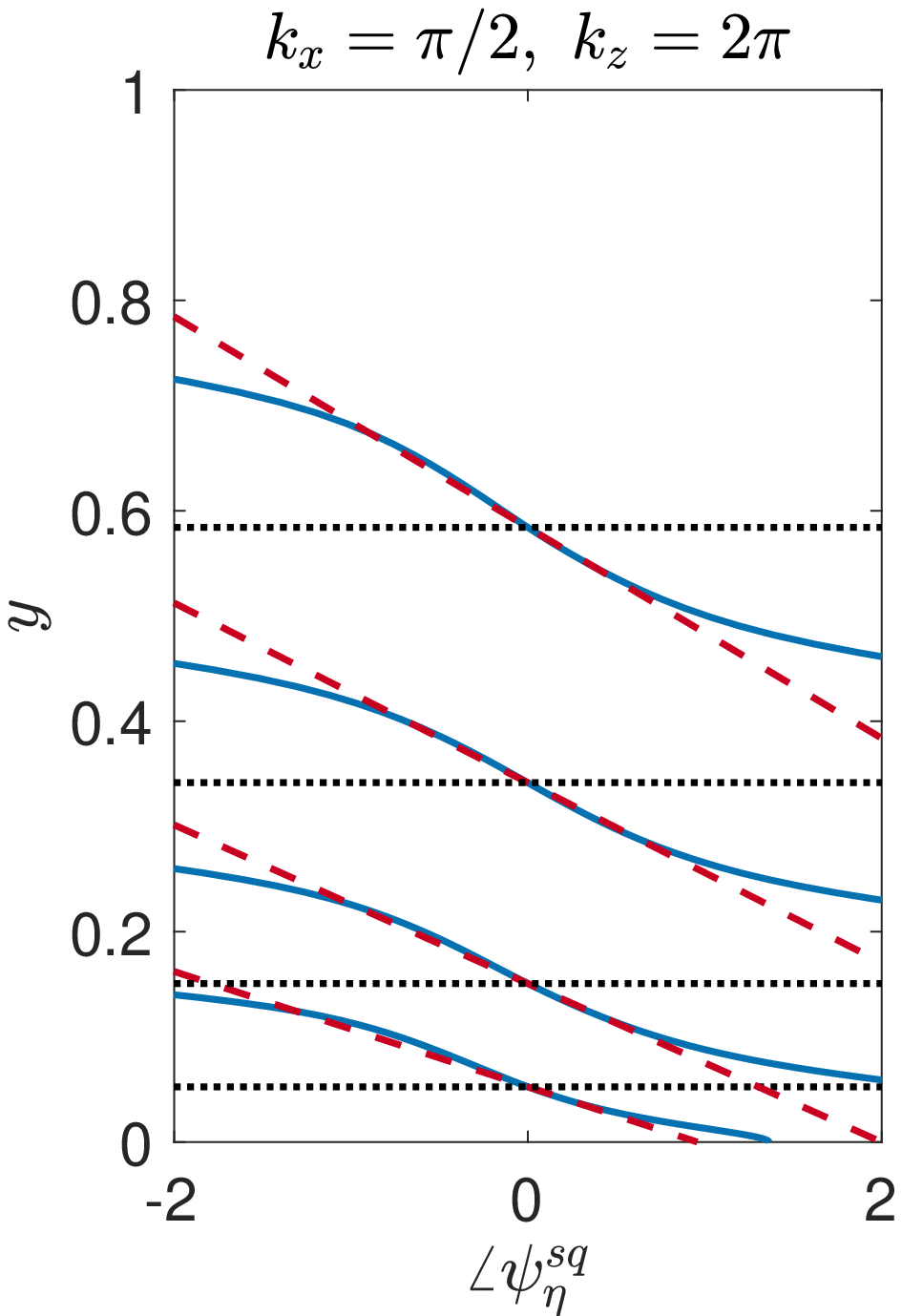}
\includegraphics[width= 0.24\textwidth]{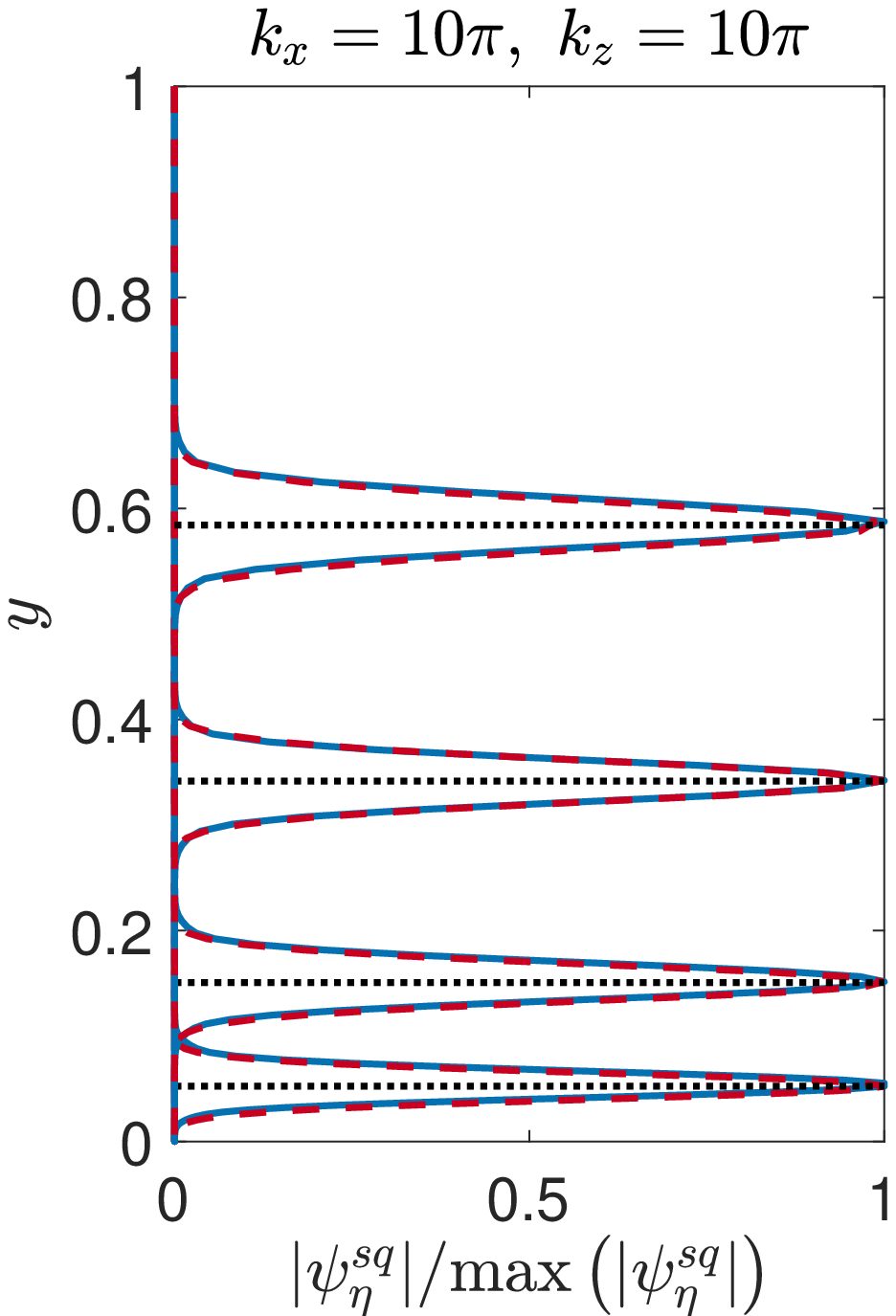}
\includegraphics[width= 0.24\textwidth]{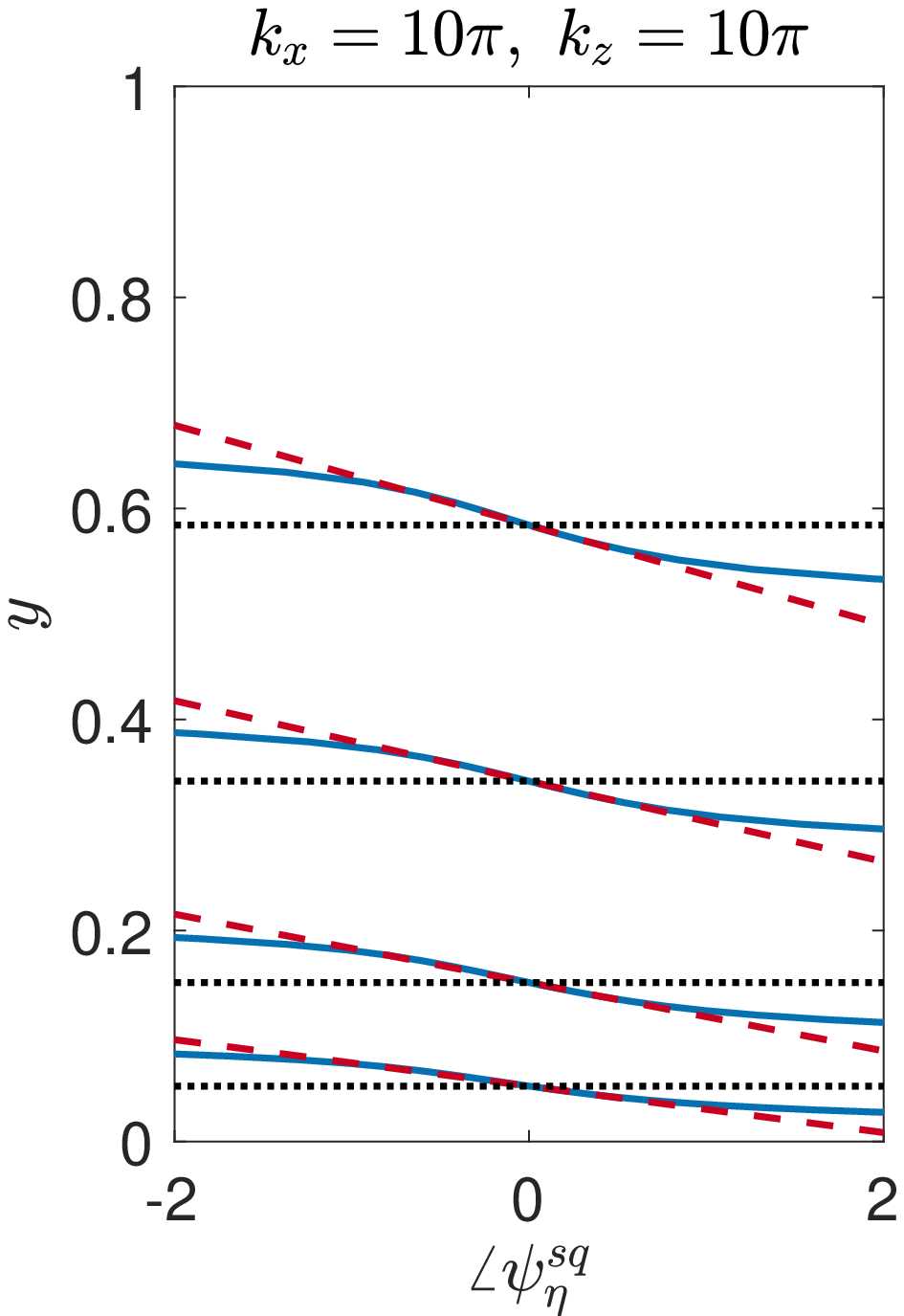}
\includegraphics[width= 0.24\textwidth]{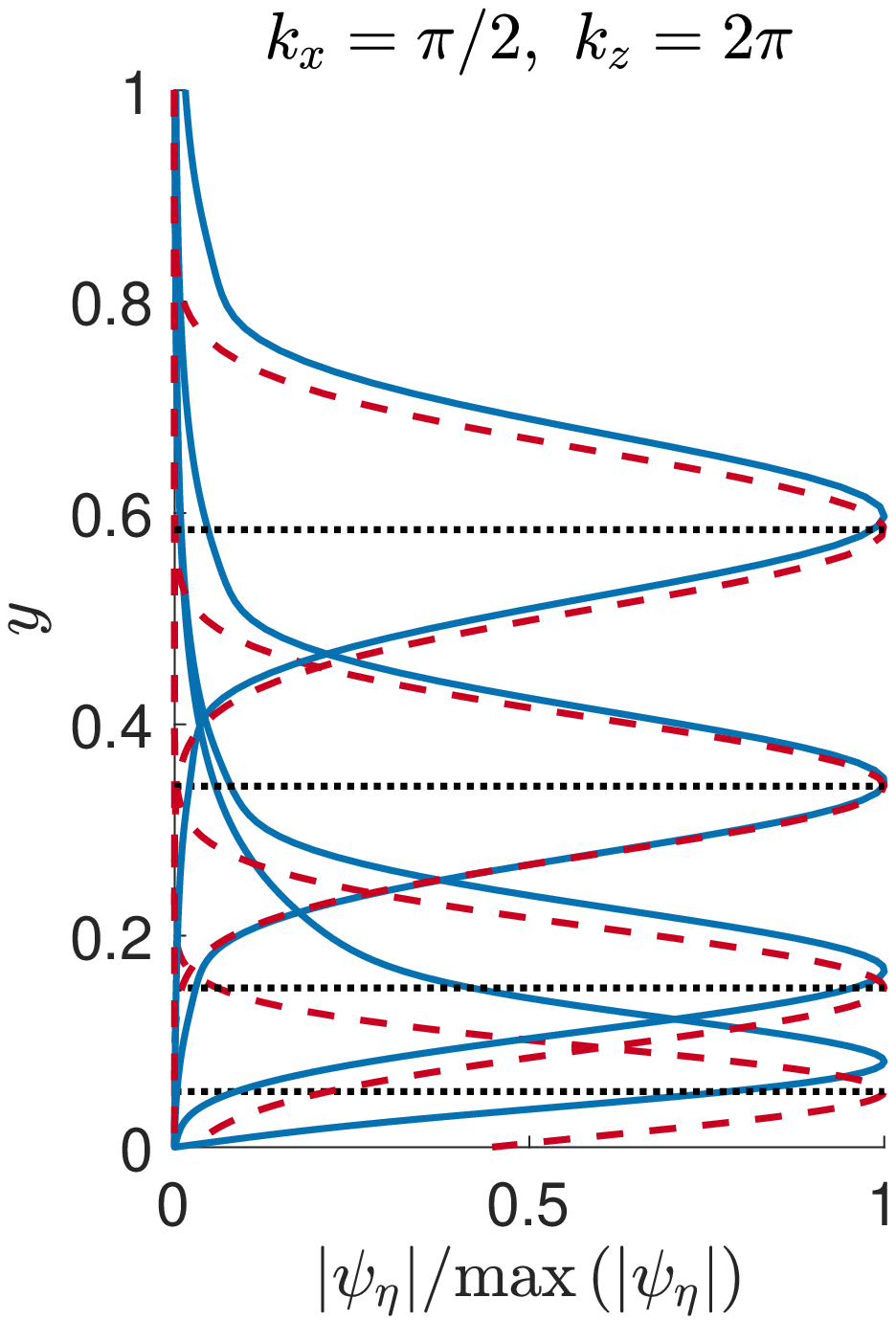}
\includegraphics[width= 0.24\textwidth]{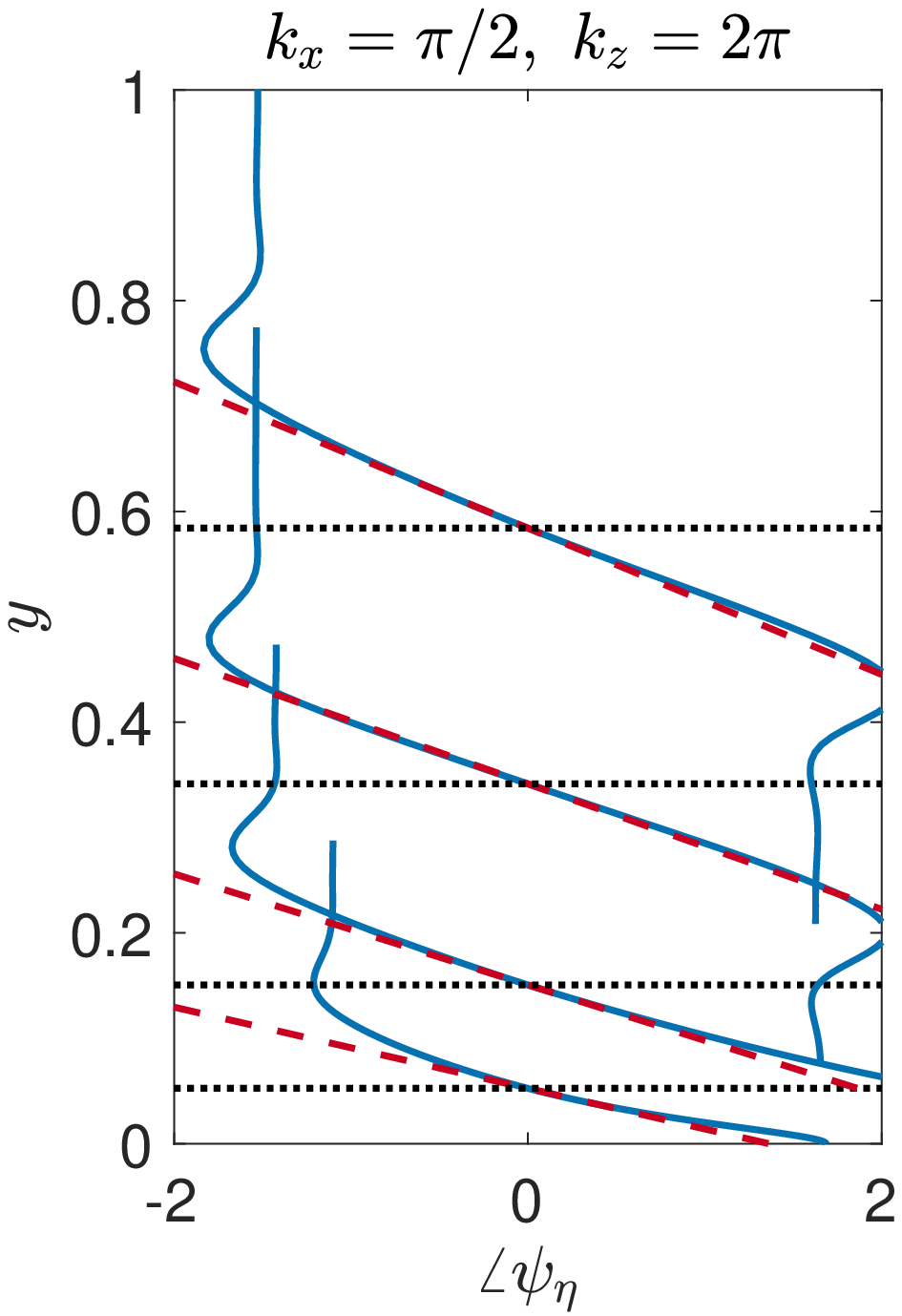}
\includegraphics[width= 0.24\textwidth]{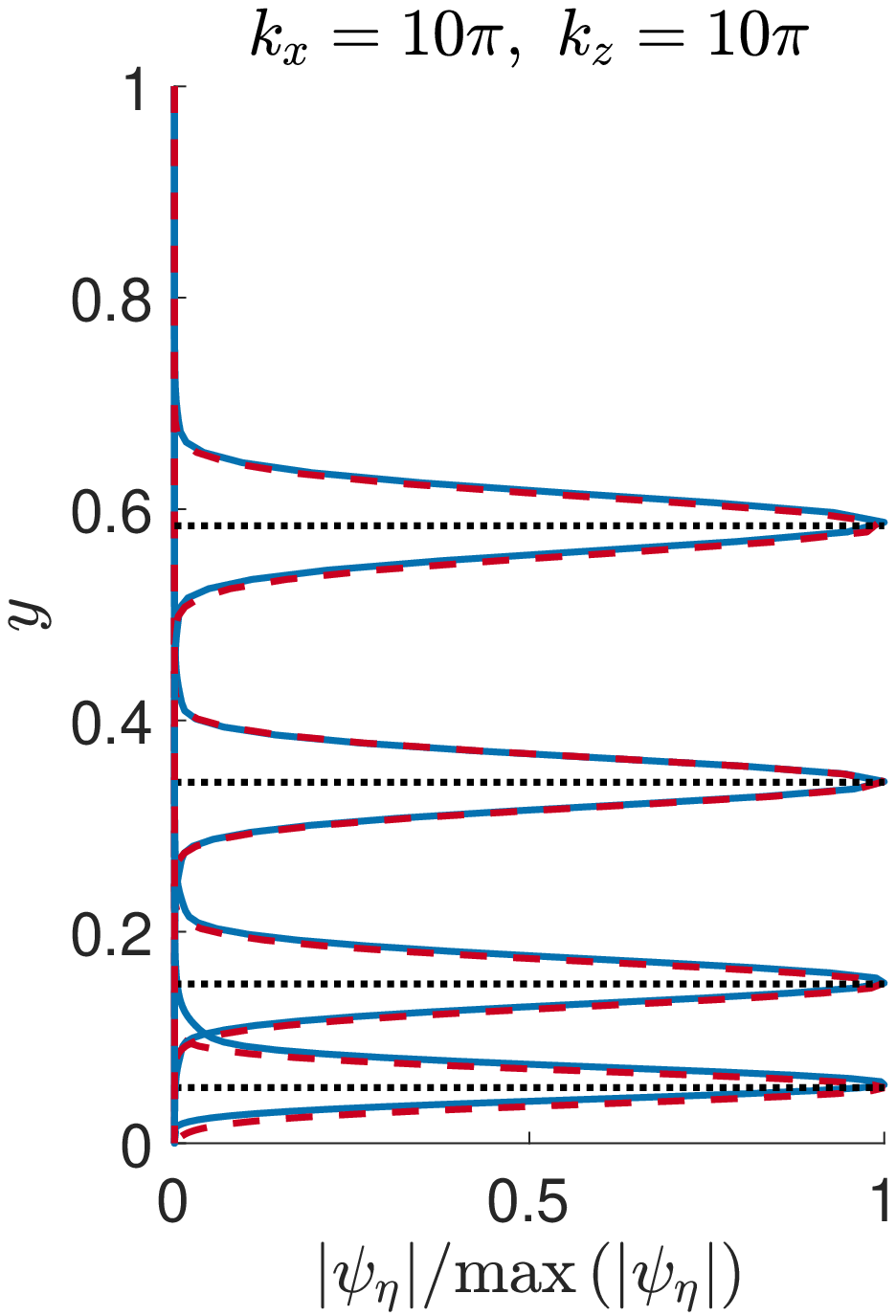}
\includegraphics[width= 0.24\textwidth]{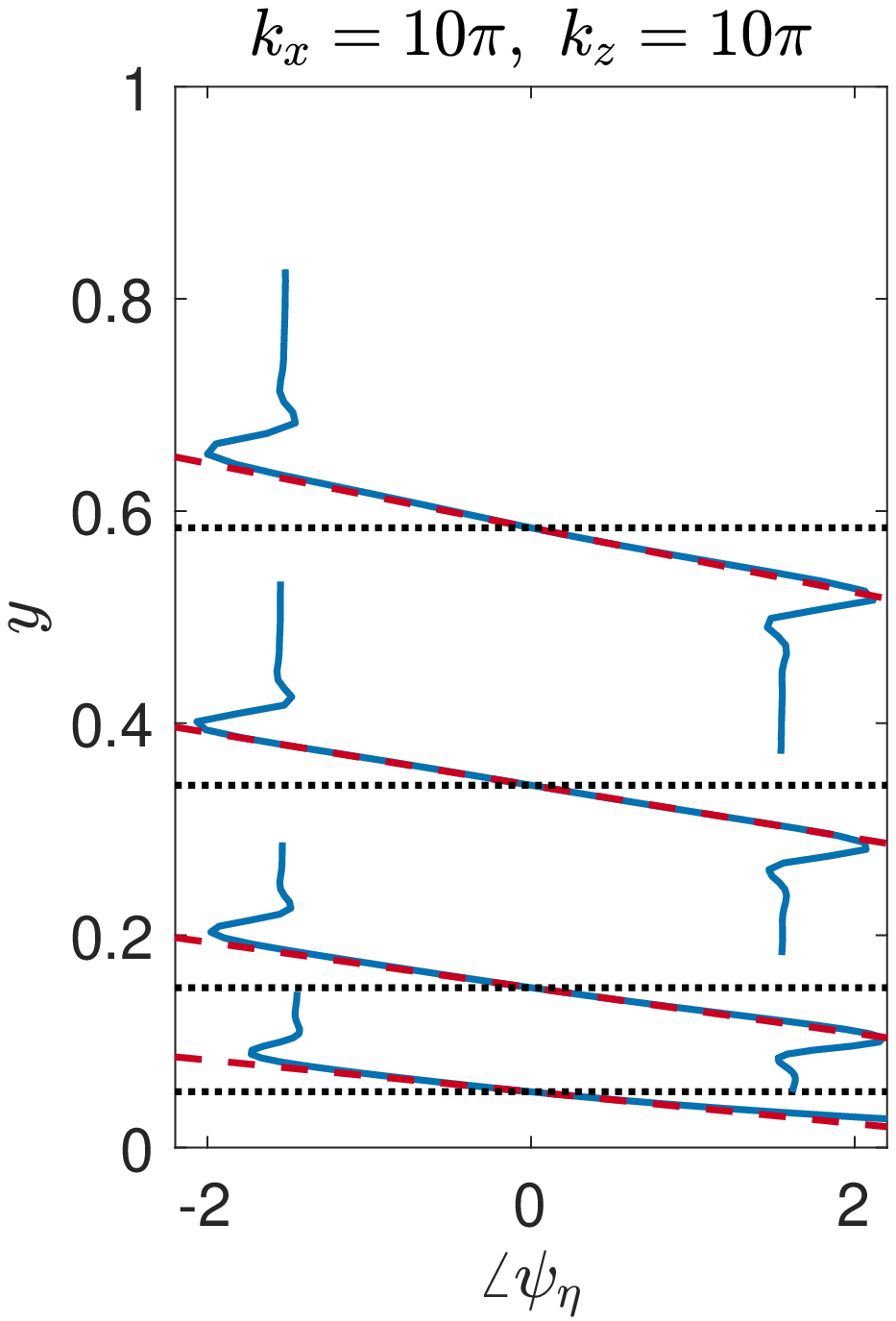}
}
\caption{True (solid lines) and predicted (dashed) leading resolvent mode amplitudes  and phases for the Squire system (top row) and wall-normal vorticity component of the Navier--Stokes system (bottom) for a turbulent boundary layer with $\Rey_\tau = 900$. Modes are shown for two wavenumber pairs, $(k_x,k_z) = (\pi/2,2\pi)$ and $(10\pi,10\pi)$, and temporal frequencies corresponding to wavespeeds of $0.6U_\infty$, $0.7U_\infty$, $0.8U_\infty$, $0.9U_\infty$, with critical layer locations indicated by dotted lines.}
\label{fig:BLpredModes}
\end{figure}

This section applies the methodology developed in sections \ref{sec:SquirePred} and \ref{sec:OSpred} to a turbulent boundary layer. 
By linearizing the mean velocity profile about the critical layer, the equations for predicting mode shape parameters are the same as those developed in section \ref{sec:OSpred} (i.e., equations \ref{eq:Jsq}-\ref{eq:Josb}), though now $U_{y_c}$ and $R = k_x U_{y_c}\Rey $ are dependent on the critical layer location. 
Figure \ref{fig:BLpredModes} shows the predicted and true response mode shapes for the Squire and Navier--Stokes resolvent operators for two pairs of spatial wavenumbers and various critical layer locations. Note that the case where $\{k_x,k_z,c\} = \{\pi/2,2\pi,0.8U_\infty \}$ as considered in figures \ref{fig:Modes1}-\ref{fig:Modes3} as representative of a typical large-scale motion, is included. The location typical of very large-scale motions ($c \approx 0.6U_\infty$) is also included, though these structures would correspond to slightly smaller streamwise wavenumbers than $k_x = \pi/2$. We observe that mode amplitude and phase variation (in the local region of high amplitude) is accurately estimated, provided the mode is not substantially affected by the presence of the wall. This shows in particular the validity of using a mean velocity profile linearised about the critical layer to estimate mode shapes. 

\begin{figure}
 \centering {
(a)\includegraphics[width= 0.45\textwidth]{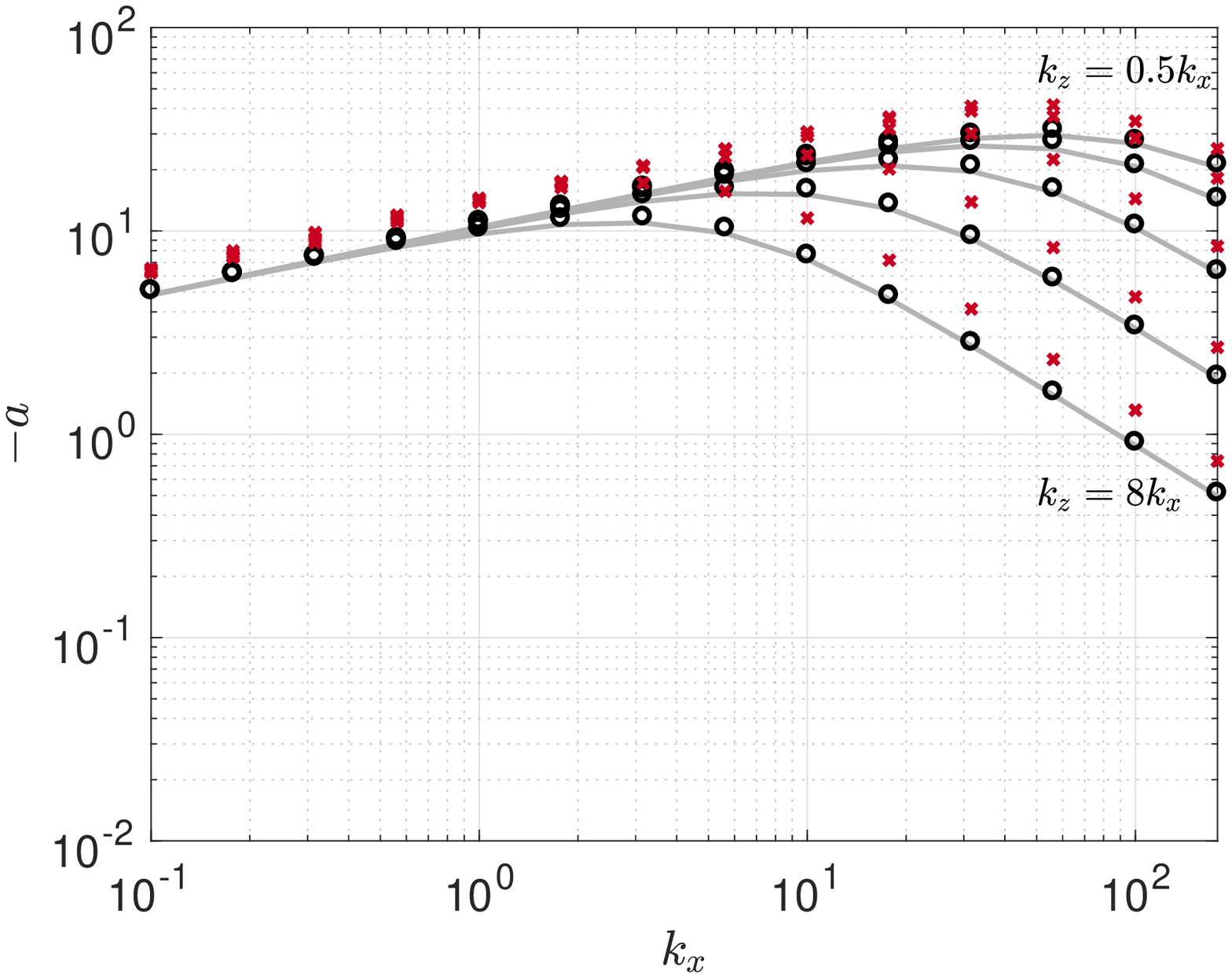}
(b)\includegraphics[width= 0.45\textwidth]{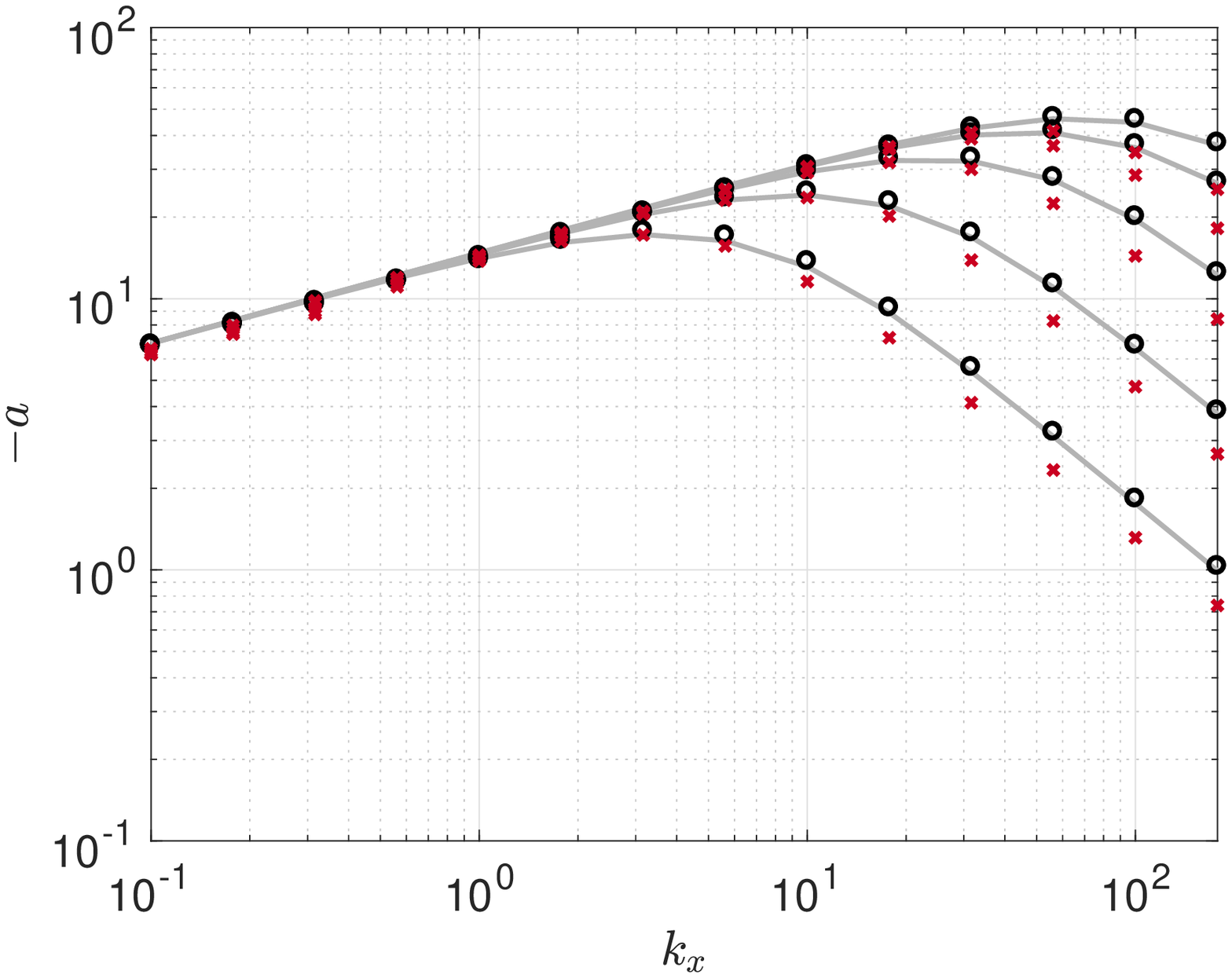}
(c)\includegraphics[width= 0.45\textwidth]{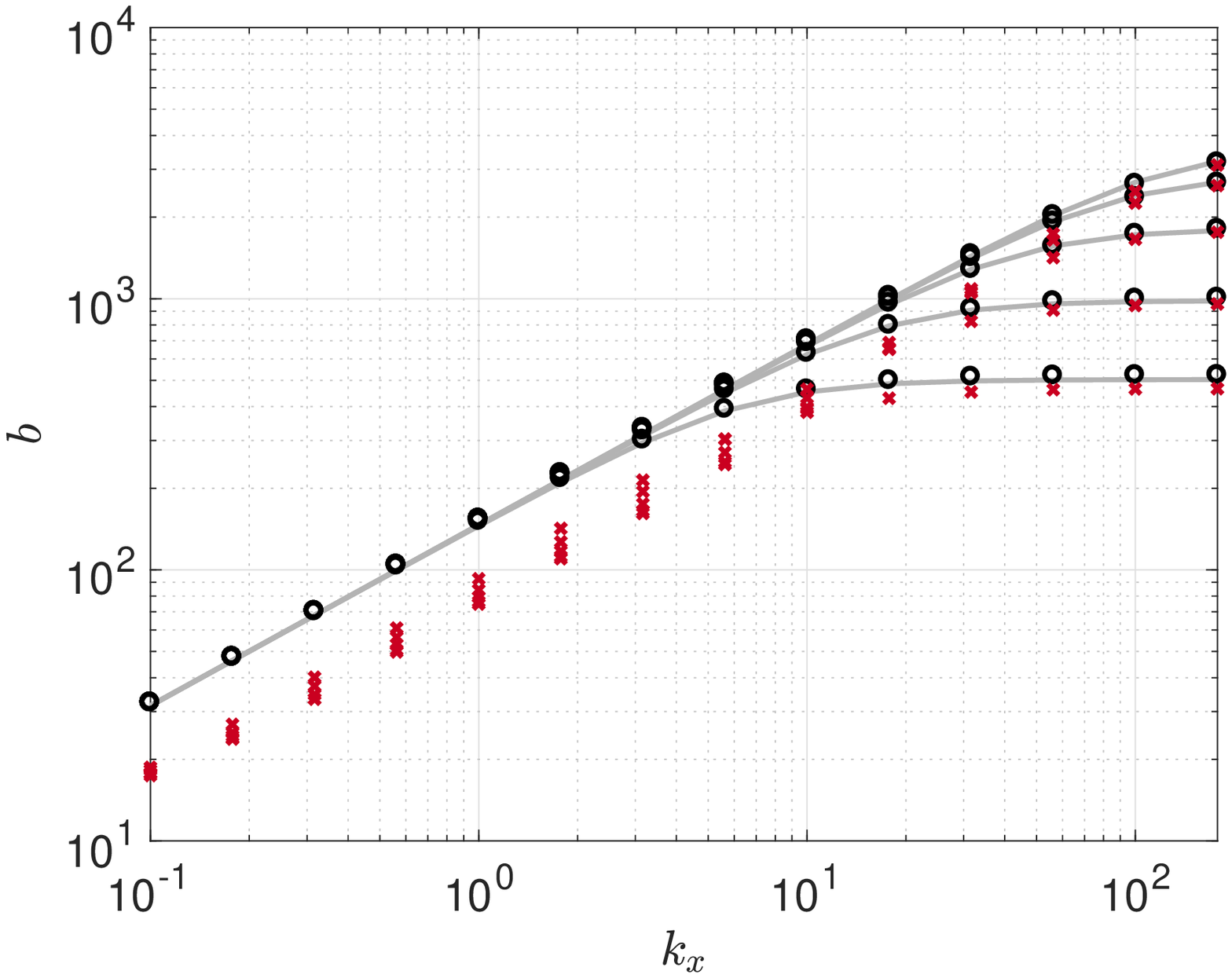}
(d)\includegraphics[width= 0.45\textwidth]{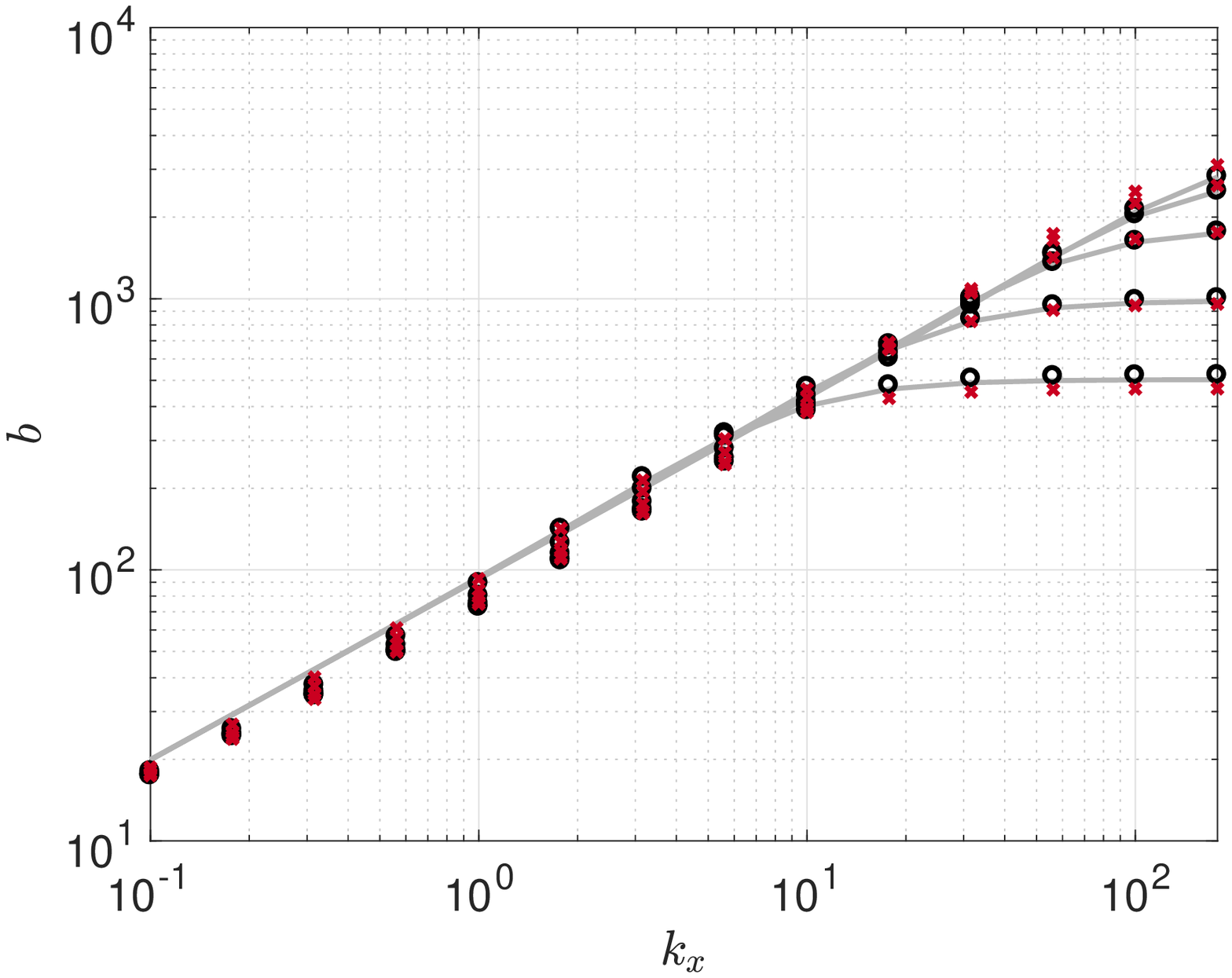}
}
\caption{Comparison between predicted ($-$) and fitted ($\circ$) mode shape parameters $a$ (subplots a,b) and $b$ (c,d)
for the Squire operator with a standard (a,c) and Laplacian (b,d) inner product, as a function of streamwise wavenumber $k_x$, for various mode aspect ratios for a turbulent boundary layer with $Re_\tau = 900$ and $c = 0.8U_\infty$. Also shown are fitted shape parameters for the $\eta$ component of the full Navier--Stokes system ($\times$'S).}
\label{fig:abkxBL}
\end{figure}

\begin{figure}
 \centering {
(a)\includegraphics[width= 0.45\textwidth]{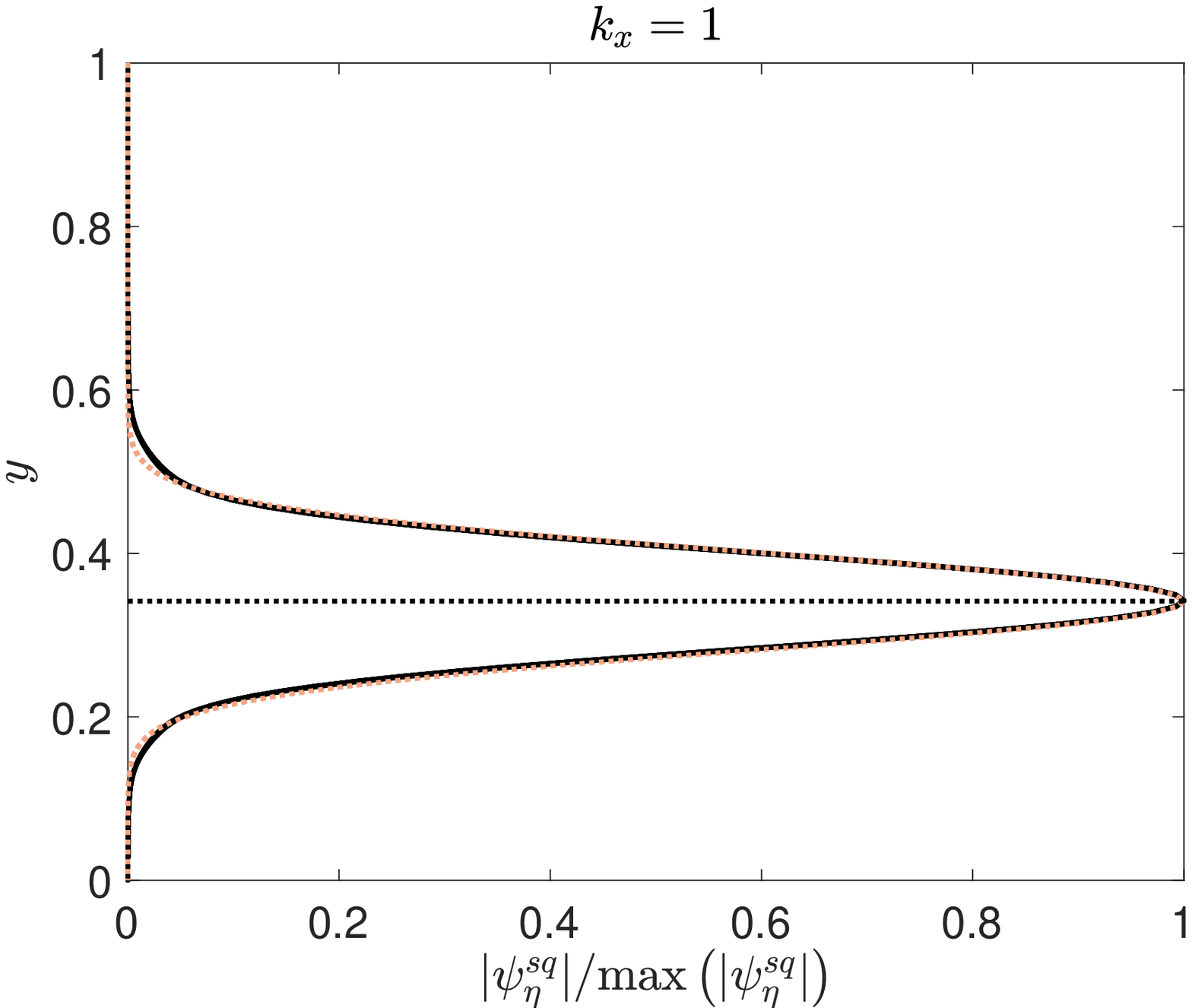}
(b)\includegraphics[width= 0.45\textwidth]{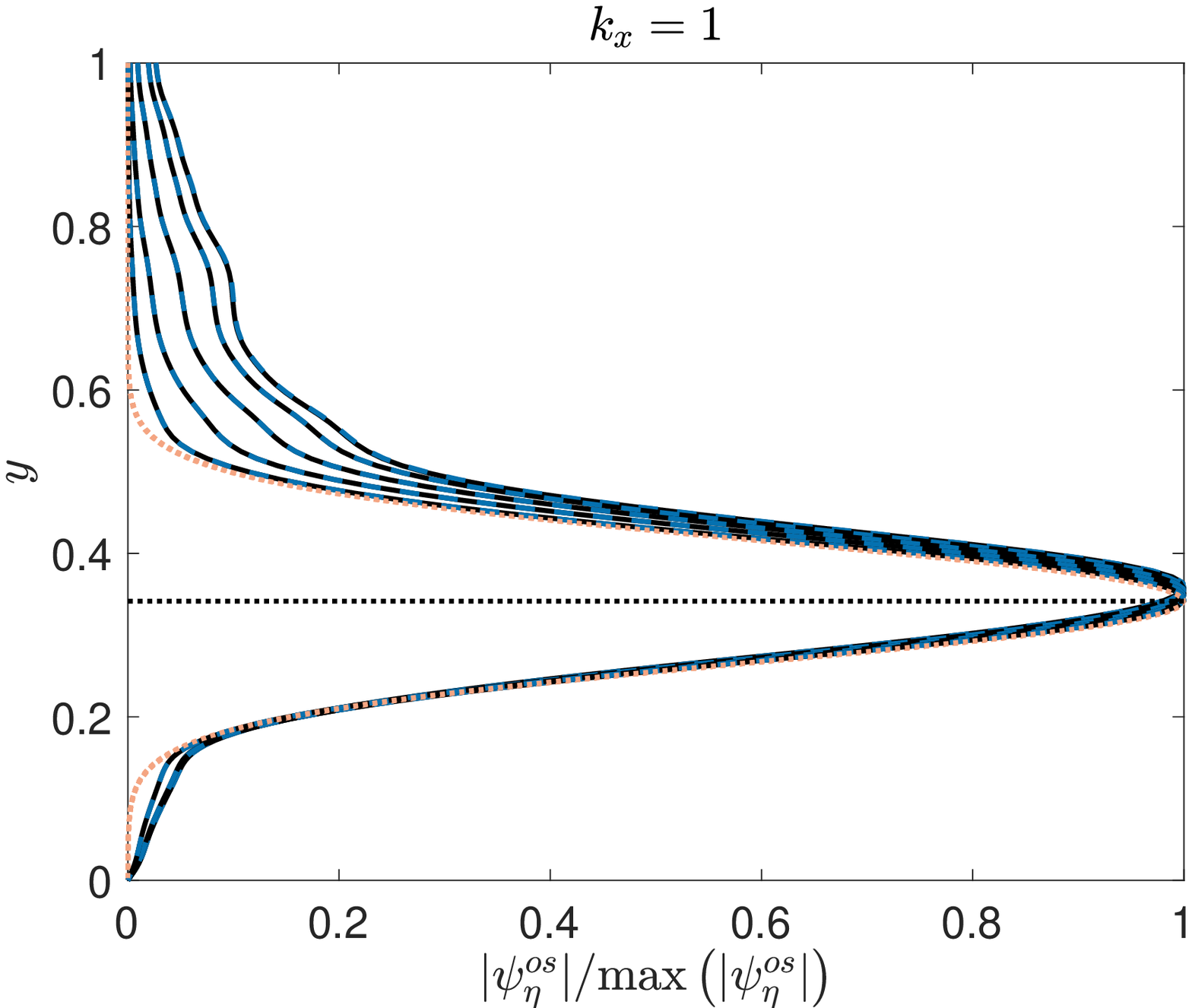}
(c)\includegraphics[width= 0.45\textwidth]{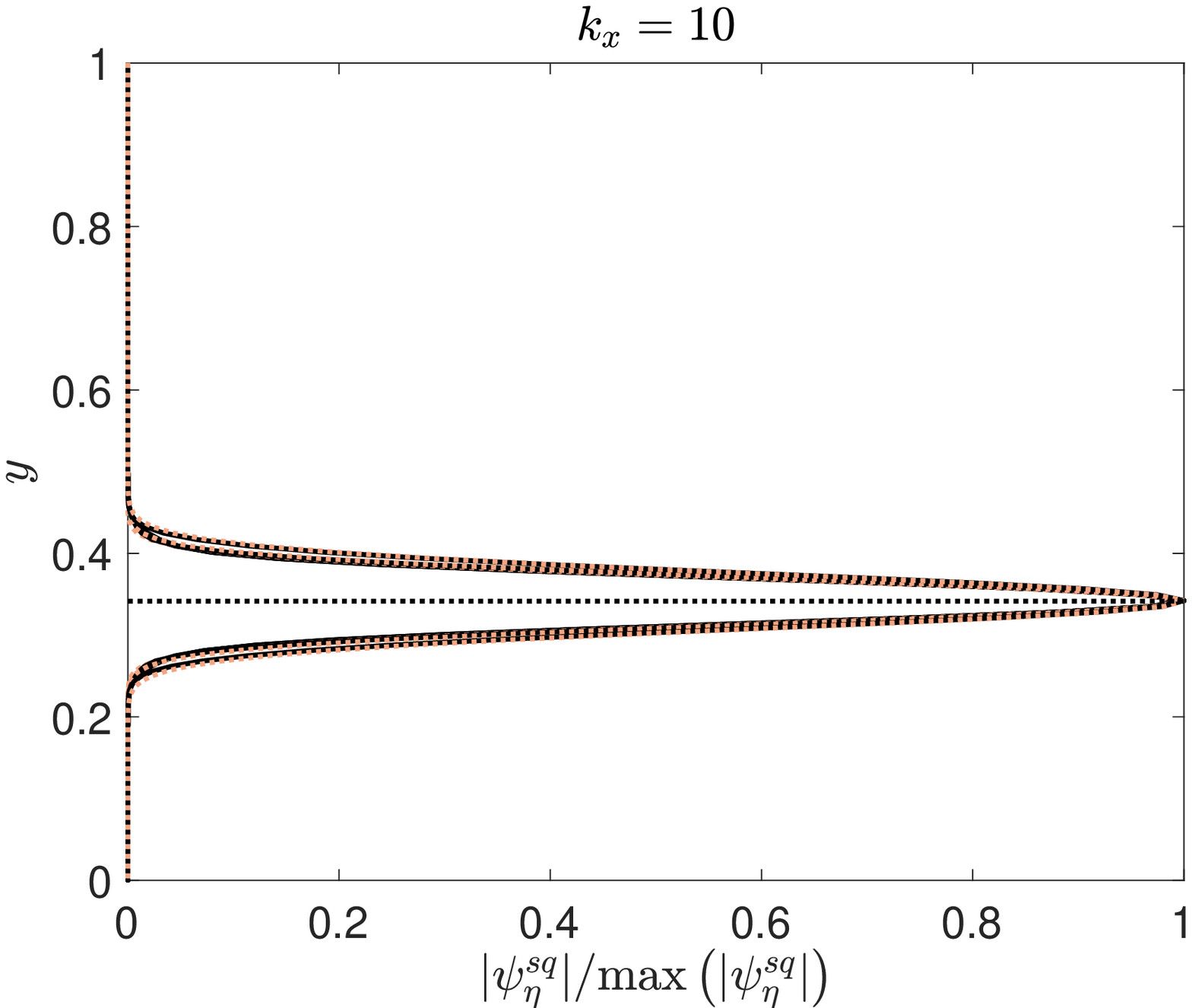}
(d)\includegraphics[width= 0.45\textwidth]{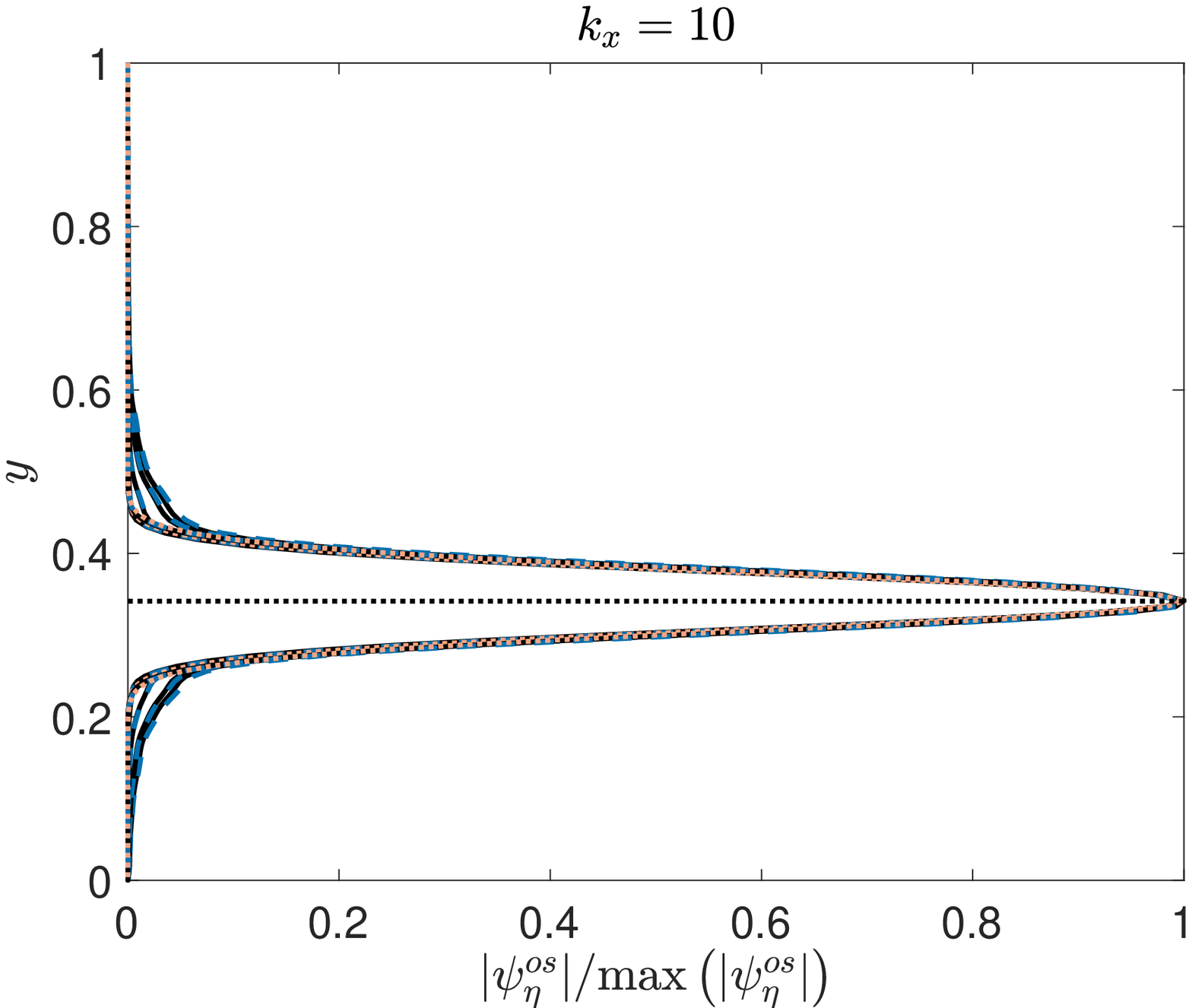}
(e)\includegraphics[width= 0.45\textwidth]{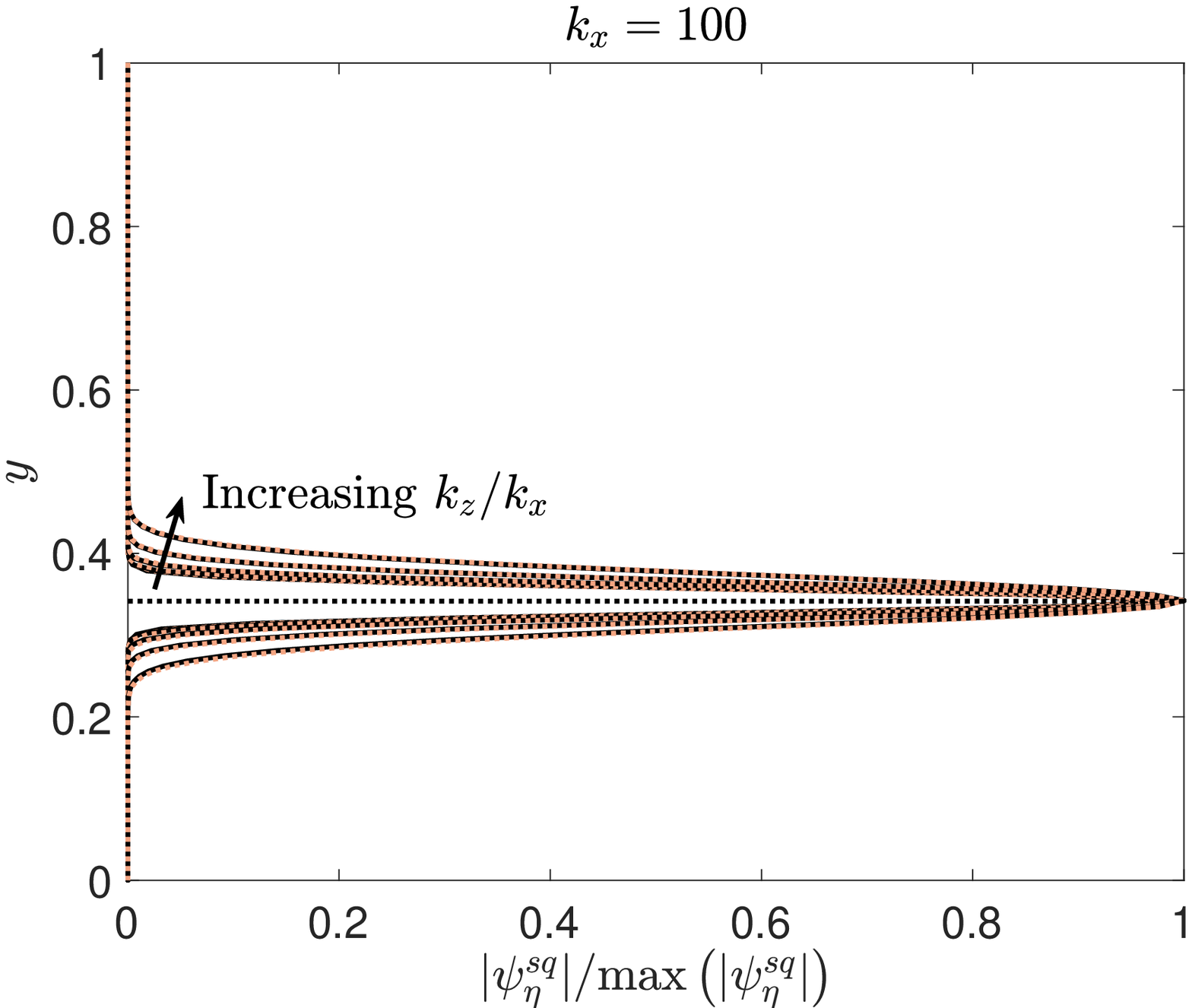}
(f)\includegraphics[width= 0.45\textwidth]{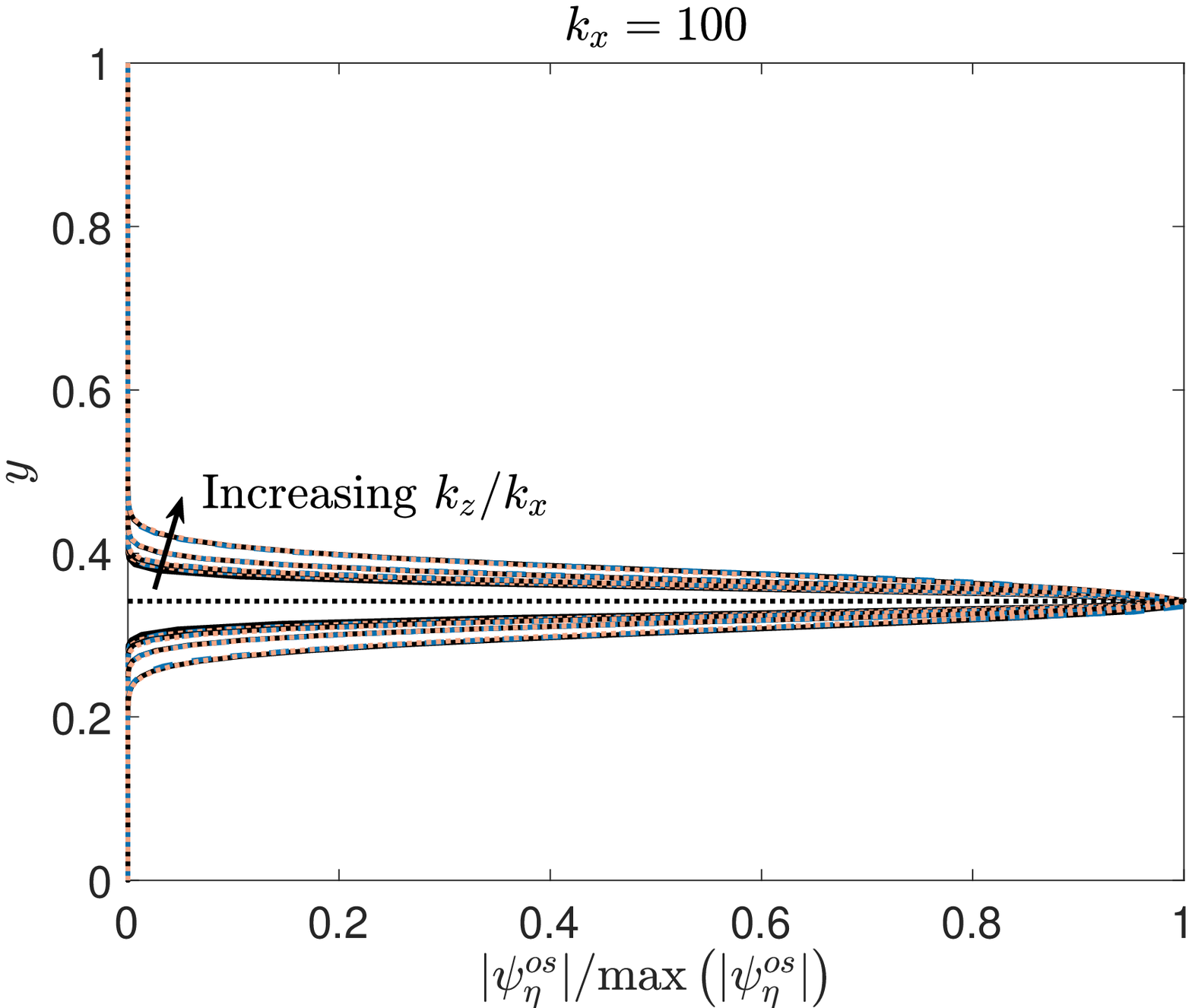}
}
\caption{Comparison between true (numerically computed, solid lines) and predicted (dotted lines) mode amplitudes for various $k_x$, and aspect ratios $k_z/k_x \in \{0.5,1,2,4,8\}$, for a turbulent boundary layer with $Re_\tau = 900$ and $c = 0.8U_\infty$. The black horizontal dotted line indicates the critical layer location. Subplots (a,c,e) show results for the Squire subsystem, while subplots (b,d,f) are for the $\eta$-component of the Navier--Stokes operator. The dashed lines in subplots (b,d,f) also show (numerically computed) mode amplitudes for the Orr-Sommerfeld subsystem.}
\label{fig:ModeShapesBL}
\end{figure}

Figure \ref{fig:abkxBL} compares predicted and fitted mode shape parameters as a function of $k_x$ for various aspect ratios $k_z/k_x$, for a wavespeed $c = 0.8U_\infty$, with predicted and true modes shapes plotted in figure \ref{fig:ModeShapesBL}. We observe the same trends, and similar accuracy in prediction of parameters as for laminar Couette flow (figure \ref{fig:abkxCouette}).

\section{Discussion and conclusions}

This work has presented a method for approximating leading resolvent response modes for quasi-parallel shear-driven flows. This method relies on the assumption that the true mode may be closely approximated by a simple template function, the general form of which can be reasoned from consideration of wavepacket pseudomode theory. 
In essence, the method reduces the space of possible mode shapes from an infinite-dimensional space (which in practice is approximated by a high dimensional space defined by the numerical discretisation) to a two-dimensional family of functions. 
 Once this template function is identified, the optimal shape parameters (which govern the width and phase variation of the mode) may be found as the minimisers of a cost function, which is directly related to the resolvent norm of the underlying operator.  In practice, this amounts to finding the roots of a pair of coupled equations, which may be arranged to be polynomials in the shape parameters. In addition, it is possible to derive differential equations in parameter space that govern the evolution of these optimal shape parameters.  Importantly, this method precludes the need for the formulation and decomposition of discretised linear operators, leading to substantial reduction in computational cost.  The extent of the reduction in computational cost is dependent on the size of the discretisation, and on the extent and resolution of the parameter space (e.g., wavenumbers and temporal frequencies) that one wishes to study. 
  The method may be readily applied to a model operator, as considered in section \ref{sec:SquirePred}, and in the analysis of the Squire operator in sections \ref{sec:OSpred}-\ref{sec:BLpred}. 
 Application for the full Navier--Stokes system relies on additional simplifications to arrive at a scalar differential operator which has a leading response mode (left singular vector)  which approximates the wall-normal vorticity component of the response mode of the Navier--Stokes resolvent operator, as detailed in section \ref{sec:simp}. This simplification is made in three steps. Firstly, the observation that the wall-normal vorticity response is dominated by the Orr-Sommerfeld component (equation \ref{eq:ResolventOSSQ}). Secondly, that this operator is in turn dominated by the vorticity response to velocity forcing, governed by the scalar operator (equation \ref{eq:scalarOS}). Finally, the leading left singular vector of this scalar operator may be approximated by that of the Squire operator furnished with the scalar Laplacian inner product (equation \ref{eq:IPscalarLap}). This results in an operator which may be studied by applying the same techniques as those used for the model/Squire system.  Note that, even without this analysis, study of the Squire operator (for which the associated cost function has a simpler form) typically gives the same qualitative behaviour observed for the full system. In particular, the Squire system obeys many of the same scaling laws of mode shape parameters with wavenumbers and Reynolds numbers. More detailed analysis of scaling and self-similarity properties of resolvent modes in wall-bounded flows are given in \cite{moarref2013channels}.

While the vorticity response mode shapes of the Squire and Orr-Sommerfeld resolvent operators are qualitatively similar, they represent quite different physical phenomena. The Squire resolvent operator describes amplification through forcing in the same component, with an upstream-leaning optimal forcing mode giving a downstream-leaning response mode. This is a manifestation of the classical Orr mechanism. Note in particular that the Squire forcing mode has the same amplitude profile as the response mode, but with an opposite phase variation.  The Orr-Sommerfeld sub-operator on the other hand typically has a leading response mode that is dominated by the wall-normal vorticity component, but is forced primarily by wall-normal velocity, in a manner resembling the lift-up mechanism. Despite this phenomenological difference, we have shown that the shape of the response may be accurately predicted from the Squire operator with a modified inner product.

This work has focused on characterising the shape of the (dominant) wall-normal vorticity component resolvent modes for parallel wall-bounded flows. It is possible that these methods could be extended for application to more complex geometries, and other mode components. 
For most of the cases considered in this work, the cost function had only one local minimum corresponding to a wavepacket mode (i.e., with $b > 0$), with the exception being for negative $\omega_c$ for the model operator considered in section \ref{sec:SquirePred}, as seen in figure \ref{fig:Jcontours}. More complex geometries might result in more complex cost functions, for which more care must be taken to select the true global optimum.  Note that the operators considered here also typically have a large spectral gap between the first and second singular value. Future work could also seek to identify modes corresponding to additional singular values, which would be of particular interest for situations where the operator does not exhibit low-rank behaviour (i.e., there  does not exist a large gap between the leading and second singular values). Similar methods could also be applied to study optimal forcing modes in more detail, and to compute nonlinear forcing terms induced from analytic approximations to wavepacket modes. This would provide an alternative route to study phenomena such as the self-similarity of the nonlinear forcing \citep{sharma2017scaling}. 

In terms of the methodology itself, there are a number of possible refinements that could be investigated. For example, nonlinear terms in the expansion of the mean velocity profile about the critical layer could be retained, and additional terms in the template function (for example, a  term could be added to allow for the phase variation to be cubic in $y-y_c$). 
The former modification might be particularly useful when dealing with mean velocity profiles that have stationary points, such as in channel flow. 
Such additions would lead to more complex cost functions, but the same techniques of analysis should be applicable. 
The accuracy of the assumed form of the modes relies on the mode being far enough away from the wall. Further extensions of the methodology could seek  to explicitly model the effect of the wall.

The authors acknowledge support from the Air Force Office of Scientific Research grant FA9550-16-1-0232 (program manager Ivett Leyva). The authors also thank Peter Schmid, Anthony Leonard, and Kevin Rosenberg for valuable discussions, and Xiaohua Wu for allowing us to use his turbulent boundary layer database.

\bibliographystyle{jfm}
\bibliography{ModeShapeRefs}


\end{document}